\documentclass[reprint, amsmath,amssymb, prb,superscriptaddress,longbibliography]{revtex4-2}

\usepackage{graphicx}
\usepackage{dcolumn}
\usepackage{bm}
\usepackage[usenames,dvipsnames]{xcolor}
\usepackage{hyperref}
\usepackage[inline]{enumitem}
\usepackage{bbold}
\usepackage{soul}
\usepackage{placeins}
\usepackage{amsmath}
\usepackage{amssymb}
\usepackage{physics}
\usepackage{placeins}
\usepackage{lipsum}
\usepackage{ulem}
\usepackage{multirow}
\usepackage{float}

\newcommand{\up}{\uparrow}

\newcommand{\dn}{\downarrow}

\newcommand{\updn}{{\uparrow\downarrow}}

\begin{document}

\title{Fate of poor man's Majoranas in the long Kitaev chain limit}

\author{Melina Luethi$^\dagger$}
\affiliation{Department of Physics, University of Basel, Klingelbergstrasse 82, CH-4056 Basel, Switzerland\\
$^\dagger$These authors contributed equally to this work.}
\author{Henry F. Legg$^\dagger$}
\email{henry.legg@unibas.ch}
\affiliation{Department of Physics, University of Basel, Klingelbergstrasse 82, CH-4056 Basel, Switzerland\\
	$^\dagger$These authors contributed equally to this work.}
\author{Daniel Loss}
\affiliation{Department of Physics, University of Basel, Klingelbergstrasse 82, CH-4056 Basel, Switzerland\\
$^\dagger$These authors contributed equally to this work.}
\author{Jelena Klinovaja}
\affiliation{Department of Physics, University of Basel, Klingelbergstrasse 82, CH-4056 Basel, Switzerland\\
$^\dagger$These authors contributed equally to this work.}

\date{\today}

\begin{abstract}
A minimal Kitaev chain, consisting of two quantum dots connected via a superconductor, can host highly localized near-zero-energy states, known as poor man's Majoranas (PMMs). These states have been proposed as promising candidates to study Majorana bound states (MBSs) in a highly tunable setup. However, it is unclear whether and how PMMs observed in real systems are actually connected to the topological phase of the full Kitaev chain. Here, we study PMMs using a microscopic model and show that, in the long chain limit, not all PMMs are related to topological states. Rather, in long chains, some PMMs evolve into trivial highly localized low-energy states. We provide an 
explanation for the occurrence of these  states and show that there is no clear conductance signature that is able to distinguish PMMs that evolve into true topological states from PMMs that evolve into trivial states. 
\end{abstract}

\maketitle

\section{\label{sec:intro}Introduction}
Majorana bound states (MBSs) are quasiparticles that emerge at the boundaries or in vortices of topological superconductors~\cite{kitaev2001unpaired} and, due to their non-Abelian exchange statistics~\cite{ivanov2001non}, have been proposed as candidates to store and manipulate quantum information in a fault-tolerant way~\cite{kitaev2003fault, nayak2008non, elliot2015colloquium}. 
They were prominently described using the Kitaev chain~\cite{kitaev2001unpaired}, a simple spinless minimal model relying on $p$-wave superconductivity. Nanowires with strong spin-orbit interaction (SOI) proximitized by superconductors have in recent years been prominent systems to realize an effective Kitaev chain~\cite{lutchyn2010majorana, oreg2010helical, stanescu2011majorana, mourik2012signatures, das2012zero, deng2012anomalous, laubscher2021majorana}. However, disorder has hindered the conclusive observation of MBSs in these heterostructures since disorder can result in signatures that mimic MBSs~\cite{kells2012near, lee2012zero, rainis2013towards, roy2013topologically, ptok2017controlling, liu2017andreev, moore2018two, moore2018quantized, reeg2018zero, vuik2019reproducing, stanescu2019robust, woods2019zero, chen2019ubiquitous, awoga2019supercurrent, prada2020andreev, yu2021non, sarma2021disorder, valentini2021nontopological, hess2021local, hess2023trivial, hess2022prevalence}. 
Due to the issues with disorder in nanowires, realizing an effective Kitaev chain as an array of quantum dots (QDs) has been proposed as an alternative platform to host MBSs with a high degree of tunability~\cite{sau2012realizing, leijnse2012parity, fulga2013adaptive, sothmann2013fractional, li2014tunable, su2017andreev, tsintzis2022creating, liu2022tunable, dvir2023realization, sanches2023fractionalization, chunxiao2023fusion, koch2023adversarial, souto2023probing, zatelli2023robust, samuelson2023minimal, tsintzis2023roadmap, pino2024minimal, luna2024fluxtunable, bordin2024signatures, boross2023braiding, liu2024coupling, vilkelis2024fermionic, ezawa2024even, bozkurt2024interaction, vandriel2024crossplatform, pino2024thermodynamics, geier2023fermionparity, haaf2023engineering, kocsis2024strong, cayao2024emergent, miles2024kitaev, cayao2024nonhermitian, liu2024protocol, alvarado2024interplay, svensson2024quantum, gomez2024high, pan2024rabi, pandey2024dynamics, liu2023enhancing, benestad2024machine, luethi2024perfect, ten2024edge}. 
In these proposals, the QDs are separated by superconducting sections, which couple the QDs through elastic cotunneling (ECT) and crossed Andreev reflection (CAR)~\cite{recher2001andreev,lesovik2001electronic,falci2001correlated, bouchiat2002single, feinberg2003andreev, hofstetter2009cooper, herrmann2010carbon}. In such setups, an inhomogeneous magnetic field or SOI can enable effective $p$-wave superconductivity using a conventional $s$-wave superconductor~\cite{sau2012realizing, leijnse2012parity}.

It has been shown that a fully spin-polarized model consisting of only two QDs with a superconductor transmitting ECT and CAR between the QDs is sufficient to obtain states similar to MBSs~\cite{leijnse2012parity}. These states share most of their properties with MBSs found in long Kitaev chains, i.e., they are at zero energy, separated from excited states by a finite gap, and have non-Abelian exchange statistics. However, they exist only at finely tuned points of parameter space, called sweet spots, and therefore they lack topological protection. Due to this lack of protection MBSs in minimal Kitaev chains are often called poor man's Majoranas (PMMs).

Recent demonstrations of control over the strength of ECT and CAR~\cite{liu2022tunable, wang2022singlet, kurtossy2022parallel, dejong2023controllable, bordin2023tunable, wang2023triplet, bordin2023crossed} have spurred great interest in PMMs. Although originally derived using the spinless Kitaev chain, PMMs have also been studied in spinful models.  
It was shown in Ref.~\cite{luethi2024perfect} that sweet spots with perfectly localized zero-energy PMMs can exist in an idealized spinful model, however, reaching these sweet spots requires complete control over many degrees of freedom. In more realistic models and experiments such control is impossible and only imperfect PMMs are possible, i.e., states that are close to zero energy and highly, but not perfectly, localized. 
Nevertheless, the study of imperfect PMMs in minimal Kitaev chains has attracted much attention due to the hope that they are the building blocks of topological states in longer QD chains~\cite{sau2012realizing, leijnse2012parity, dvir2023realization, zatelli2023robust, samuelson2023minimal, tsintzis2023roadmap, pino2024minimal, luna2024fluxtunable, bordin2024signatures, ezawa2024even, miles2024kitaev, liu2024protocol, alvarado2024interplay, svensson2024quantum}.
Furthermore, in recent experiments, conductance signatures associated with PMMs have been found~\cite{dvir2023realization, haaf2023engineering, zatelli2023robust, bordin2024signatures, vandriel2024crossplatform, ten2024edge}. 
It is therefore important to clarify whether imperfect PMMs evolve into topological or trivial states in the long chain limit.

\begin{figure}[t]
	\centering
	\includegraphics[width=\linewidth]{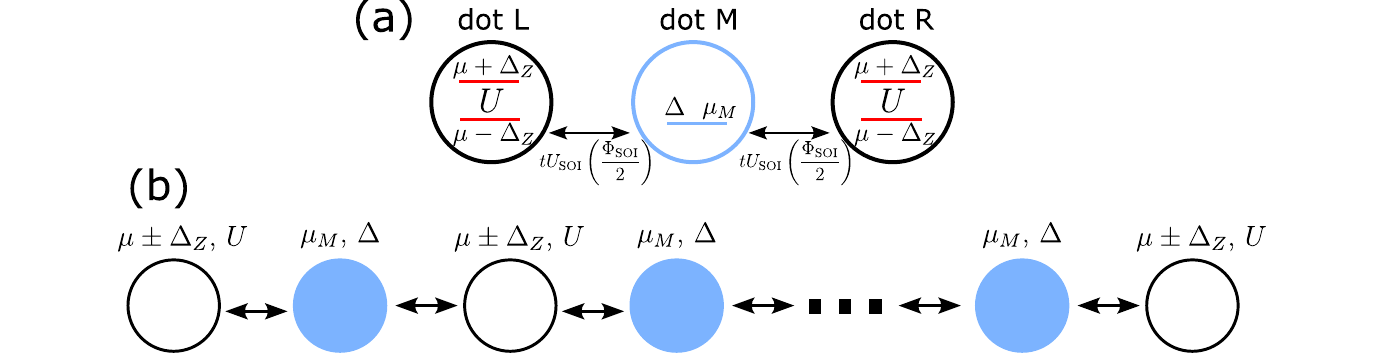}
	\caption{ {\bf Microscopic model of a minimal Kitaev chain and the long chain limit.} 
		(a) Minimal Kitaev chain, consisting of three QDs. The left and right QDs, representing two Kitaev sites, have a chemical potential $\mu$, Zeeman energy $\Delta_Z$, and on-site Coulomb interaction $U$. The middle QD M has a superconducting pairing potential $\Delta$ and, due to the screening by the superconductor, we assume no Zeeman energy or Coulomb interaction. This middle QD models an 
		ABS that transmits the effective couplings between the two outer QDs. The hopping amplitude is $t$ and $U_\mathrm{SOI}(\Phi_\mathrm{SOI}/2)$ characterizes the SOI.
	(b) Long chain limit. Normal QDs are white, they are equivalent to the QDs L and R in (a) and each represents one normal site. The QDs with a pairing potential are blue and they are equivalent to the QD M in~(a).	
}
	\label{fig:setup}
\end{figure}

In this work, we demonstrate that imperfect PMMs in a microscopic model of the minimal Kitaev chain are neither necessary nor sufficient to obtain MBSs in the long chain limit. We explain 
the origin of these imperfect PMMs and demonstrate that there is no clear conductance signature in a minimal Kitaev chain that can distinguish PMMs that evolve into true MBSs from PMMs that evolve into trivial states.

This paper is structured as follows.
The model and all quantities used to characterize PMMs are introduced in Sec.~\ref{sec:model}. The threshold region, which defines imperfect PMMs, is defined in Sec.~\ref{sec:tr}. We study the connection between imperfect PMMs and topology in Sec.~\ref{sec:connection_to_topo}, and their conductance signatures in Sec.~\ref{sec:conductance}. We give an explanation for the origin of false PMMs in Sec.~\ref{sec:origin_false_PMMs}. False PMMs in longer chains are studied in Sec.~\ref{sec:longer_chain}. We conclude in Sec.~\ref{sec:conclusion}. 
For an overview on the different types of PMMs, Appendix~\ref{app:nomenclature} summarizes all PMM definitions.
The rest of the Appendix contains more information on the calculations, further examples of true and false PMMs, and a study of the energy spectrum as the QD chain is extended.

\vspace*{-0.2cm}\section{\label{sec:model}Model}\vspace*{-0.2cm}
To model a minimal Kitaev chain microscopically we consider a system of three QDs, see Fig.~\ref{fig:setup}(a), based on the model introduced in Ref.~\cite{tsintzis2022creating}. The left and right QDs, modeling the two sites of a minimal Kitaev chain, have chemical potentials $\mu_L$ and $\mu_R$, respectively (if not stated otherwise, we assume $\mu_L = \mu_R \equiv \mu$), Zeeman energy $\Delta_Z$, and on-site Coulomb repulsion $U$. 
The middle QD is proximitized by a superconductor and models an Andreev bound state (ABS) that effectively transmits ECT and CAR between the outer QDs~\cite{tsintzis2022creating, liu2022tunable, miles2024kitaev, souto2023probing, liu2023enhancing, zatelli2023robust, luna2024fluxtunable, valentini2024subgap}. 
The middle QD also causes local Andreev reflection (LAR) and renormalizes other parameters of the outer QDs~\cite{luethi2024perfect}. 
This middle dot has a superconducting pairing potential $\Delta$ and chemical potential $\mu_M$. Due to screening by the superconductor we assume that the middle QD has no Zeeman energy and no Coulomb repulsion. We note, however, that it has been shown in Ref.~\cite{tsintzis2022creating} that, even without this assumption, PMMs persist. 
The hopping between the QDs is determined by the hopping amplitude $t$ and the SOI angle $\Phi_\mathrm{SOI}$. The Hamiltonian describing the system is given by
\begin{align}  
	&  H =  \sum_{j=L,R} \left[
	\sum_{\sigma=\up,\dn} 
	(\mu_j + \sigma \Delta_Z) n_{j,\sigma}
	+ U n_{j,\up} n_{j,\dn}  \right]
	\nonumber \\ & 
	\!\!  + \sum_{\sigma=\up,\dn}  \mu_M c_\sigma^\dagger c_\sigma
	+ \Delta (c_\up^\dagger c_\dn^\dagger + c_\dn c_\up)	
	\nonumber \\ &
	\!\! + t \!\!\!\!\! \sum_{\sigma,\sigma'=\up,\dn} \! \left[ U_\mathrm{SOI} \! \left(\frac{\Phi_\mathrm{SOI}}{2}\right)_{\!\!\!\sigma\sigma'} \!\! (
	c_\sigma^\dagger d_{1,\sigma'} \! + \! d_{2,\sigma}^\dagger c_{\sigma'}
	)
	\! + \! \mathrm{H.c.} \right] \!,
	\label{eq:full_hamiltonian}
\end{align}
where $n_{j,\sigma}=d_{j,\sigma}^\dagger d_{j,\sigma}$, $d_{j,\sigma}^\dagger$ ($c_\sigma^\dagger$) creates a particle on dot $j \in \{L,R\}$ (the middle QD M) 
and
the notation $\sigma \Delta_Z$ means $+\Delta_Z$ for $\sigma=\up$ and $-\Delta_Z$ for $\sigma=\dn$.  The chemical potentials are measured with respect to the chemical potential of the superconductor.
Assuming that SOI is perpendicular to the magnetic field, the SOI matrix is 
\begin{align}
U_\mathrm{SOI} \left(\frac{\Phi_\mathrm{SOI}}{2}\right) = \cos \left(\frac{\Phi_\mathrm{SOI}}{2}\right) + i \sin \left(\frac{\Phi_\mathrm{SOI}}{2}\right) \sigma_y,
\end{align}
where $\sigma_y$ is the second Pauli matrix. 
We will use the second quantized form and, if $U=0$, also the BdG form of the Hamiltonian given in Eq.~\eqref{eq:full_hamiltonian} for the calculations in this work, see Appendix~\ref{app:hamiltonians}.

\begin{figure*}
	\centering
	\includegraphics[width=\linewidth]{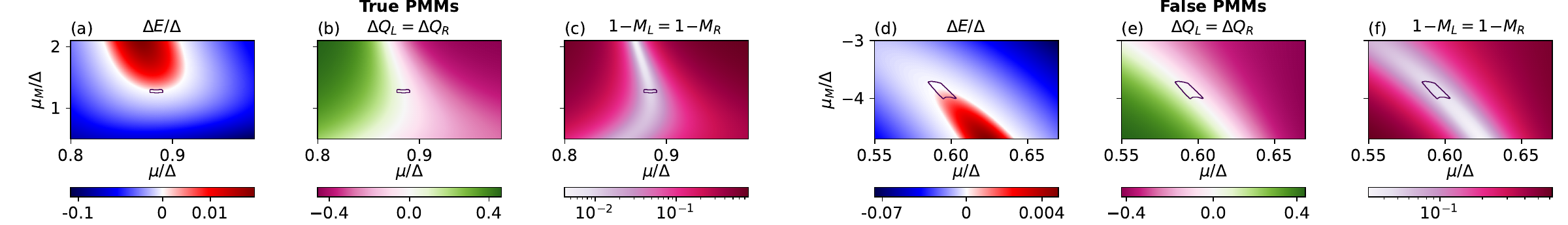}
	\caption{
	{\bf Quality measures for true PMMs versus false PMMs.}
	Energy difference $\Delta E$ [(a) and (d)], charge difference $\Delta Q_j$ [(b) and (e)], and the Majorana polarization $M_j$ [(c) and (f)] for two parameter sets. The area circled in black is the TR.
	Both sets of parameters lead to qualitatively equal results, and both lead to a TR with highly localized near-zero-energy states, i.e., PMMs. However, only (a)--(c) are connected to true MBSs in the long chain limit, whereas (d)--(f) are connected to trivial states in the long chain limit, see Fig.~\ref{fig:energy_vs_delta_z_combined}.
	The parameters for (a)--(c) are $t/\Delta = 0.42$, $\Phi_\mathrm{SOI} = 0.26 \pi$, $U=0$, and $\Delta_Z/\Delta=0.8$. The largest excitation gap in the TR is $E_\mathrm{ex}/\Delta = 0.103$.
	The parameters for (d)--(f) are $t/\Delta=0.99$, $\Phi_\mathrm{SOI}=0.44\pi$, $U=0$, and $\Delta_Z/\Delta=0.8$. The largest excitation gap in the TR is $E_\mathrm{ex}/\Delta = 0.084$. 
	The TR threshold values for all panels are $\Delta E_\mathrm{th}/\Delta=10^{-3}$, $\Delta Q_\mathrm{th}=0.05$, and $M_\mathrm{th}=0.05$.
	}
	\label{fig:energy_pol_charge_combined}
\end{figure*}

The Hamiltonian given in Eq.~\eqref{eq:full_hamiltonian} conserves the total particle number parity. Therefore, the second quantized Hamiltonian is block diagonal. We label the Hamiltonian coupling states with an even (odd) particle number as $H_\mathrm{even}$ ($H_\mathrm{odd}$), its eigenstates as $\ket{\Psi_{a}^{\mathrm{even}}}$ ($\ket{\Psi_{a}^{\mathrm{odd}}}$) and the corresponding eigenvalues as $E_{a}^{\mathrm{even}}$ ($E_{a}^{\mathrm{odd}}$), where $a$ numbers the eigenstates and eigenvalues such that $E_{0}^{\mathrm{even}} \leq E_{1}^{\mathrm{even}} \leq \dots$ ($E_{0}^{\mathrm{odd}} \leq E_{1}^{\mathrm{odd}} \leq \dots$). We introduce the energy difference~\cite{leijnse2012parity}
\begin{equation} \label{eq:definition_dE}
\Delta E = E_{0}^{\mathrm{even}} - E_{0}^{\mathrm{odd}},
\end{equation}
the charge difference on dot $j \in \{L,R\}$~\cite{tsintzis2022creating}
\begin{equation} \label{eq:definition_dQ}
\!\! \Delta Q_j \! = \!\!\! \sum_{\sigma=\up,\dn} \!\!\!
\left(
\bra{\Psi_{0}^{\mathrm{even}}} \! n_{j,\sigma} \! \ket{\Psi_{0}^{\mathrm{even}}}
\! - \! \bra{\Psi_{0}^{\mathrm{odd}}} \! n_{j,\sigma} \! \ket{\Psi_{0}^{\mathrm{odd}}}
\right),
\end{equation}
the Majorana polarization (MP) on dot $j \in \{L,R\}$~\cite{tsintzis2022creating}
\begin{subequations} \label{eq:definition_M}
\begin{align}
M_j =& \frac{
\left| \sum_{\sigma=\up,\dn} ( w_{+,\sigma}^2 - w_{-,\sigma}^2 ) \right|
}{
\sum_{\sigma=\up,\dn} ( w_{+,\sigma}^2 + w_{-,\sigma}^2 )
} ,  \\
w_{\pm,\sigma} =& \bra{\Psi_{0}^{\mathrm{odd}}} (d_{j,\sigma} \pm d_{j,\sigma}^\dagger ) \ket{\Psi_{0}^{\mathrm{even}}}, 
\end{align}
\end{subequations} 
and the excitation gap~\cite{luethi2024perfect}
\begin{equation} \label{eq:definition_gap}
E_\mathrm{ex} = \min \{
E_{1}^{\mathrm{even}} - E_{0}^{\mathrm{even}}, 
E_{1}^{\mathrm{odd}} - E_{0}^{\mathrm{odd}}
\} .
\end{equation}

\section{\label{sec:tr}Threshold region}
A sweet spot is a point in parameter space that fulfills $\Delta E=0$, $\Delta Q_L = \Delta Q_R = 0$, $M_L=M_R=1$, and $E_\mathrm{ex} > 0$. The corresponding perfect PMMs to these sweet spots are perfectly localized and at zero energy.
It was shown in Ref.~\cite{luethi2024perfect} that, for the microscopic Hamiltonian of a minimal Kitaev chain in Eq.~\eqref{eq:full_hamiltonian}, no sweet spots could be found. Instead, the notion of a threshold region (TR) was introduced. It is a region of parameter space where the PMM conditions are approximately realized, i.e.,
\begin{align} \label{eq:definition_ROT}
&|\Delta E| < \Delta E_\mathrm{th}
\text{ and }
E_\mathrm{ex} > E_\mathrm{ex,th} 
\nonumber \\ &
\text{ and }
|\Delta Q_L| < \Delta Q _\mathrm{th}
\text{ and }
|\Delta Q_R| < \Delta Q _\mathrm{th}
\nonumber \\ &
\text{ and }
M_L > 1 - M_\mathrm{th}
\text{ and }
M_R > 1 - M_\mathrm{th},
\end{align}
where $\Delta E_\mathrm{th}$, $E_\mathrm{ex,th}$,  $\Delta Q _\mathrm{th}$, and $M_\mathrm{th}$ are threshold values that need to be chosen. 
The corresponding imperfect PMMs are highly, but not perfectly, localized near-zero-energy states. This highlights that it is necessary to have a degree of arbitrariness in the definition of PMMs in realistic systems that will always depend on the threshold values chosen in Eq.~\eqref{eq:definition_ROT}.

\section{\label{sec:connection_to_topo}Connection to topology}
Although the requirement to define a TR creates some ambiguity, it is often implicitly assumed that imperfect PMMs in more realistic microscopic models, such as that shown in Fig.~\ref{fig:setup}(a), will still evolve into topological MBSs in the long chain limit, i.e., in the system shown in  Fig.~\ref{fig:setup}(b). In the following, we demonstrate that this is not necessarily the case and show that none of the previously used experimental signatures for PMMs can clearly distinguish between PMMs that would evolve into topological states, from those that evolve into highly localized trivial states.

\begin{figure}
	\centering
	\includegraphics[width=\linewidth]{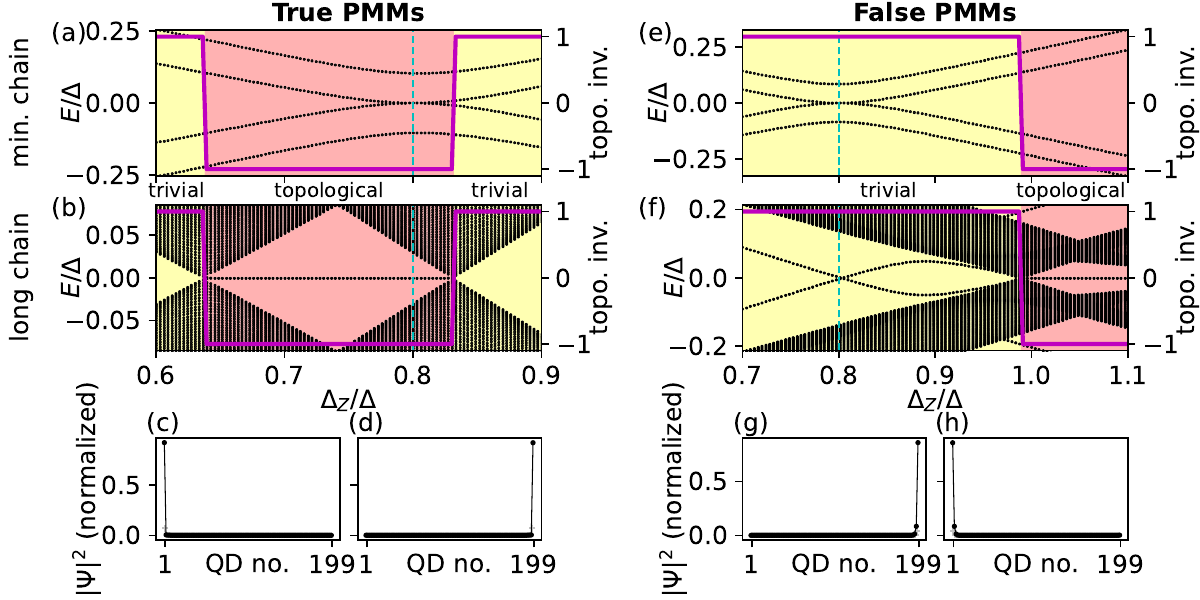}
	\caption{
		{\bf True and false PMMs in the long chain limit.} Energy spectrum $E$ (black dots) as a function of the Zeeman energy $\Delta_Z$ for the minimal chain [(a) and~(e)] and in the long chain limit [(b) and~(f)].
		The magenta line indicates the topological invariant, calculated in the infinite chain limit, see Appendix~\ref{app:topo_inv}. A red (yellow) background in the spectrum indicates that the system is topological (trivial).
		(c) and (d): Probability density of the two zero-energy states at $\Delta_Z/\Delta=0.8$ from (b). The weight on the normal (superconducting) QDs are indicated by black dots (gray crosses). These states are MBSs.
		(g) and (h): Probability density of two zero-energy states at $\Delta_Z/\Delta = 0.8$ in (f). Although these states are at zero-energy and mainly localized on the first and last QD, they are trivial and not related to MBSs.
		Hence, although in the minimal chain [(a) and (e)], both sets of parameters result in qualitatively similar results, only one of the parameter sets is connected to a topological state in the long chain limit. The chain of (f) could be brought into the topological phase by setting $\Delta_Z/\Delta = 1.05$, but then the states in the minimal chain are no longer PMMs, as their energy is nonzero, see (e). 
		The parameters for (a)--(d) are as for (a)--(c) in Fig.~\ref{fig:energy_pol_charge_combined} with $\mu/\Delta = 0.884$ and $\mu_M/\Delta = 1.275$. 
		The parameters for (e)--(h) are as for (d)--(f) in Fig.~\ref{fig:energy_pol_charge_combined} with $\mu/\Delta = 0.593$ and $\mu_M/\Delta = -3.836$. 	
		The long chain consists of 100 normal QDs and 99 QDs with a pairing potential.
	}
	\label{fig:energy_vs_delta_z_combined}
\end{figure}

\begin{figure*}
	\centering
	\includegraphics[width=\linewidth]{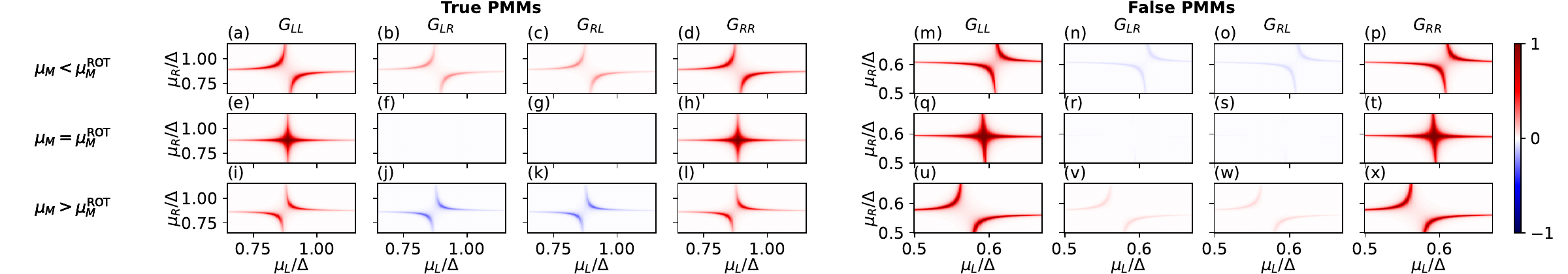}
	\caption{ {\bf Zero-bias conductance signatures of true PMMs versus false PMMs.}
		Normalized zero-energy conductance $G_{LL}$ [(a), (e), (i), (m), (q), and (u)], $G_{LR}$ [(b), (f), (j), (n), (r), and (v)], $G_{RL}$ [(c), (g), (k), (o), (s), and (w)], and $G_{RR}$ [(d), (h), (l), (p), (t), and (x)], as $\mu_L$ and $\mu_R$ are varied independently, for different values of $\mu_M$. For each setup all conductance values are normalized with respect to the maximum value.
		In (a)--(l) the same parameters as for (a)--(c) in Fig.~\ref{fig:energy_pol_charge_combined} are used, with $\mu_M/\Delta = 0.975$ for (a)--(d), $\mu_M/\Delta = 1.275$ (corresponding to the value for the TR) for (e)--(h), and $\mu_M/\Delta = 1.975$ for (i)--(l). The coupling to the leads is $t_l/\Delta = 0.005$.
		In (m)--(x) the same parameters as for (d)--(f) in Fig.~\ref{fig:energy_pol_charge_combined} are used, with $\mu_M/\Delta = -4.236$ for (m)--(p), $\mu_M/\Delta = -3.836$ (corresponding to the value for the TR) for (q)--(t), and $\mu_M/\Delta = -3.336$ for (u)--(x). The coupling amplitude to the leads is $t_l/\Delta = 0.001$.
		Although both parameter sets give qualitatively similar results, only the parameters for (a)--(l) are related to true MBSs in the long chain limit, whereas (m)--(x) are related to trivial states.	}
	\label{fig:zero_energy_conductance_combined}
\end{figure*}

For instance, in Fig.~\ref{fig:energy_pol_charge_combined}, we consider two sets of parameters, both result in a TR and therefore corresponding to imperfect PMMs. In the minimal Kitaev chain, the ground state energy difference $\Delta E$, the charge difference $\Delta Q_j$, and the MP $M_j$ (all shown in Fig.~\ref{fig:energy_pol_charge_combined}) give qualitatively similar results for both parameter sets. 
Since the topological character of these states cannot be judged in a minimal chain, we extend the setup to a long uniform chain, keeping all parameters unchanged. We observe that the PMMs of Figs.~\ref{fig:energy_pol_charge_combined}(a)--\ref{fig:energy_pol_charge_combined}(c) [\ref{fig:energy_pol_charge_combined}(d)--\ref{fig:energy_pol_charge_combined}(f)] evolve into MBSs (trivial states), see Fig.~\ref{fig:energy_vs_delta_z_combined}, and verify this in the infinite chain limit using the topological invariant $Q_{\mathbb{Z}_2}$, see Appendix~\ref{app:topo_inv}. If $Q_{\mathbb{Z}_2}=1$ ($Q_{\mathbb{Z}_2}=-1$), then the system is trivial (topological). The highly localized trivial state is associated with two doubly degenerate states that split from the bulk states and cross zero energy in the long chain limit. We emphasize that these states become degenerate only in the long chain limit, i.e., in short chains the excitation energy is finite. 
We call the PMMs that evolve into topological (trivial) states in the long chain limit true PMMs (false PMMs). 
False PMMs are unrelated to MBSs in the sense that they do not act as the minimal building block for MBSs, i.e., by extending the minimal chain, the resulting state is trivial. One could tune parameters to bring the long chain into the topological phase, e.g., set $\Delta_Z/\Delta=1.05$ in Fig.~\ref{fig:energy_vs_delta_z_combined}(f). In this case, however, the corresponding states in the minimal chain are no longer PMMs, as their energy is nonzero, see Fig.~\ref{fig:energy_vs_delta_z_combined}(e). Therefore, a PMM is neither necessary nor sufficient to ensure that the state evolves into an MBS in the long chain limit.
We note that we set $U=0$ here, but false PMMs also exist for finite $U$, see Appendix~\ref{app:examples}.

\section{\label{sec:conductance}Conductance signatures}
Zero-bias conductance is often used as a signature to search for PMMs in a minimal Kitaev chain~\cite{dvir2023realization, zatelli2023robust, haaf2023engineering, vandriel2024crossplatform, bordin2024signatures}. 
We use the Python package Kwant~\cite{groth2014kwant} to calculate the conductance $G_{\alpha\beta} = d I_\alpha / d V_\beta$, attaching leads to the two outer QDs and introducing the coupling $t_l$ between the QDs and the leads.
Varying $\mu_L$ and $\mu_R$, for different values of $\mu_M$, we expect a clear signature in the TR associated with PMMs. Within the TR, there is a crossing of the conductance at $\mu_L=\mu_R$. As $\mu_M$ is varied away from its value in the TR, the zero-energy conductance exhibits an anti-crossing and the nonlocal conductance changes its sign~\cite{tsintzis2022creating, dvir2023realization, zatelli2023robust, haaf2023engineering}. 
We find this signature for both parameter sets, see Fig.~\ref{fig:zero_energy_conductance_combined}. Therefore,  zero-energy conductance cannot clearly distinguish true from false PMMs.

\begin{figure}
	\centering
	\includegraphics[width=\linewidth]{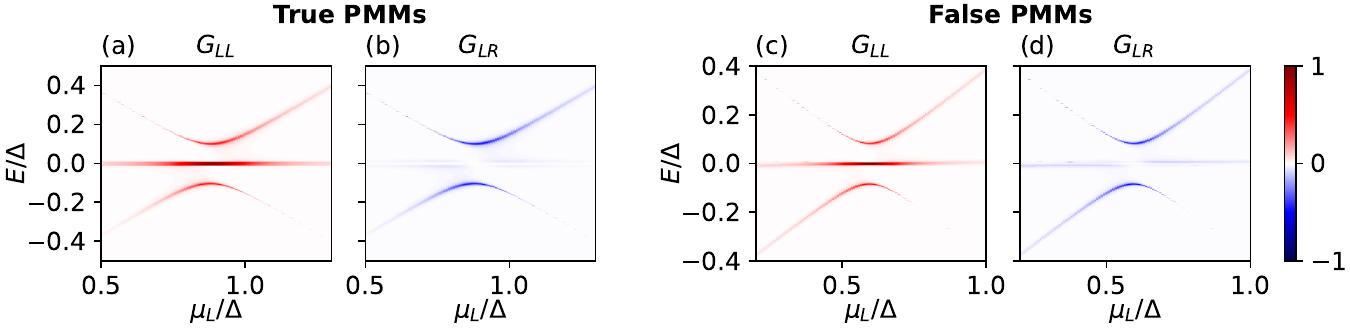}
	\caption{ {\bf Finite-energy conductance of true PMMs versus false PMMs.}
	Normalized conductance at energy $E$, as the chemical potential of the left QD, $\mu_L$, is varied, while the chemical potential of the right QD remains at its value in the TR. (a) and (c) show the local conductance $G_{LL}$, (b) and (d) show the nonlocal conductance $G_{LR}$. 
	For each setup, all conductance values are normalized with respect to the maximum value.
	The parameters for (a) and (b) are the same as for (a)--(c) in Fig.~\ref{fig:energy_pol_charge_combined}, with $t_l/\Delta = 0.01$. The parameters for (c) and (d) are the same as for (d)--(f) in Fig.~\ref{fig:energy_pol_charge_combined}, with $t_l/\Delta = 0.006$.
	Both sets of parameters result in qualitatively similar results, but only (a) and (b) are related to true MBSs in the long chain limit, whereas (c) and (d) are related to trivial states in a long chain.
	}
	\label{fig:finite_energy_conductance_combined}
\end{figure}

Another signature, which has been used to identify PMMs, is the finite-energy conductance when tuning the chemical potential of only one, e.g., the left, QD away from the TR~\cite{dvir2023realization, haaf2023engineering, bozkurt2024interaction}. 
However, we again find that there is no clear qualitative difference in this signature between a false and a true PMM, see Fig.~\ref{fig:finite_energy_conductance_combined}. In Appendix~\ref{app:examples}, we also show finite-energy conductance results when both QDs are tuned away from the TR. This, too, is inadequate to distinguish false from true PMMs.

\section{\label{sec:origin_false_PMMs}Origin of false PMMs}
As shown in Fig.~\ref{fig:energy_vs_delta_z_combined}, in the long chain limit,  the false PMMs discussed here are related to degenerate energy levels that are split from the bulk states. These false PMMs can be associated with localized zero-energy states in a system consisting of only two QDs, see Appendix~\ref{app:analytics}. In particular, these false PMMs arise because the first and last normal QDs of a chain couple directly to only one superconducting QD, but the bulk normal QDs couple directly to two superconducting QDs. 

Using the microscopic model defined in Eq.~\eqref{eq:full_hamiltonian}
we have found many examples of false PMMs, see Appendix~\ref{app:examples}. Most of these false PMMs are related to trivial zero-energy states in the long chain limit, which was our focus here. However, some false PMMs are not related to zero-energy states in the long chain limit. 
Furthermore, it might seem that starting and ending the long chain with superconducting QDs makes the outer normal QDs equivalent to the bulk normal QDs. This, however, is not the case, and we still find that false PMMs exist in this modified model, see Appendix~\ref{app:SNSNS}.

\section{\label{sec:longer_chain}False PMMs in longer chains}
Although we focused here on minimal chains with just two QDs, we also find that chains consisting of a few QDs~\cite{bordin2024signatures, ten2024edge} can host Majorana-like states, which we call scaled PMMs, that do not evolve into MBSs in the long chain limit, see Appendix~\ref{app:scaled_pmms}. This demonstrates that the generalization of the PMM concept to finite chains also does not guarantee that these states will eventually evolve into MBSs in the long chain limit, even in setups where the chain can be uniformly extended.

\section{\label{sec:conclusion}Conclusion}
We have demonstrated that the observation of PMMs in minimal Kitaev chains is neither necessary nor sufficient to ensure that these states evolve into MBSs in longer (uniform) chains and conductance signatures so far associated with such states can also originate from false PMMs, i.e., highly localized near-zero-energy states in the minimal chain that evolve into trivial states in the long chain limit.
One possibility to distinguish true from false PMMs might be the study of excited states in a longer chain, see Appendix~\ref{app:spectrum_vs_N}. However, 
throughout we assumed uniform parameters in the long chain limit and it is important to emphasize that this is a theoretical tool to determine the nature of PMMs. In realistic experimental chains, there is no reason to assume uniform parameters. For instance, although gates can tune chemical potentials, it is not expected that other parameters, such as $g$ factor, SOI strength and direction, on-site $U$, and many more, can be uniform, which could cause problems with the identification of MBSs in longer chains~\cite{hess2022prevalence, hess2023trivial, legg2024reply}.
It is therefore interesting to consider if true and false PMMs can be distinguished before a long uniform chain is assembled. 

\begin{acknowledgments}
This work was supported by the Swiss National Science Foundation and NCCR SPIN (Grant No. 51NF40-180604). This project received funding from the European Union’s Horizon 2020 research and innovation program (ERC Starting Grant, Grant Agreement No. 757725).  H.F.L. acknowledges support by the Georg H. Endress Foundation. 
\end{acknowledgments}

\FloatBarrier

\appendix

\begin{table*}
	\centering
	\begin{tabular}{ccccccc}
		\hline\hline
		& Perfect PMMs & Imperfect PMMs & Scaled PMMs & True PMMs &  False PMMs & MBS \\
		\hline 
		\multirow{2}{*}{Two-site chain }& $\Delta E = 0$, $\Delta Q_{L/R} = 0$, & \multirow{2}{*}{Eq.~\eqref{eq:definition_ROT}}  & \multirow{2}{*}{-}  &  \multirow{2}{*}{Eq.~\eqref{eq:definition_ROT}}  & \multirow{2}{*}{Eq.~\eqref{eq:definition_ROT}} & \multirow{2}{*}{-} \\
		& $M_{L/R}=1$, $E_\mathrm{ex}>0$ &  &  & &  &  \\
		Few-site chain & - & - & Eq.~\eqref{eq:definition_ROT} & - & - & - \\
		Long chain & - & - & - & topological & trivial & topological \\
		\hline\hline
	\end{tabular}
	\caption{
		Summary of the terminology used to describe states in this work. For each type of state (perfect, imperfect, scaled, true, and false PMMs, and MBS) we give the behavior of the state in the two-site chain (minimal chain), in a chain consisting of few sites, and in the long chain (hundreds of sites). The entry ``Eq.~\eqref{eq:definition_ROT}'' indicates that the state must satisfy the TR condition defined in Eq.~\eqref{eq:definition_ROT}. The entry ``-'' means that the behavior of the state in the corresponding system is irrelevant for its definition.   
	}
	\label{tab:nomenclature}
\end{table*}

\section{\label{app:nomenclature}Terminology of states}
	For reference, we summarize the terminology used to describe the different types of states defined in this work in Table~\ref{tab:nomenclature}.

\section{\label{app:hamiltonians}Hamiltonians used for the calculations}
In this Appendix, we present all Hamiltonians used for this work. They are all based on Eq.~\eqref{eq:full_hamiltonian}.

\subsection{\label{app:second_quant_ham}Second quantized form of the Hamiltonian}
Since the total electron parity is conserved by the Hamiltonian given in Eq.~\eqref{eq:full_hamiltonian}, we consider the total even and total odd parity sectors separately. For the even parity sector, the basis is
\begin{widetext}
\begin{align} \label{eq:even_basis}
	&\Psi_\mathrm{even} = \big(
	\ket{0, 0, 0},
	\ket{\updn, 0, 0},
	\ket{\up, \up, 0},
	\ket{\dn, \up, 0},
	\ket{\up, \dn, 0},
	\ket{\dn, \dn, 0},
	\ket{0, \updn, 0},
	\ket{\updn, \updn, 0},
	\ket{\up, 0, \up},
	\ket{\dn, 0, \up},
	\ket{0, \up, \up},
	\nonumber \\ &
	\ket{\updn, \up, \up},
	\ket{0, \dn, \up},
	\ket{\updn, \dn, \up},
	\ket{\up, \updn, \up},
	\ket{\dn, \updn, \up},
	\ket{\up, 0, \dn},
	\ket{\dn, 0, \dn},
	\ket{0, \up, \dn},
	\ket{\updn, \up, \dn},
	\ket{0, \dn, \dn},
	\ket{\updn, \dn, \dn},
	\ket{\up, \updn, \dn},
	\nonumber \\ &
	\ket{\dn, \updn, \dn},
	\ket{0, 0, \updn},
	\ket{\updn, 0, \updn},
	\ket{\up, \up, \updn},
	\ket{\dn, \up, \updn},
	\ket{\up, \dn, \updn},
	\ket{\dn, \dn, \updn},
	\ket{0, \updn, \updn},
	\ket{\updn, \updn, \updn}
	\big)
\end{align}
and for the odd parity sector, the basis is
\begin{align} \label{eq:odd_basis}
	&\Psi_\mathrm{odd} = \big(
	\ket{\up, 0, 0},
	\ket{\dn, 0, 0},
	\ket{0, \up, 0},
	\ket{\updn, \up, 0},
	\ket{0, \dn, 0},
	\ket{\updn, \dn, 0},
	\ket{\up, \updn, 0},
	\ket{\dn, \updn, 0},
	\ket{0, 0, \up},
	\ket{\updn, 0, \up},
	\ket{\up, \up, \up},
	\nonumber \\ &
	\ket{\dn, \up, \up},
	\ket{\up, \dn, \up},
	\ket{\dn, \dn, \up},
	\ket{0, \updn, \up},
	\ket{\updn, \updn, \up},
	\ket{0, 0, \dn},
	\ket{\updn, 0, \dn},
	\ket{\up, \up, \dn},
	\ket{\dn, \up, \dn},
	\ket{\up, \dn, \dn},
	\ket{\dn, \dn, \dn},
	\ket{0, \updn, \dn},
	\nonumber \\ &
	\ket{\updn, \updn, \dn},
	\ket{\up, 0, \updn},
	\ket{\dn, 0, \updn},
	\ket{0, \up, \updn},
	\ket{\updn, \up, \updn},
	\ket{0, \dn, \updn},
	\ket{\updn, \dn, \updn},
	\ket{\up, \updn, \updn},
	\ket{\dn, \updn, \updn}
	\big).
\end{align}
\end{widetext}
In total, there are $4^3=64$ states, thus $32$ states per parity sector. The corresponding Hamiltonians $\mathcal{H}_\mathrm{even}$ and $\mathcal{H}_\mathrm{odd}$ are thus $32 \times 32$ matrices and we do not give their explicit form here. However, their matrix elements $(\mathcal{H}_\mathrm{even/odd})_{ij}$ are calculated as follows:
\begin{subequations} \label{eqSM:definition_h_even_odd}
	\begin{align}
		(\mathcal{H}_\mathrm{even})_{ij} &= \bra{\Psi_{\mathrm{even},i}} H \ket{\Psi_{\mathrm{even},j}}, \\
		(\mathcal{H}_\mathrm{odd})_{ij} &= \bra{\Psi_{\mathrm{odd},i}} H \ket{\Psi_{\mathrm{odd},j}} ,
	\end{align}
\end{subequations}
where $H$ is the Hamiltonian defined in Eq.~\eqref{eq:full_hamiltonian} and $\Psi_{\mathrm{even},i}$ ($\Psi_{\mathrm{odd},i}$) is the $i$th element of the even (odd) basis defined in Eq.~\eqref{eq:even_basis} [Eq.~\eqref{eq:odd_basis}].
The eigenvectors of the matrices defined in Eq.~\eqref{eqSM:definition_h_even_odd} are labeled as $\ket{\Psi_{a}^{\mathrm{even}}}$ and $\ket{\Psi_{a}^{\mathrm{odd}}}$, respectively, and the corresponding eigenvalues are $E_{a}^{\mathrm{even}}$ and $E_{a}^{\mathrm{odd}}$, where $a$ numbers the eigenstates and eigenvalues. 
For short chains, the eigenvectors and eigenvalues are calculated using exact diagonalization. For the long chain limit, the approach of the density matrix renormalization group (DMRG)~\cite{steven1992density, steven1993density, schollwock2005density} is used.
The eigenvectors and eigenvalues are used to calculate $\Delta E$ [defined in Eq.~\eqref{eq:definition_dE}], $\Delta Q_{L/R}$ [defined in Eq.~\eqref{eq:definition_dQ}], $M_{L/R}$ [defined in Eq.~\eqref{eq:definition_M}], and $E_\mathrm{ex}$ [defined in Eq.~\eqref{eq:definition_gap}].

\subsection{\label{app:BdG_ham}BdG Hamiltonian}
If $U=0$, then one can work with the BdG form of the Hamiltonian given in Eq.~\eqref{eq:full_hamiltonian}. This form can also be trivially extended to the long chain limit with $M$ sites [$M$ must be odd, $(M+1)/2$ of these sites are normal, $(M-1)/2$ are superconducting sites]. Introducing the Nambu basis
\begin{equation} \label{eq:nambu_basis}
	\Psi_j = (c_{j,\up}, c_{j,\dn}, c_{j,\up}^\dagger, c_{j,\dn}^\dagger ),
\end{equation}
the Hamiltonian is given by
\begin{align}
	& H_\mathrm{BdG} = 
	\frac{1}{2} \sum_{j=1}^M 
	\Psi_j^\dagger [ 
	( \mu_j + \Delta_{Z,j} \sigma_z ) \tau_z
	+ \Delta_j \sigma_y \tau_y ] \Psi_j
	\nonumber \\&\quad 
	+ t \sum_{j=1}^{M-1}
	\Psi_{j+1}^\dagger \left[
	\cos \left(\frac{\Phi_\mathrm{SOI}}{2}\right) 
	+ i \sin \left(\frac{\Phi_\mathrm{SOI}}{2}\right) \sigma_y 
	\right] \tau_z \Psi_j
	\nonumber \\&\quad 
	+ \mathrm{H.c.},
\end{align}
where the chemical potential, Zeeman energy, and superconducting gap are position dependent as follows:
\begin{align}
	\mu_j =&
	\begin{cases}
		\mu & \mathrm{if} \, j \,\mathrm{mod}\, 2 = 1, \\
		\mu_M & \mathrm{if} \, j \,\mathrm{mod}\, 2 = 0 ,
	\end{cases} \\
	\Delta_{Z,j} =&
	\begin{cases}
		\Delta_Z & \mathrm{if} \, j \,\mathrm{mod}\, 2 = 1, \\
		0 & \mathrm{if} \, j \,\mathrm{mod}\, 2 = 0 ,
	\end{cases} \\
	\Delta_j =&
	\begin{cases}
		0 & \mathrm{if} \, j \,\mathrm{mod}\, 2 = 1 ,\\
		\Delta & \mathrm{if} \, j \,\mathrm{mod}\, 2 = 0 ,
	\end{cases} 
\end{align}
where $\mu$, $\mu_M$, $\Delta_Z$, and $\Delta$ are as defined in the main text.
If $U=0$, then the BdG Hamiltonian is used to calculate the long chain limit (shown, e.g., in Fig.~\ref{fig:energy_vs_delta_z_combined}), topological invariants (see Appendix~\ref{app:topo_inv}), and all conductance data (shown, e.g., in Figs.~\ref{fig:zero_energy_conductance_combined} and~\ref{fig:finite_energy_conductance_combined}).

\section{\label{app:topo_inv}Topological invariant}
To calculate the topological invariant, one needs to consider an infinitely long chain. This is possible only for the case $U=0$. The unit cell consist of a normal QD [a white dot in Fig.~\ref{fig:setup}(b)] and a superconducting QD [a gray dot in Fig.~\ref{fig:setup}(b)]. The basis is thus given by
\begin{equation}
	\Psi_k = (
	d_{\up k},
	d_{\dn k},
	d_{\up -k}^\dagger,
	d_{\dn -k}^\dagger,
	c_{\up k},
	c_{\dn k},
	c_{\up -k}^\dagger,
	c_{\dn -k}^\dagger
	)^T \!\!\! ,
\end{equation}
where $d_{\sigma k}^\dagger$ ($c_{\sigma k}^\dagger$) creates a particle of spin $\sigma$ and momentum $k$ on the normal (superconducting) QD.
The infinite chain Hamiltonian in momentum space is given by
\begin{subequations}
	\begin{align} 
		&H_\mathrm{inf}(k) = \Psi_k^\dagger \mathcal{H}_\mathrm{inf}(k) \Psi_k ,\\
		&\mathcal{H}_\mathrm{inf}(k) =
		(\mu + \Delta_Z \sigma_z ) \tau_z \frac{\eta_0 + \eta_z}{2}
		+ \mu_M \tau_z \frac{\eta_0 - \eta_z}{2}
		\nonumber \\ &
		\quad + t \cos \left(\frac{\Phi_\mathrm{SOI}}{2} \right) \tau_z \{ \eta_x [1 + \cos(ka)] + \eta_y \sin(ka)\}
		\nonumber \\ &
		\quad - t \sin \left( \frac{\Phi_\mathrm{SOI}}{2} \right) \sigma_y \tau_z \{\eta_x \sin(ka) + \eta_y [1- \cos (ka) ]\}
		\nonumber \\ &
		\quad - \Delta \sigma_y \tau_y \frac{\eta_0 - \eta_z}{2},
	\end{align}
\end{subequations}
where $\sigma_\alpha$ are the Pauli matrices acting in spin space, $\tau_\alpha$ the Pauli matrices acting in particle-hole space, $\eta_\alpha$ the Pauli matrices acting in the first QD-second QD space, $k$ is the momentum, and $a$ is the unit cell spacing. The topological invariant $Q_{\mathbb{Z}_2}$ is then given by~\cite{chiu2016classification}
\begin{equation} \label{eq:topo_inv}
	Q_{\mathbb{Z}_2} = \mathrm{sgn} \{
	\mathrm{Pf}[H_\mathrm{inf}(ka=0)] \mathrm{Pf}[H_\mathrm{inf}(ka=\pi)]
	\}.
\end{equation}
If $Q_{\mathbb{Z}_2}=1$, then the system is trivial and if $Q_{\mathbb{Z}_2}=-1$, then the system is topological.

\subsection{\label{app:majorana_number}Majorana number}
In Ref.~\cite{kitaev2001unpaired}, the Majorana number $\mathcal{M}(H) = \pm 1$ was introduced as
\begin{equation} \label{eq:majorana_number}
	P[
	H(L_1 + L_2)
	]
	= \mathcal{M} (H)
	P [
	H (L_1)
	]
	P [
	H (L_2)
	],
\end{equation}
where $P[H(L)]$ is the particle number parity of the ground state of the Hamiltonian $H(L)$ that describes a system of length $L$ with periodic boundary conditions. The length $L$ should be big compared to the MBS localization length. A Majorana number  $\mathcal{M}(H) = -1$ ($+1$) means that the system is in the topological (trivial) phase. It was also shown in Ref.~\cite{kitaev2001unpaired} that, for a system without interactions, the Majorana number is equal to the topological invariant $Q_\mathbb{Z}$ defined in Eq.~\eqref{eq:topo_inv} and we have checked that these two measures agree for all examples shown in this work with $U=0$. Furthermore, the classification in true and false PMMs according to the Majorana number for the examples with $U>0$ agrees with the classification we have derived from inspecting the spectrum $E(\Delta_Z)$ in Fig.~\ref{fig:dmrg_results}.

{\raggedbottom
\section{\label{app:examples}Further examples of true and false PMMs}

In the main text, we have given just one example of a false PMM, i.e., a near-zero-energy state that is highly localized but that is not related to a topological MBS in the long chain limit, and one example of a true PMM, i.e., a near-zero-energy state that is highly localized and that is related to a topological MBS in the long chain limit.
In this Appendix, we give additional examples of a true and several false PMMs. We note that we present mainly false PMMs here and leave the discussion of the prevalence of true versus false PMMs for future work.

\FloatBarrier
\subsection{Energy difference, charge difference, and Majorana polarization}
In Figs.~\ref{fig:NSN false no U fish before 2 energy_pol_charge}--\ref{fig:NSN false with U 3 energy_pol_charge}, we present the energy difference $\Delta E$ [defined in Eq.~\eqref{eq:definition_dE}, panels (a) and (d)], the charge difference $\Delta Q_j$ [defined in Eq.~\eqref{eq:definition_dQ}, panels (b) and (e)], the Majorana polarization [defined in Eq.~\eqref{eq:definition_M}, panels (c) and (f)], and the excitation gap $E_\mathrm{ex}$ [defined in Eq.~\eqref{eq:definition_gap}] for true and false PMMs. All calculations are done in the second quantized formulation of the Hamiltonian given in Eq.~\eqref{eq:full_hamiltonian}. The black circled areas correspond to the TR, which is defined in Eq.~\eqref{eq:definition_ROT}.  We show both examples where the on-site Coulomb repulsion is zero, i.e., $U=0$, and examples where $U>0$. Values for $t$ and $\Phi_\mathrm{SOI}$ were set using random values chosen from a uniform distribution in the $t$-$\Phi_\mathrm{SOI}$ parameter space. Likewise, for the examples with $U>0$, values for $U$ were set using random values from a uniform distribution.

\begin{figure}[H]
	\centering
	\includegraphics[width=\linewidth]{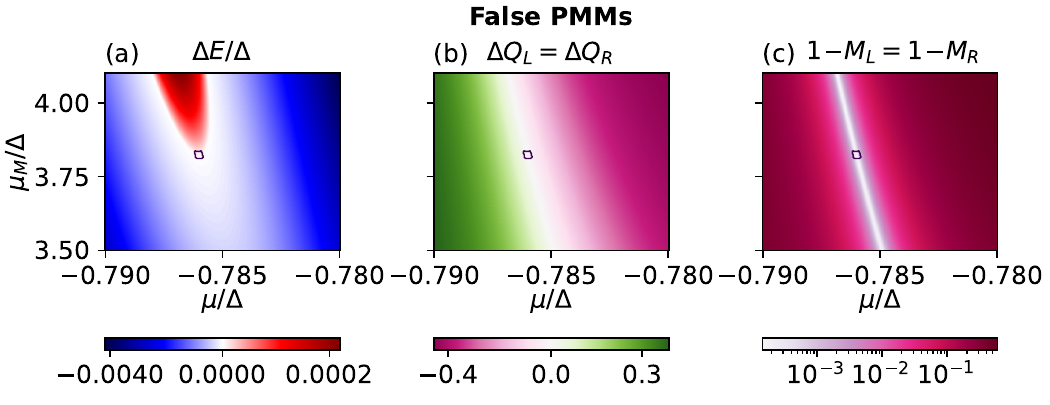}
	\caption{
		The parameters are $t/\Delta=0.24$, $\Phi_\mathrm{SOI} = 0.42 \pi$, $U=0$, and $\Delta_Z/\Delta = 0.8$.
		The threshold values for the TR are $\Delta E_\mathrm{th}/\Delta = 10^{-5}$, $\Delta Q_\mathrm{th}=0.02$, $M_\mathrm{th}=0.02$, and the maximum excitation gap in the TR is $E_\mathrm{ex}/\Delta=0.007$.
	}
	\label{fig:NSN false no U fish before 2 energy_pol_charge}
\end{figure}

\begin{figure}[H]
	\centering
	\includegraphics[width=\linewidth]{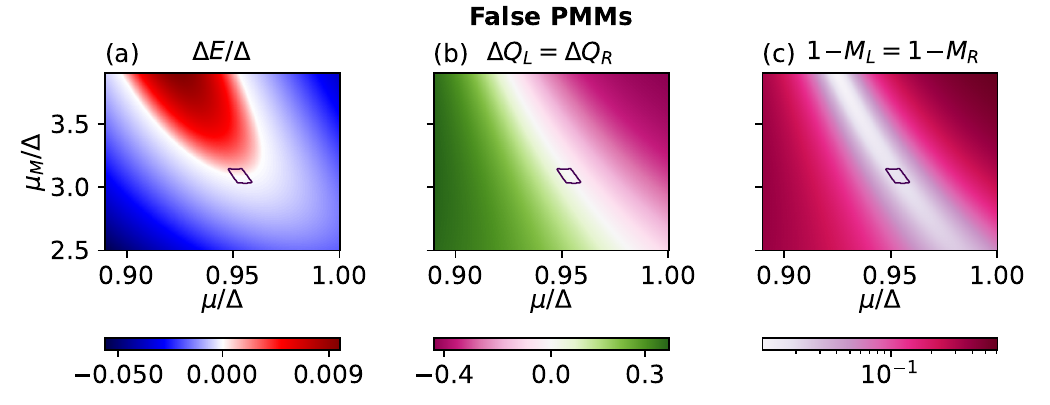}
	\caption{
		The parameters are $t/\Delta=0.7$, $\Phi_\mathrm{SOI} = 0.38 \pi$, $U=0$, and $\Delta_Z/\Delta = 0.8$.
		The threshold values for the TR are $\Delta E_\mathrm{th}/\Delta = 10^{-3}$, $\Delta Q_\mathrm{th}=0.03$, $M_\mathrm{th}=0.03$, and the maximum excitation gap in the TR is $E_\mathrm{ex}/\Delta=0.098$.
	}
	\label{fig:NSN false no U fish after 1 energy_pol_charge}
\end{figure}

\begin{figure}[H] 
	\centering
	\includegraphics[width=\linewidth]{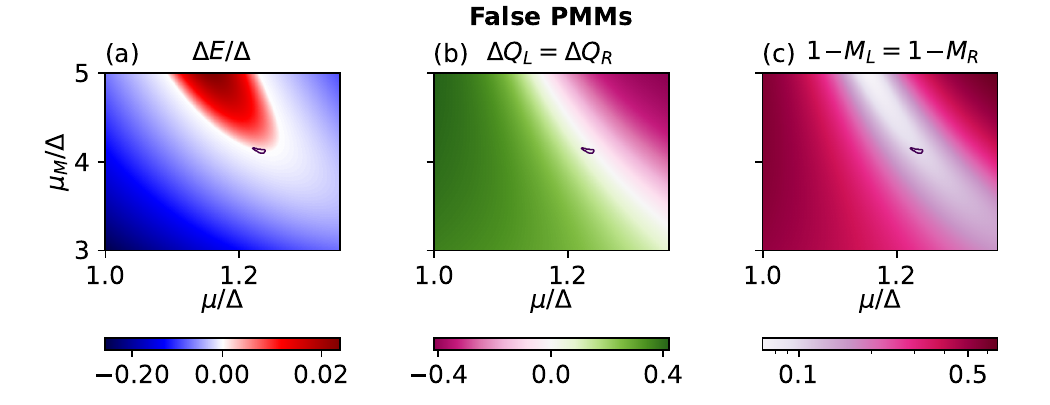}
	\caption{
		The parameters are $t/\Delta=1.27$, $\Phi_\mathrm{SOI} = 0.38 \pi$, $U=0$, and $\Delta_Z/\Delta = 0.8$.
		The threshold values for the TR are $\Delta E_\mathrm{th}/\Delta = 10^{-3}$, $\Delta Q_\mathrm{th}=0.1$, $M_\mathrm{th}=0.1$, and the maximum excitation gap in the TR is $E_\mathrm{ex}/\Delta=0.218$.
	}
	\label{fig:NSN false no U fish after 2 energy_pol_charge}
\end{figure}

\begin{figure}[H] 
	\centering
	\includegraphics[width=\linewidth]{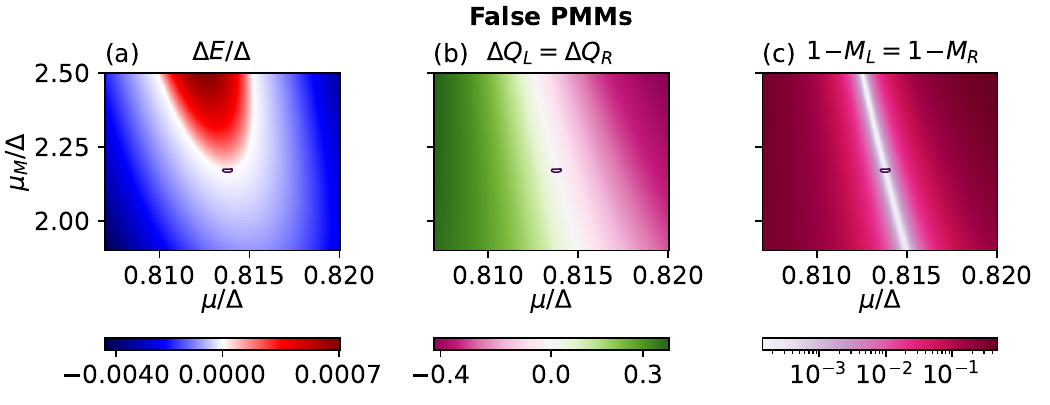}
	\caption{
		The parameters are $t/\Delta=0.19$, $\Phi_\mathrm{SOI} = 0.36 \pi$, $U=0$, and $\Delta_Z/\Delta = 0.8$.
		The threshold values for the TR are $\Delta E_\mathrm{th}/\Delta = 10^{-5}$, $\Delta Q_\mathrm{th}=0.02$, $M_\mathrm{th}=0.02$, and the maximum excitation gap in the TR is $E_\mathrm{ex}/\Delta=0.012$.
	}
	\label{fig:NSN false no U fish after 3 energy_pol_charge}
\end{figure}

\begin{figure}[H] 
	\centering
	\includegraphics[width=\linewidth]{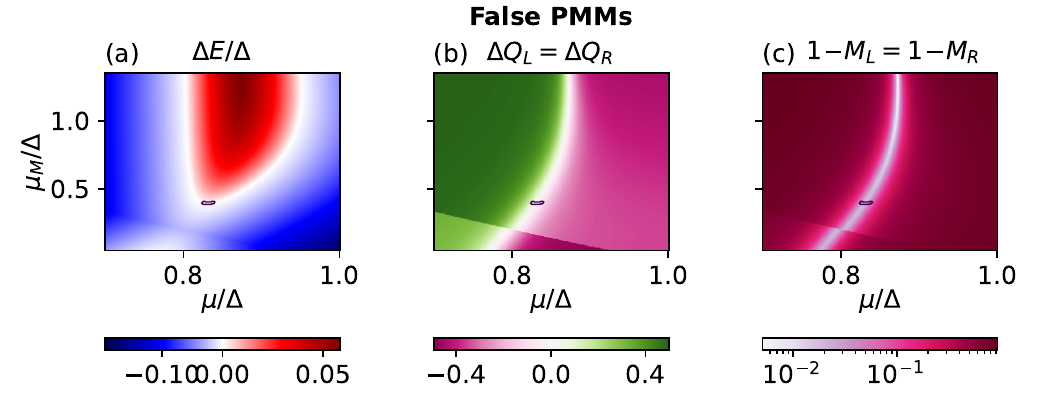}
	\caption{
		The parameters are $t/\Delta=0.41$, $\Phi_\mathrm{SOI} = 0.06 \pi$, $U=0$, and $\Delta_Z/\Delta = 0.8$.
		The threshold values for the TR are $\Delta E_\mathrm{th}/\Delta = 10^{-3}$, $\Delta Q_\mathrm{th}=0.1$, $M_\mathrm{th}=0.1$, and the maximum excitation gap in the TR is $E_\mathrm{ex}/\Delta=0.059$.
	}
	\label{fig:NSN false no U no fish 1 energy_pol_charge}
\end{figure}

\begin{figure}[H] 
	\centering
	\includegraphics[width=\linewidth]{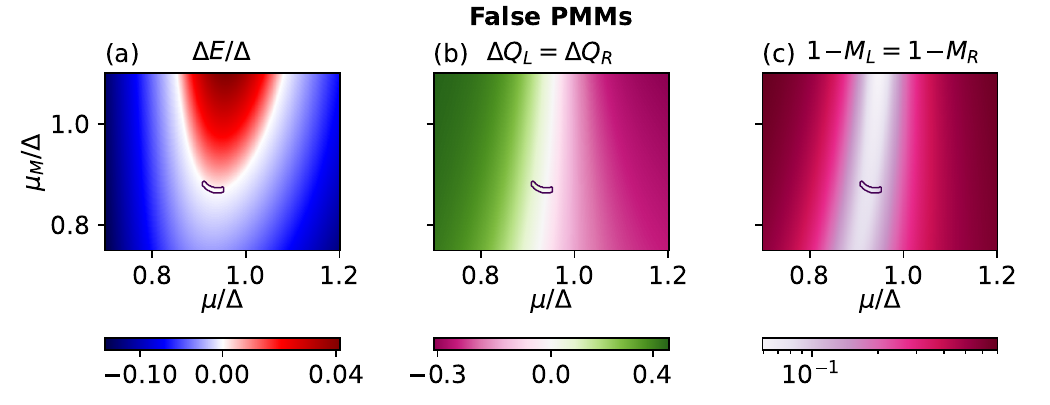}
	\caption{
		The parameters are $t/\Delta=0.64$, $\Phi_\mathrm{SOI} = 0.12 \pi$, $U=0$, and $\Delta_Z/\Delta = 0.8$.
		The threshold values for the TR are $\Delta E_\mathrm{th}/\Delta = 10^{-3}$, $\Delta Q_\mathrm{th}=0.1$, $M_\mathrm{th}=0.1$, and the maximum excitation gap in the TR is $E_\mathrm{ex}/\Delta=0.195$.
	}
	\label{fig:NSN false no U no fish 2 energy_pol_charge}
\end{figure}

\begin{figure}[H]
	\centering
	\includegraphics[width=\linewidth]{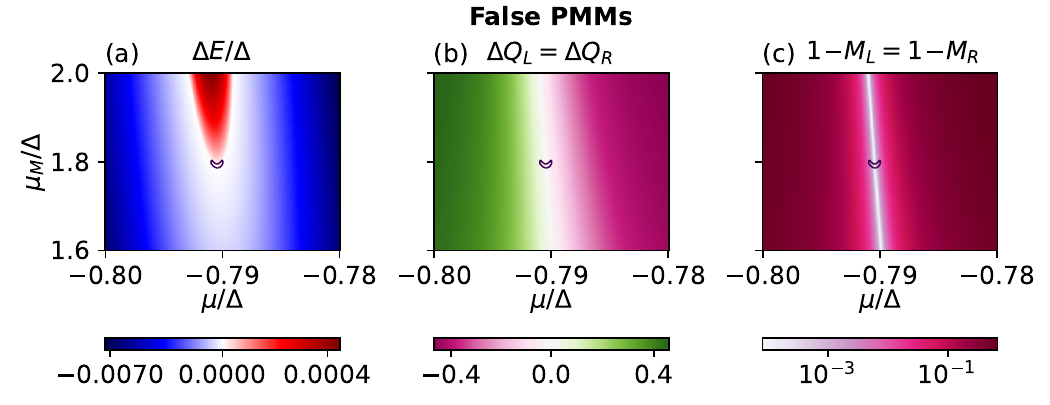}
	\caption{
		The parameters are $t/\Delta=0.15$, $\Phi_\mathrm{SOI} = 0.34 \pi$, $U=0$, and $\Delta_Z/\Delta = 0.8$.
		The threshold values for the TR are $\Delta E_\mathrm{th}/\Delta = 10^{-5}$, $\Delta Q_\mathrm{th}=0.05$, $M_\mathrm{th}=0.05$, and the maximum excitation gap in the TR is $E_\mathrm{ex}/\Delta=0.009$.
	}
	\label{fig:NSN false no U no fish 3 energy_pol_charge}
\end{figure}

\begin{figure}[H] 
	\centering
	\includegraphics[width=\linewidth]{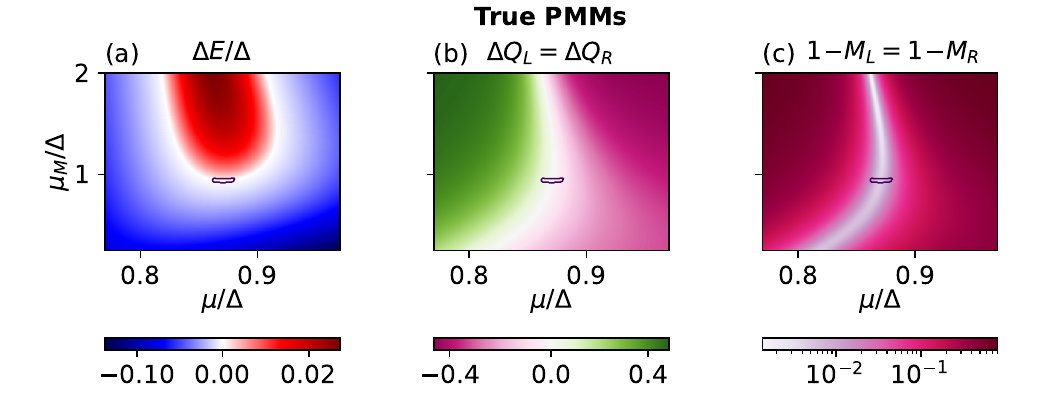}
	\caption{
		The parameters are $t/\Delta=0.4$, $\Phi_\mathrm{SOI} = 0.22 \pi$, $U/\Delta=4.69$, and $\Delta_Z/\Delta = 0.8$.
		The threshold values for the TR are $\Delta E_\mathrm{th}/\Delta = 10^{-3}$, $\Delta Q_\mathrm{th}=0.07$, $M_\mathrm{th}=0.07$, and the maximum excitation gap in the TR is $E_\mathrm{ex}/\Delta=0.108$.
	}
	\label{fig:NSN real with U energy_pol_charge}
\end{figure}

\begin{figure}[H] 
	\centering
	\includegraphics[width=\linewidth]{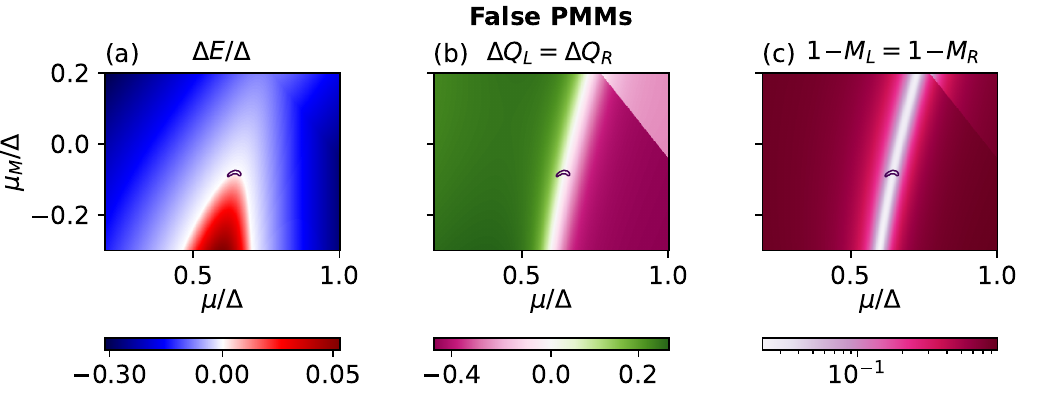}
	\caption{
		The parameters are $t/\Delta=0.6$, $\Phi_\mathrm{SOI} = 0.06 \pi$, $U/\Delta=4.68$, and $\Delta_Z/\Delta = 0.8$.
		The threshold values for the TR are $\Delta E_\mathrm{th}/\Delta = 10^{-3}$, $\Delta Q_\mathrm{th}=0.1$, $M_\mathrm{th}=0.1$, and the maximum excitation gap in the TR is $E_\mathrm{ex}/\Delta=0.109$.
	}
	\label{fig:NSN false with U 1 energy_pol_charge}
\end{figure}

\begin{figure}[H] 
	\centering
	\includegraphics[width=\linewidth]{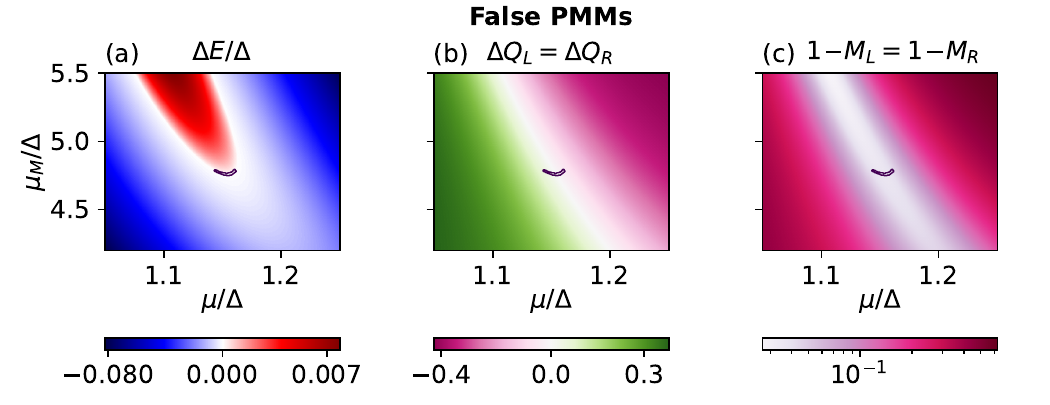}
	\caption{
		The parameters are $t/\Delta=1.29$, $\Phi_\mathrm{SOI} = 0.46 \pi$, $U/\Delta=4.31$, and $\Delta_Z/\Delta = 0.8$.
		The threshold values for the TR are $\Delta E_\mathrm{th}/\Delta = 10^{-4}$, $\Delta Q_\mathrm{th}=0.05$, $M_\mathrm{th}=0.05$, and the maximum excitation gap in the TR is $E_\mathrm{ex}/\Delta=0.070$.
	}
	\label{fig:NSN false with U 2 energy_pol_charge}
\end{figure}

\begin{figure}[H] 
	\centering
	\includegraphics[width=\linewidth]{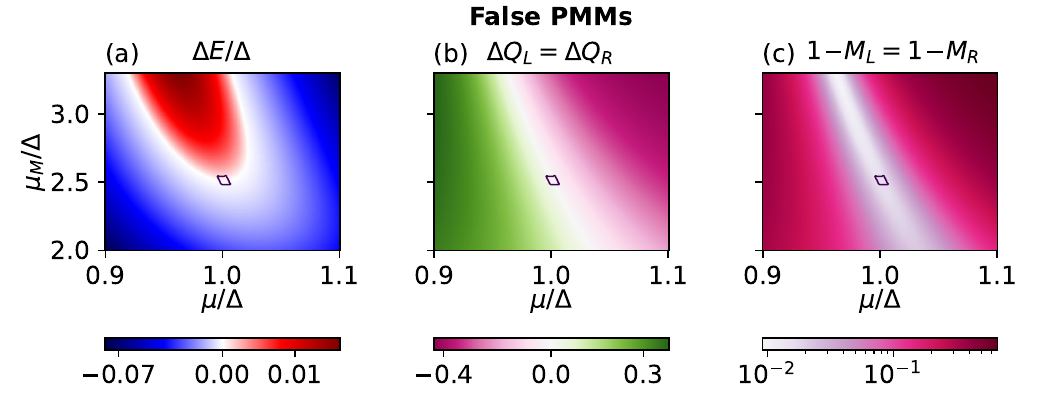}
	\caption{
		The parameters are $t/\Delta=0.76$, $\Phi_\mathrm{SOI} = 0.38 \pi$, $U=3.81$, and $\Delta_Z/\Delta = 0.8$.
		The threshold values for the TR are $\Delta E_\mathrm{th}/\Delta = 10^{-3}$, $\Delta Q_\mathrm{th}=0.02$, $M_\mathrm{th}=0.02$, and the maximum excitation gap in the TR is $E_\mathrm{ex}/\Delta=0.129$.
	}
	\label{fig:NSN false with U 3 energy_pol_charge}
\end{figure}

\FloatBarrier

\subsection{Energy spectrum vs. Zeeman energy}
In Figs.~\ref{fig:zeeman_first_figure}--\ref{fig:zeeman_last_figure}, we present the energy spectrum (black dots) of the Hamiltonian as a function of the Zeeman energy $\Delta_Z$ in the long chain limit, and the topological invariant (magenta line) in the infinite chain limit (see also Appendix~\ref{app:topo_inv}). With this, we determine if a TR is connected to topological MBSs or if it is connected to highly localized trivial states. The Zeeman energy value $\Delta_Z/\Delta=0.8$ signifies the Zeeman energy at which we consider all TR in Figs.~\ref{fig:NSN false no U fish before 2 energy_pol_charge}--\ref{fig:NSN false with U 3 energy_pol_charge}. All calculations for $U=0$ are done using the BdG formulation of the Hamiltonian given in Eq.~\eqref{eq:full_hamiltonian}. For the data sets with finite $U$, we use DMRG~\cite{steven1992density, steven1993density, schollwock2005density}, see the next section.

For the parameter sets with $U=0$, we see that at $\Delta_Z/\Delta = 0.8$, none of the parameter sets presented in this section are related to topological MBSs, instead they all result in false PMMs.

\begin{figure}[H]
	\centering
	\includegraphics[width=0.8\linewidth]{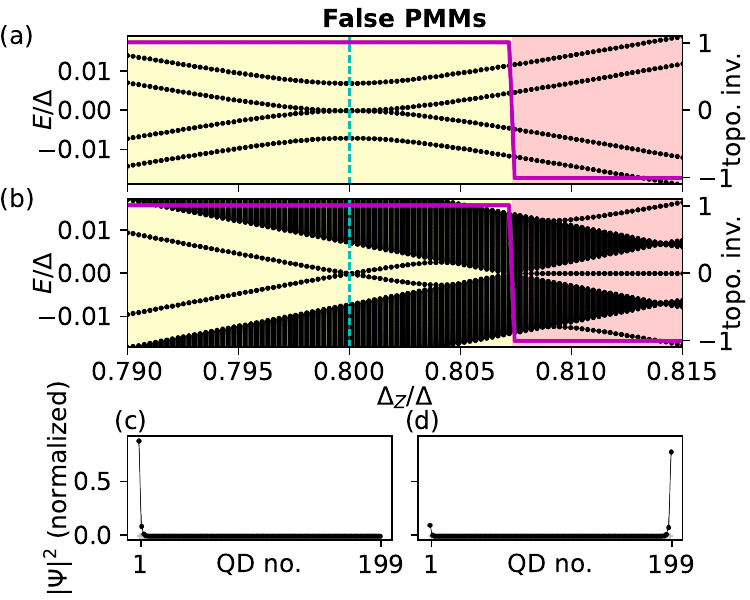}
	\caption{The parameters are as given in Fig.~\ref{fig:NSN false no U fish before 2 energy_pol_charge}, i.e., $t/\Delta=0.24$, $\Phi_\mathrm{SOI} = 0.42 \pi$, $U=0$, $\mu/\Delta = -0.786$, and $\mu_M/\Delta= 3.843$. In (c) and (d) $\Delta_Z/\Delta = 0.8$.
		\label{fig:zeeman_first_figure}
	}
\end{figure}

\begin{figure}[H]
	\centering
	\includegraphics[width=0.8\linewidth]{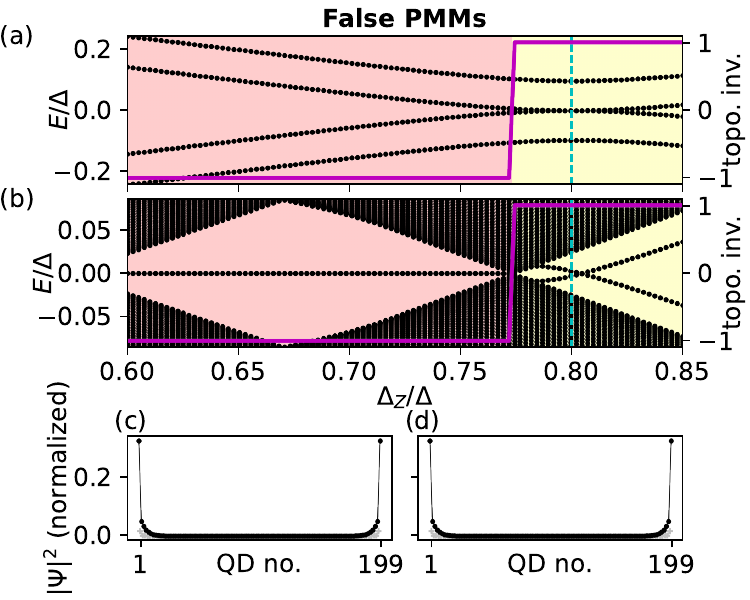}
	\caption{The parameters are as given in Fig.~\ref{fig:NSN false no U fish after 1 energy_pol_charge}, i.e.,
		$t/\Delta=0.7$, $\Phi_\mathrm{SOI} = 0.38 \pi$, $U=0$,
		$\mu/\Delta = 0.953$, and $\mu_M/\Delta= 3.086$. In (c) and (d) $\Delta_Z/\Delta = 0.8$.
	}
	\label{fig:NSN fake no U fish after 1 energy vs delta_z}
\end{figure}

\begin{figure}[H]
	\centering
	\includegraphics[width=0.8\linewidth]{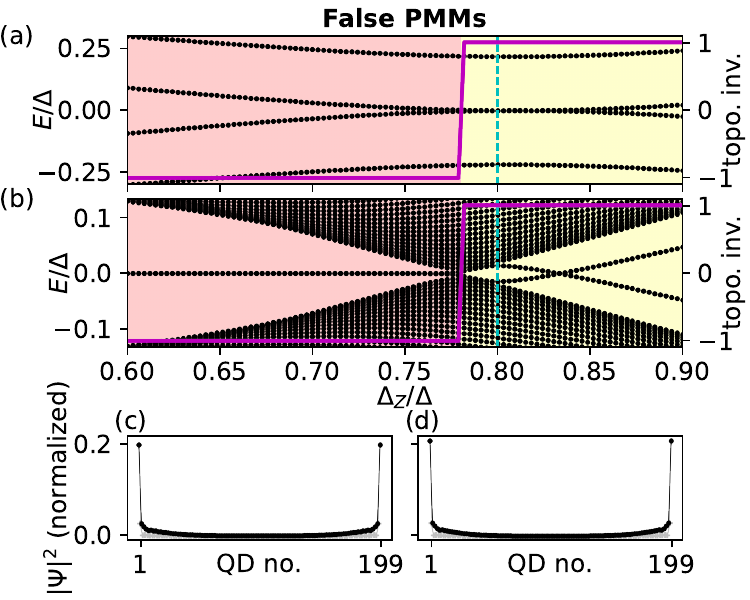}
	\caption{The parameters are as given in Fig.~\ref{fig:NSN false no U fish after 2 energy_pol_charge}, i.e.,
		$t/\Delta=1.27$, $\Phi_\mathrm{SOI} = 0.38 \pi$, $U=0$, 
		$\mu/\Delta = 1.239$, and $\mu_M/\Delta= 4.110$. In (c) and (d) $\Delta_Z/\Delta = 0.8$.
	}
	\label{fig:NSN fake no U fish after 2 energy vs delta_z}
\end{figure}

\begin{figure}[H]
	\centering
	\includegraphics[width=0.8\linewidth]{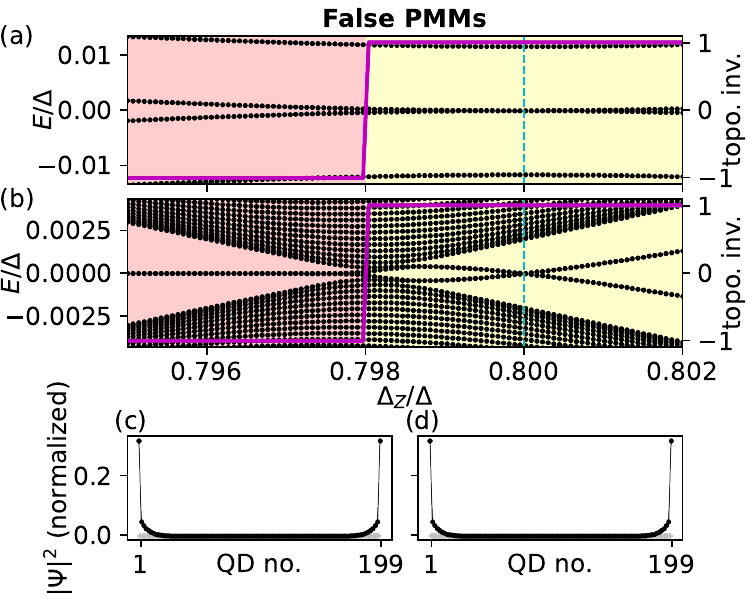}
	\caption{The parameters are as given in Fig.~\ref{fig:NSN false no U fish after 3 energy_pol_charge}, i.e.,
		$t/\Delta=0.19$, $\Phi_\mathrm{SOI} = 0.36 \pi$, $U=0$, 
		$\mu/\Delta = 0.814$, and $\mu_M/\Delta= 2.170$.
		In (c) and (d) $\Delta_Z/\Delta = 0.8$.
	}
	\label{fig:NSN fake no U fish after 3 energy vs delta_z}
\end{figure}

\begin{figure}[H]
	\centering
	\includegraphics[width=0.8\linewidth]{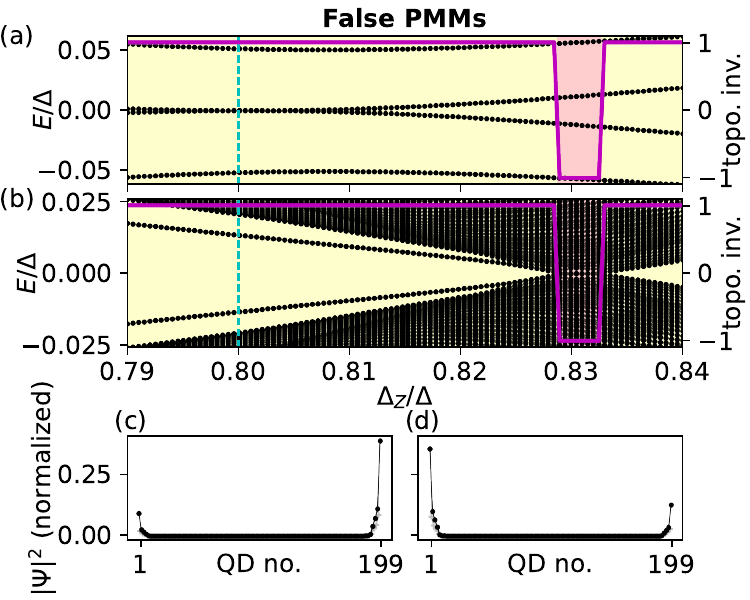}
	\caption{The parameters are as given in Fig.~\ref{fig:NSN false no U no fish 1 energy_pol_charge}, with 
		$t/\Delta=0.41$, $\Phi_\mathrm{SOI} = 0.06 \pi$, $U=0$, 
		$\mu/\Delta = 0.831$, and $\mu_M/\Delta= 0.397$.
		In (c) and (d) $\Delta_Z/\Delta = 0.8$.
	}
\end{figure}

\begin{figure}[H]
	\centering
	\includegraphics[width=0.8\linewidth]{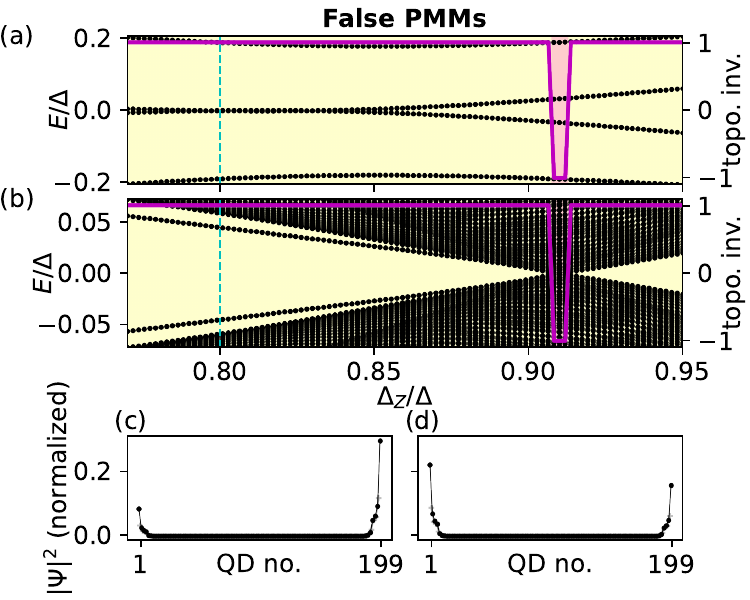}
	\caption{The parameters are as given in Fig.~\ref{fig:NSN false no U no fish 2 energy_pol_charge}, i.e.,
		$t/\Delta=0.64$, $\Phi_\mathrm{SOI} = 0.12 \pi$, $U=0$,
		$\mu/\Delta = 0.936$, and $\mu_M/\Delta= 0.870$.
		In (c) and (d) $\Delta_Z/\Delta = 0.8$.	 
	}
\end{figure}

\begin{figure}[H]
	\centering
	\includegraphics[width=0.8\linewidth]{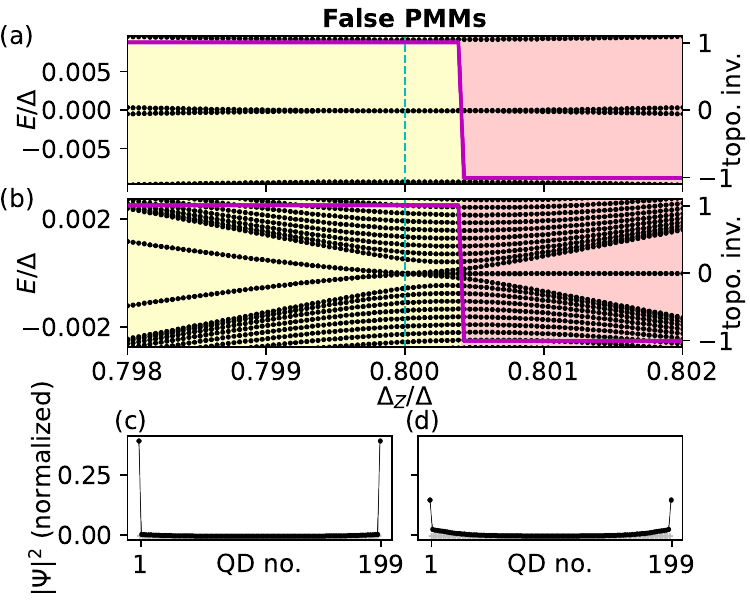}
	\caption{The parameters are as given in Fig.~\ref{fig:NSN false no U no fish 3 energy_pol_charge}, i.e., 
		$t/\Delta=0.15$, $\Phi_\mathrm{SOI} = 0.34 \pi$, $U=0$, 
		$\mu/\Delta = -0.790$ and $\mu_M/\Delta= 1.793$.
		In (c) and (d) $\Delta_Z/\Delta = 0.8$. 
	}
	\label{fig:zeeman_last_figure}
\end{figure}

\subsubsection{\label{app:dmrg} DMRG}
For a finite on-site Coulomb interaction $U>0$, one can no longer use the BdG Hamiltonian to calculate the energy spectrum. Instead, the second quantized form of Eq.~\eqref{eq:full_hamiltonian} must be used. The corresponding Hamiltonian, however, scales exponentially with the number of sites in the system, thus making the long chain limit computationally expensive. We therefore use DMRG~\cite{steven1992density, steven1993density, schollwock2005density}, more specifically, the ITensor library~\cite{ITensor}, to calculate the energy difference $\Delta E$ [defined in Eq.~\eqref{eq:definition_dE}] as a function of the Zeeman energy $\Delta_Z$, see Fig.~\ref{fig:dmrg_results}. This does not allow for a study of the closing and reopening of the bulk gap, or calculating the topological invariant. However, it still gives an idea of whether the corresponding state is related to a topological MBS, or if it is a false PMMs. 
If it is a true PMMs, then in the long chain limit, the state is at zero energy and there is a finite range of $\Delta_Z$ values around it which also exhibits zero-energy states, see Fig.~\ref{fig:dmrg_results}(a), which shows a true PMMs. 
If the state is a false PMMs, then in the long chain limit, the state is either no longer at zero energy, or there is no finite range of $\Delta_Z$ values around it which exhibits zero-energy states, see Figs.~\ref{fig:dmrg_results}(f)--\ref{fig:dmrg_results}(h), which show false PMMs.

\begin{figure}[H]
	\centering
	\includegraphics[width=\linewidth]{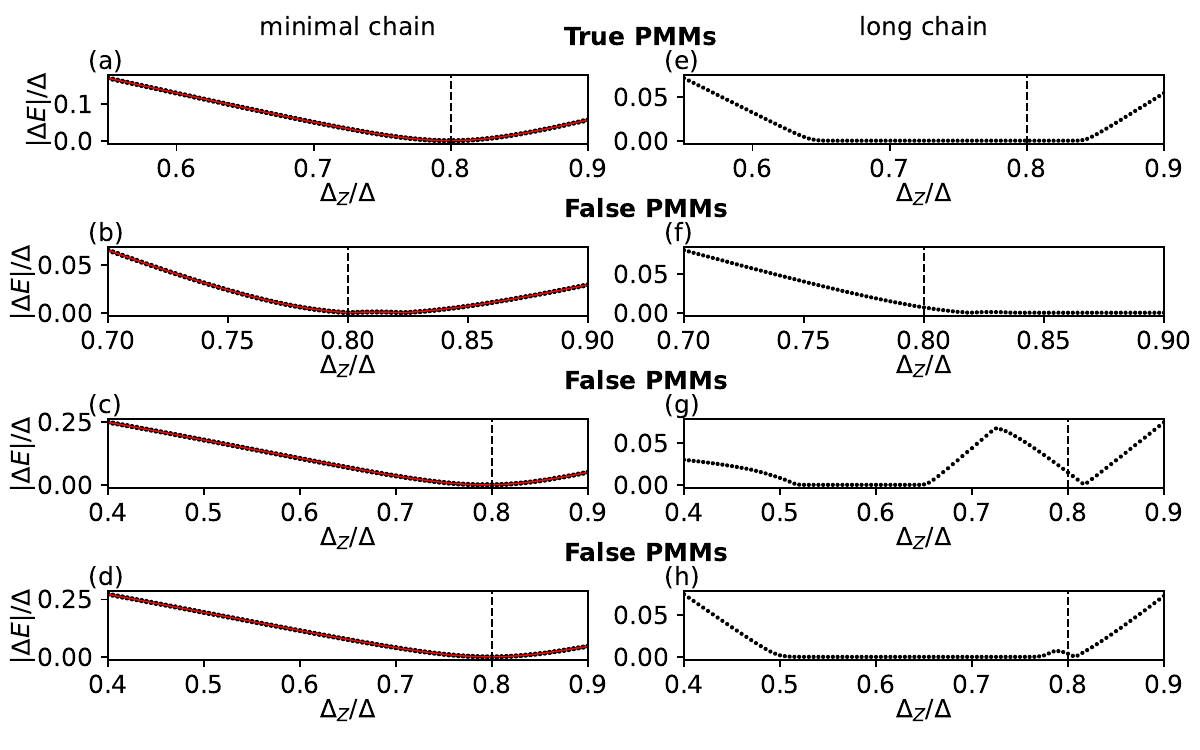}
	\caption{
		The energy difference $\Delta E$ [defined in Eq.~\eqref{eq:definition_dE}] between the even and odd ground state, calculated in the minimal chain [(a)--(d), setup as in Fig.~\ref{fig:setup}(a)] and the long chain limit [(e)--(h), setup as in Fig.~\ref{fig:setup}(b), consisting of 101 QDs, 51 of which are normal QDs and 50 are QDs proximitized by a superconductor] using DMRG (black dots) and exact diagonalization in (a)--(d) (red line). 
		For all panels, we are interested in the state at $\Delta_Z/\Delta=0.8$. In the minimal chain, all four examples have $\Delta E \approx 0$, thus these states are PMMs. In (e), the long chain limit for the first example, the state remains at zero energy and there is a finite range of $\Delta_Z$ values around it that also exhibits zero-energy states, thus the state is a true PMM. In contrast, in the long chain limit for the second to fourth examples [panels (f)--(h)], the state is either no longer at zero energy, or there is no finite range of $\Delta_Z$ values around it which exhibit zero-energy states. Therefore, these examples show false PMMs. The parameters for (a) and (e) are as for Fig.~\ref{fig:NSN real with U energy_pol_charge}: $t/\Delta=0.4$, $\Phi_\mathrm{SOI} = 0.22 \pi$, $U/\Delta=4.69$, $\mu/\Delta = 0.870$, and $\mu_M/\Delta=0.939$. 
		The parameters for (b) and (f) are as for Fig.~\ref{fig:NSN false with U 1 energy_pol_charge}: $t/\Delta=0.6$, $\Phi_\mathrm{SOI} = 0.06 \pi$, $U/\Delta=4.68$, $\mu/\Delta = 0.657$, and $\mu_M/\Delta=-0.0825$.
		The parameters for (c) and (g) are as for Fig.~\ref{fig:NSN false with U 2 energy_pol_charge}: $t/\Delta=1.29$, $\Phi_\mathrm{SOI} = 0.46 \pi$, $U/\Delta=4.31$, $\mu/\Delta = 1.151$, and $\mu_M/\Delta=4.768$.
		The parameters for (d) and (h) are as for Fig.~\ref{fig:NSN false with U 3 energy_pol_charge}: $t/\Delta=0.76$, $\Phi_\mathrm{SOI}=0.38\pi$, $U/\Delta=3.81$, $\mu/\Delta = 1.000$, and $\mu_M/\Delta=2.515$.
	}
	\label{fig:dmrg_results}
\end{figure}

\FloatBarrier
\subsection{Zero-energy conductance}
In Figs.~\ref{fig:zero_energy_conductance_first_fig}--\ref{fig:zero_energy_conductance_last_fig}, we present the zero-energy conductance as the chemical potentials $\mu_L$ and $\mu_R$ are varied independently, for different values of $\mu_M$. Panels (a), (e), and (i) show the local conductance $G_{LL}$, panels (b), (f), and (j) show the nonlocal conductance $G_{LR}$, panels (c), (g), and (k) show the nonlocal conductance $G_{RL}$, and panels (d), (h), and (l) show the local conductance $G_{RR}$. Furthermore, panels (a)--(d) show the conductance for a chemical potential $\mu_{M,1}$ that is smaller than the value of $\mu_M$ in the TR, panels (e)--(h) show the conductance for $\mu_{M,2}$ in the TR, and panels (i)--(l) show the conductance for a chemical potential $\mu_{M,3}$ that is larger than the value of $\mu_M$ in the TR. All conductance values are normalized to the maximum conductance value for this parameter set.
The conductance is calculated using the Python package Kwant~\cite{groth2014kwant}, where $t_l$ is the hopping amplitude from the outer QDs to the lead. This hopping is assumed to have no SOI. Since the BdG formulation of the Hamiltonian is used for the calculations, we show only conductance measurements for parameter sets where $U=0$.

\begin{figure}[H]
	\centering
	\includegraphics[width=\linewidth]{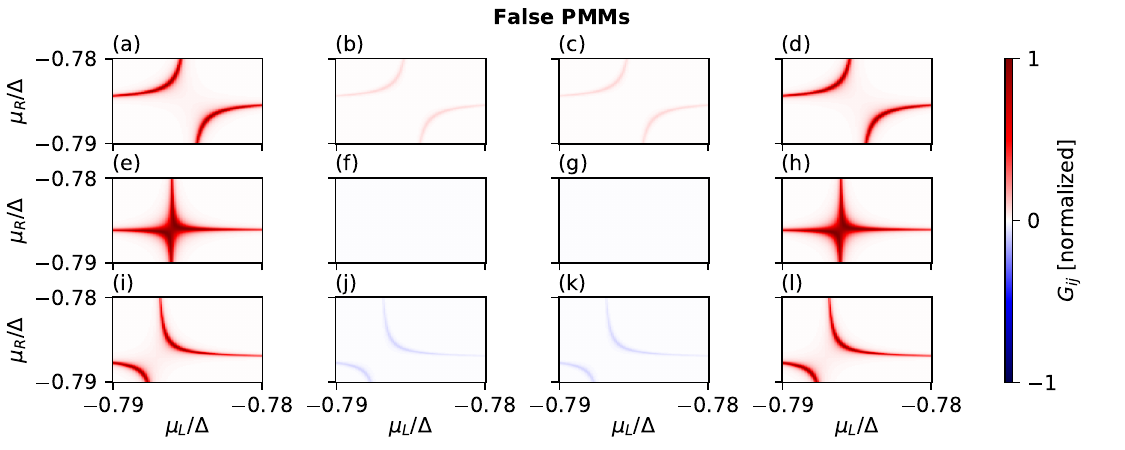}
	\caption{The parameters are as given in Fig.~\ref{fig:NSN false no U fish before 2 energy_pol_charge}, i.e., 
		$t/\Delta=0.24$, $\Phi_\mathrm{SOI} = 0.42 \pi$, $U=0$, and $\Delta_Z/\Delta = 0.8$, $t_l/\Delta = 5 \times 10^{-5}$, $\mu_{M,1}/\Delta = 3.500$, $\mu_{M,2}/\Delta = 3.843$, and $\mu_{M,3}/\Delta = 4.187$.
	}
	\label{fig:zero_energy_conductance_first_fig}
\end{figure}

\begin{figure}[H]
	\centering
	\includegraphics[width=\linewidth]{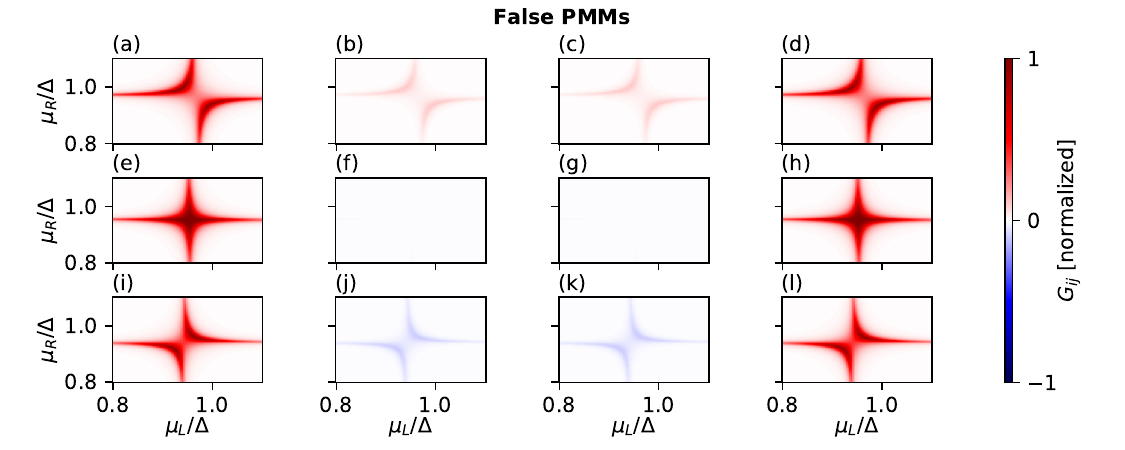}
	\caption{The parameters are as given in Fig.~\ref{fig:NSN false no U fish after 1 energy_pol_charge}, i.e.,
		$t/\Delta=0.7$, $\Phi_\mathrm{SOI} = 0.38 \pi$, $U=0$, $\Delta_Z/\Delta = 0.8$,
		$t_l/\Delta =  4 \times 10^{-3}$, $\mu_{M,1}/\Delta = 2.793$, $\mu_{M,2}/\Delta = 3.086$, and $\mu_{M,3}/\Delta = 3.379$.
	}
\end{figure}

\begin{figure}[H]
	\centering
	\includegraphics[width=\linewidth]{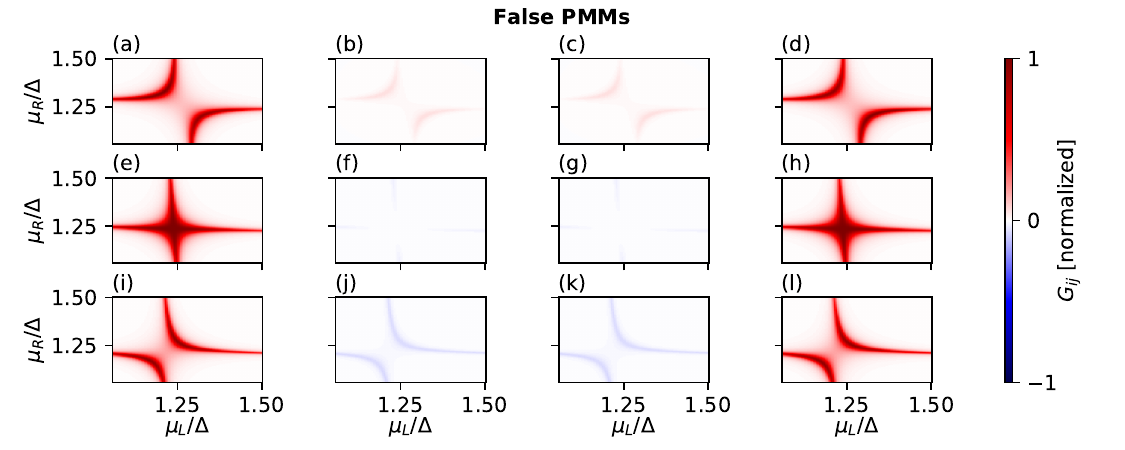}
	\caption{The parameters are as given in Fig.~\ref{fig:NSN false no U fish after 2 energy_pol_charge}, i.e.,
		$t/\Delta=1.27$, $\Phi_\mathrm{SOI} = 0.38 \pi$, $U=0$, $\Delta_Z/\Delta = 0.8$,
		$t_l/\Delta =  4 \times 10^{-3}$, $\mu_{M,1}/\Delta = 3.888$, $\mu_{M,2}/\Delta = 4.110$, and $\mu_{M,3}/\Delta = 4.333$.
	}
	
\end{figure}

\begin{figure}[H]
	\centering
	\includegraphics[width=\linewidth]{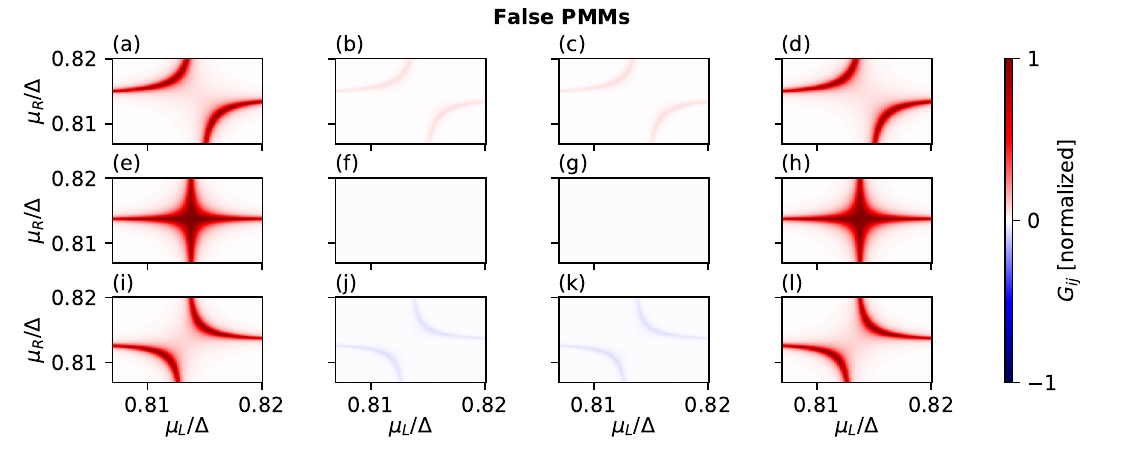}
	\caption{The parameters are as given in Fig.~\ref{fig:NSN false no U fish after 3 energy_pol_charge}, i.e.,
		$t/\Delta=0.19$, $\Phi_\mathrm{SOI} = 0.36 \pi$, $U=0$, $\Delta_Z/\Delta = 0.8$,
		$t_l/\Delta =  10^{-4}$, $\mu_{M,1}/\Delta = 2.035$, $\mu_{M,2}/\Delta = 2.170$, and $\mu_{M,3}/\Delta = 2.305$.
	}
\end{figure}

\begin{figure}[H]
	\centering
	\includegraphics[width=\linewidth]{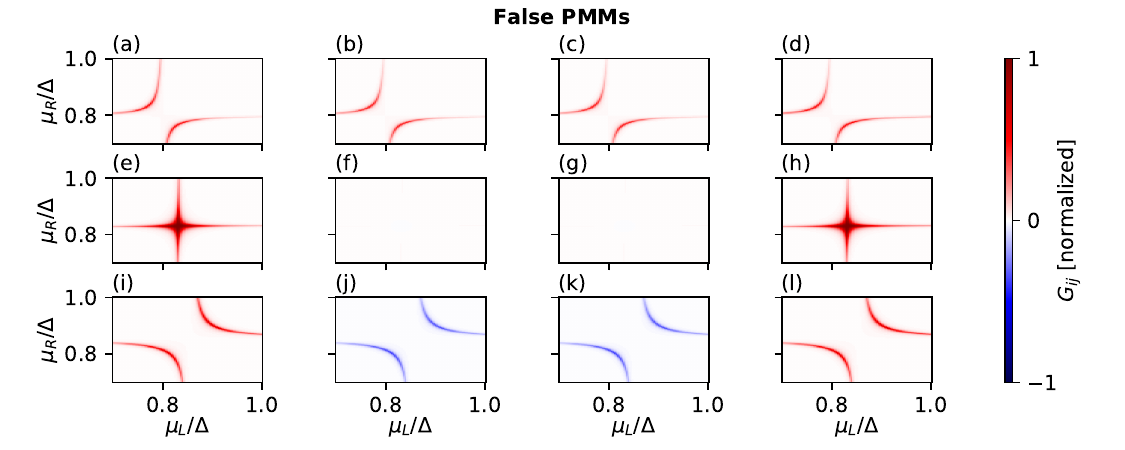}
	\caption{The parameters are as given in Fig.~\ref{fig:NSN false no U no fish 1 energy_pol_charge}, i.e.,
		$t/\Delta=0.41$, $\Phi_\mathrm{SOI} = 0.06 \pi$, $U=0$, $\Delta_Z/\Delta = 0.8$,
		$t_l/\Delta = 2 \times 10^{-3}$, $\mu_{M,1}/\Delta = 0.199$, $\mu_{M,2}/\Delta = 0.397$, and $\mu_{M,3}/\Delta = 0.596$.}
\end{figure}

\begin{figure}[H]
	\centering
	\includegraphics[width=\linewidth]{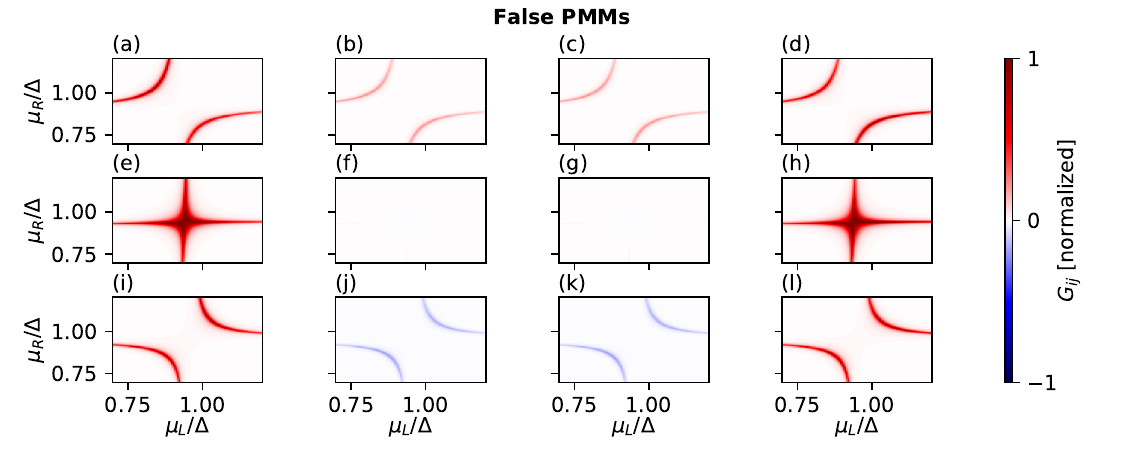}
	\caption{The parameters are as given in Fig.~\ref{fig:NSN false no U no fish 2 energy_pol_charge}, i.e.,
		$t/\Delta=0.64$, $\Phi_\mathrm{SOI} = 0.12 \pi$, $U=0$, and $\Delta_Z/\Delta = 0.8$,
		$t_l/\Delta =  3 \times 10^{-3}$, $\mu_{M,1}/\Delta = 0.75$, $\mu_{M,2}/\Delta = 0.870$, and $\mu_{M,3}/\Delta = 0.990$.}
\end{figure}

\begin{figure}[H]
	\centering
	\includegraphics[width=\linewidth]{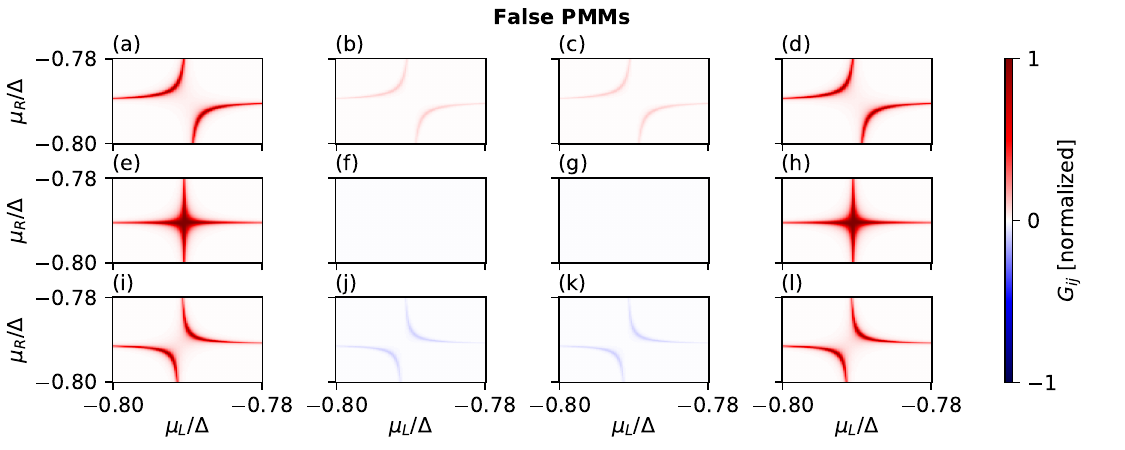}
	\caption{The parameters are as given in Fig.~\ref{fig:NSN false no U no fish 3 energy_pol_charge}, i.e.,
		$t/\Delta=0.15$, $\Phi_\mathrm{SOI} = 0.34 \pi$, $U=0$, $\Delta_Z/\Delta = 0.8$,
		$t_l\Delta =  10^{-4}$, $\mu_{M,1}/\Delta = 1.600$, $\mu_{M,2}/\Delta = 1.793$, and $\mu_{M,3}/\Delta = 1.987$.
	}
	\label{fig:zero_energy_conductance_last_fig}
\end{figure}

\subsection{\label{app:conductance_tuning_one_QD}Finite-energy conductance, tuning one QD}
In Figs.~\ref{fig:finite_energy_conductance_one_QD_first_fig}--\ref{fig:finite_energy_conductance_one_QD_last_fig}, we show the conductance at finite energy. All parameters are tuned to the values of the corresponding TR, then the chemical potential of the left QD is tuned away from its TR value, while the chemical potential of the right QD remains unchanged. This data is used to study excited states. Panels (a) show the local conductance $G_{LL}$ and panels (b) show the nonlocal conductance $G_{LR}$. All conductance values are normalized to the maximum conductance value for this parameter set.

\begin{figure}[H]
	\centering
	\includegraphics[width=\linewidth]{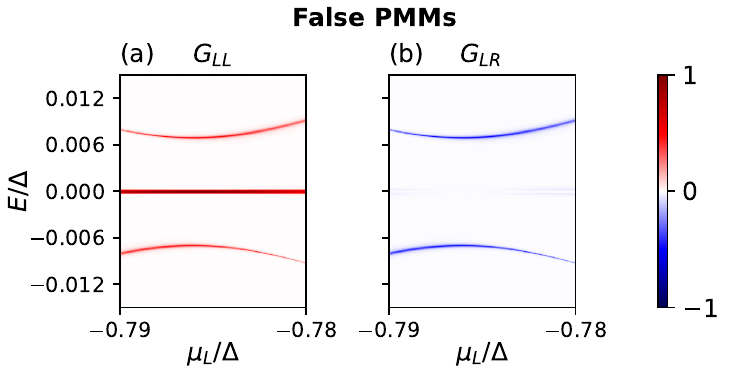}
	\caption{The parameters are as given in Fig.~\ref{fig:NSN false no U fish before 2 energy_pol_charge}, i.e., 
		$t/\Delta=0.24$, $\Phi_\mathrm{SOI} = 0.42 \pi$, $U=0$, and $\Delta_Z/\Delta = 0.8$, $\mu_R/\Delta=-0.786$, $\mu_M/\Delta=3.843$, and
		$t_l/\Delta = 2 \times 10^{-4}$.
	}
	\label{fig:finite_energy_conductance_one_QD_first_fig}
\end{figure}

\begin{figure}[H]
	\centering
	\includegraphics[width=\linewidth]{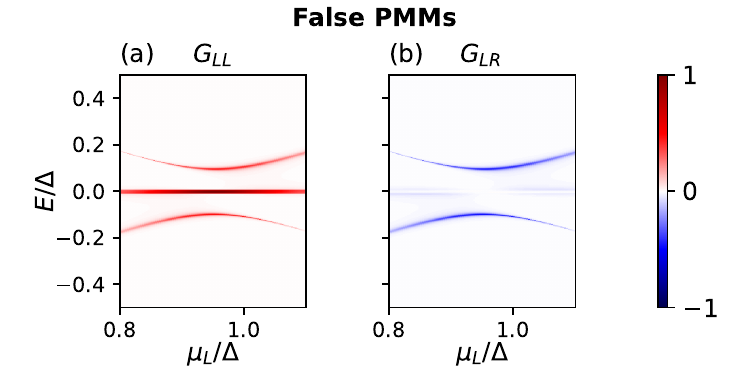}
	\caption{The parameters are as given in Fig.~\ref{fig:NSN false no U fish after 1 energy_pol_charge}, i.e.,
		$t/\Delta=0.7$, $\Phi_\mathrm{SOI} = 0.38 \pi$, $U=0$, $\Delta_Z/\Delta = 0.8$, $\mu_R/\Delta=0.953$, $\mu_M/\Delta=3.086$, and
		$t_l/\Delta =  8 \times 10^{-3}$.
	}
\end{figure}

\begin{figure}[H]
	\centering
	\includegraphics[width=\linewidth]{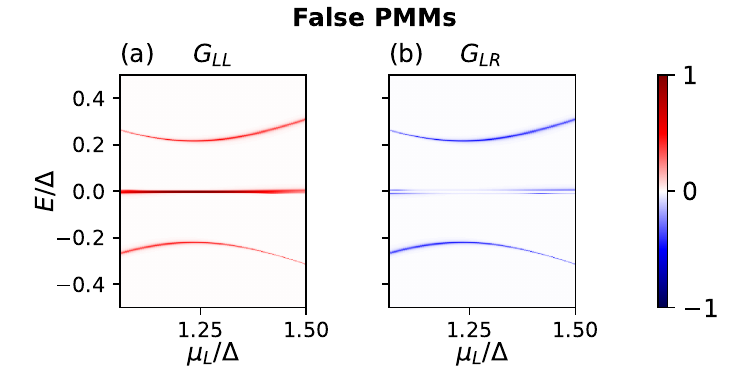}
	\caption{The parameters are as given in Fig.~\ref{fig:NSN false no U fish after 2 energy_pol_charge}, i.e.,
		$t/\Delta=1.27$, $\Phi_\mathrm{SOI} = 0.38 \pi$, $U=0$, $\Delta_Z/\Delta = 0.8$, $\mu_R/\Delta=1.239$, $\mu_M/\Delta=4.110$, and 
		$t_l/\Delta =  6 \times 10^{-3}$.
	}
	
\end{figure}

\begin{figure}[H]
	\centering
	\includegraphics[width=\linewidth]{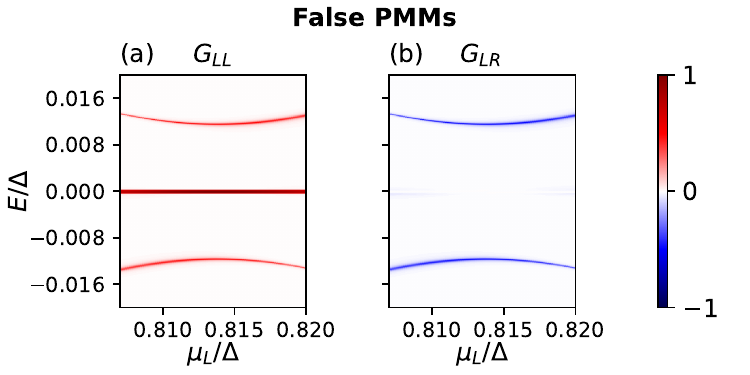}
	\caption{The parameters are as given in Fig.~\ref{fig:NSN false no U fish after 3 energy_pol_charge}, i.e.,
		$t/\Delta=0.19$, $\Phi_\mathrm{SOI} = 0.36 \pi$, $U=0$, $\Delta_Z/\Delta = 0.8$, $\mu_R/\Delta=0.814$, $\mu_M/\Delta=2.170$, and 
		$t_l/\Delta =  3 \times 10^{-4}$.
	}
\end{figure}

\begin{figure}[H]
	\centering
	\includegraphics[width=\linewidth]{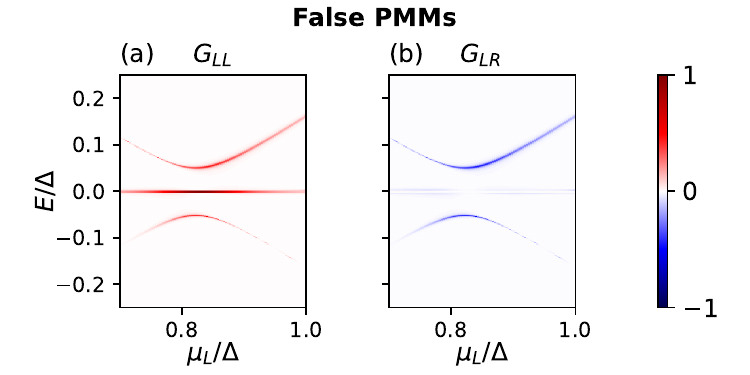}
	\caption{The parameters are as given in Fig.~\ref{fig:NSN false no U no fish 1 energy_pol_charge}, i.e.,
		$t/\Delta=0.41$, $\Phi_\mathrm{SOI} = 0.06 \pi$, $U=0$, $\Delta_Z/\Delta = 0.8$, $\mu_R/\Delta=0.831$, $\mu_M/\Delta=0.397$, and 
		$t_l/\Delta = 3 \times 10^{-3}$.
	}
\end{figure}

\begin{figure}[H]
	\centering
	\includegraphics[width=\linewidth]{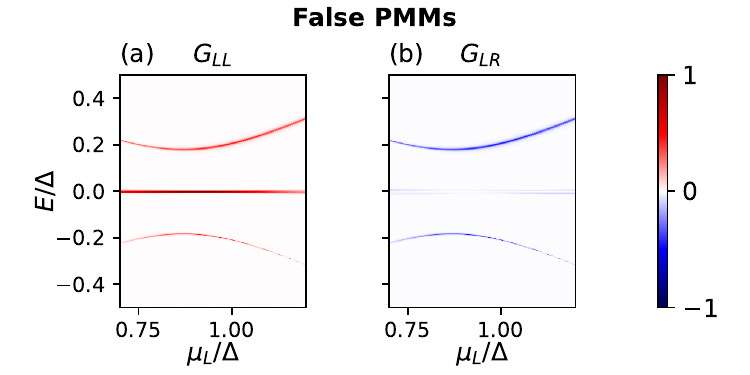}
	\caption{The parameters are as given in Fig.~\ref{fig:NSN false no U no fish 2 energy_pol_charge}, i.e.,
		$t/\Delta=0.64$, $\Phi_\mathrm{SOI} = 0.12 \pi$, $U=0$, and $\Delta_Z/\Delta = 0.8$, $\mu_R/\Delta=0.936$, $\mu_M/\Delta=0.870$, and 
		$t_l/\Delta =  4.5 \times 10^{-3}$.
	}
\end{figure}

\begin{figure}[H]
	\centering
	\includegraphics[width=\linewidth]{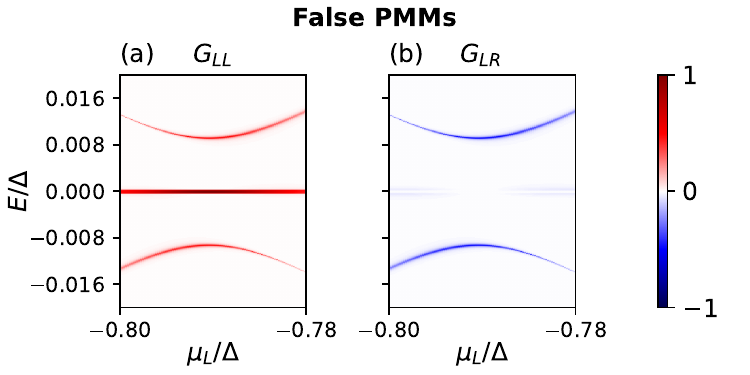}
	\caption{The parameters are as given in Fig.~\ref{fig:NSN false no U no fish 3 energy_pol_charge}, i.e.,
		$t/\Delta=0.15$, $\Phi_\mathrm{SOI} = 0.34 \pi$, $U=0$, $\Delta_Z/\Delta = 0.8$, $\mu_R/\Delta=-0.790$, $\mu_M/\Delta=1.793$, and 
		$t_l/\Delta =  3 \times 10^{-4}$.
	}
	\label{fig:finite_energy_conductance_one_QD_last_fig}
\end{figure}

\subsection{Finite-energy conductance, tuning both QDs}
In Figs.~\ref{fig:finite_energy_conductance_both_QD_first_fig}--\ref{fig:finite_energy_conductance_both_QD_last_fig}, we show the conductance at finite energy. All parameters are tuned to the values of the corresponding TR, then the chemical potentials of both QDs are tuned away from the TR value to $\mu=\mu_L=\mu_R$. Panels (a) show the local conductance $G_{LL}$ and panels (b) show the nonlocal conductance $G_{LR}$. All conductance values are normalized to the maximum conductance value for this parameter set.

\begin{figure}[H]
	\centering
	\includegraphics[width=\linewidth]{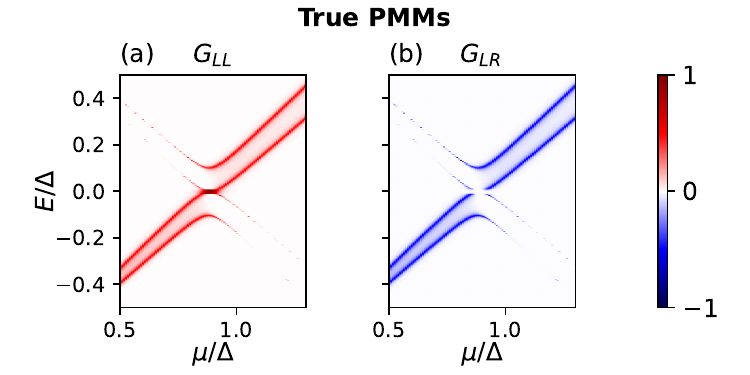}
	\caption{The parameters are as given in Fig.~\ref{fig:finite_energy_conductance_combined}(a)--\ref{fig:finite_energy_conductance_combined}(b), i.e.,
		$t/\Delta=0.42$, $\Phi_\mathrm{SOI}=0.26\pi$, $U=0$, $\Delta_Z/\Delta=0.8$, $\mu_M/\Delta=1.275$, $t_l/\Delta=0.01$.
	}
	\label{fig:finite_energy_conductance_both_QD_first_fig}
\end{figure}

\begin{figure}[H]
	\centering
	\includegraphics[width=\linewidth]{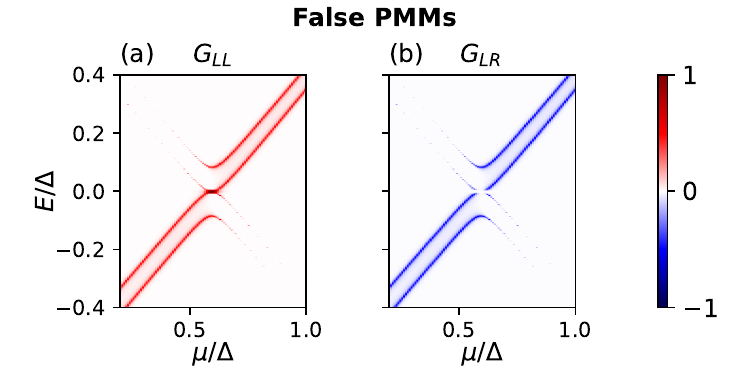}
	\caption{The parameters are as given in Fig.~\ref{fig:finite_energy_conductance_combined}(c)--~\ref{fig:finite_energy_conductance_combined}(d), i.e.,
		$t/\Delta=0.99$, $\Phi_\mathrm{SOI}=0.44\pi$, $U=0$, $\Delta_Z/\Delta=0.8$, $\mu_M/\Delta=-3.836$, $t_l/\Delta=0.006$.
	}
\end{figure}

\begin{figure}[H]
	\centering
	\includegraphics[width=\linewidth]{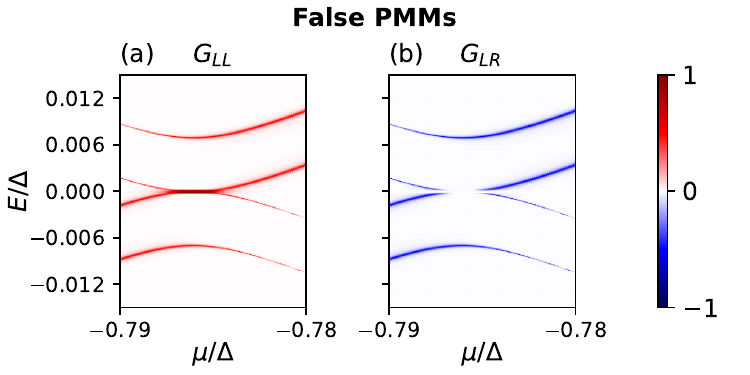}
	\caption{The parameters are as given in Fig.~\ref{fig:NSN false no U fish before 2 energy_pol_charge}, i.e.,
		$t/\Delta=0.24$, $\Phi_\mathrm{SOI} = 0.42 \pi$, $U=0$, and $\Delta_Z/\Delta = 0.8$, $\mu_M/\Delta=3.843$, 
		$t_l/\Delta = 2 \times 10^{-4}$.
	}
\end{figure}

\begin{figure}[H]
	\centering
	\includegraphics[width=\linewidth]{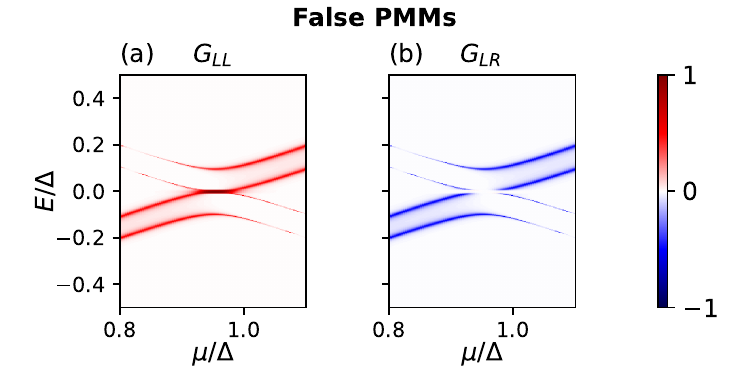}
	\caption{The parameters are as given in Fig.~\ref{fig:NSN false no U fish after 1 energy_pol_charge}, i.e.,
		$t/\Delta=0.7$, $\Phi_\mathrm{SOI} = 0.38 \pi$, $U=0$, $\Delta_Z/\Delta = 0.8$, $\mu_M/\Delta=3.086$, 
		$t_l/\Delta =  8 \times 10^{-3}$.
	}
\end{figure}

\begin{figure}[H]
	\centering
	\includegraphics[width=\linewidth]{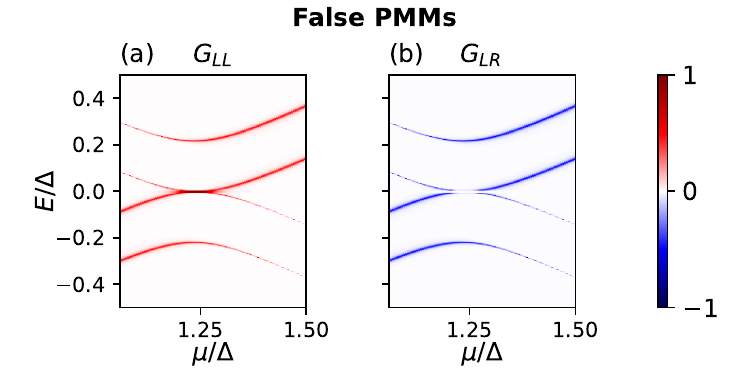}
	\caption{The parameters are as given in Fig.~\ref{fig:NSN false no U fish after 2 energy_pol_charge}, i.e.,
		$t/\Delta=1.27$, $\Phi_\mathrm{SOI} = 0.38 \pi$, $U=0$, $\Delta_Z/\Delta = 0.8$, $\mu_M/\Delta=4.110$, and 
		$t_l/\Delta =  6 \times 10^{-3}$.
	}
	
\end{figure}

\begin{figure}[H]
	\centering
	\includegraphics[width=\linewidth]{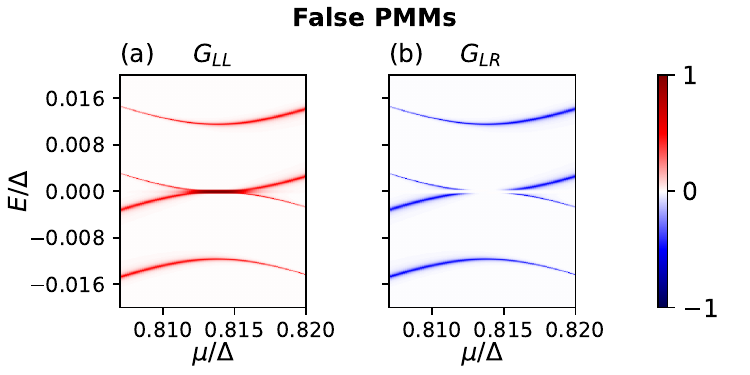}
	\caption{The parameters are as given in Fig.~\ref{fig:NSN false no U fish after 3 energy_pol_charge}, i.e.,
		$t/\Delta=0.19$, $\Phi_\mathrm{SOI} = 0.36 \pi$, $U=0$, $\Delta_Z/\Delta = 0.8$, $\mu_M/\Delta=2.170$, and 
		$t_l/\Delta =  3 \times 10^{-4}$.
	}
\end{figure}

\begin{figure}[H]
	\centering
	\includegraphics[width=\linewidth]{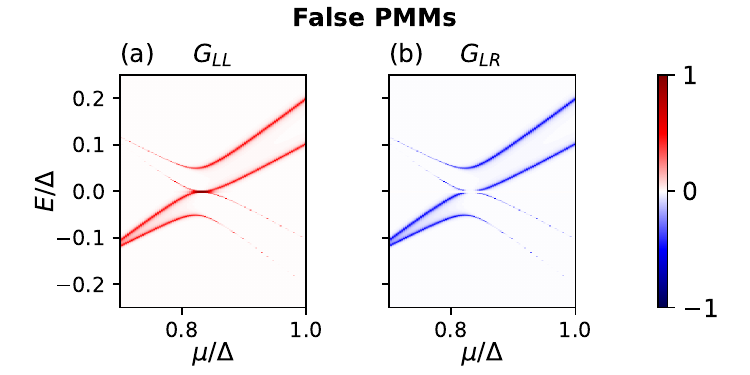}
	\caption{The parameters are as given in Fig.~\ref{fig:NSN false no U no fish 1 energy_pol_charge}, i.e.,
		$t/\Delta=0.41$, $\Phi_\mathrm{SOI} = 0.06 \pi$, $U=0$, $\Delta_Z/\Delta = 0.8$, $\mu_M/\Delta=0.397$, and 
		$t_l/\Delta = 3 \times 10^{-3}$.
	}
	
\end{figure}

\begin{figure}[H]
	\centering
	\includegraphics[width=\linewidth]{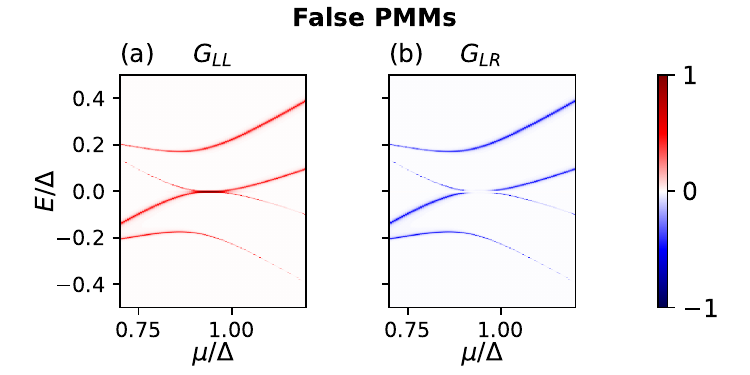}
	\caption{The parameters are as given in Fig.~\ref{fig:NSN false no U no fish 2 energy_pol_charge}, i.e.,
		$t/\Delta=0.64$, $\Phi_\mathrm{SOI} = 0.12 \pi$, $U=0$, and $\Delta_Z/\Delta = 0.8$, $\mu_M/\Delta=0.870$, and 
		$t_l/\Delta =  4.5 \times 10^{-3}$.
	}
	
\end{figure}

\begin{figure}[H]
	\centering
	\includegraphics[width=\linewidth]{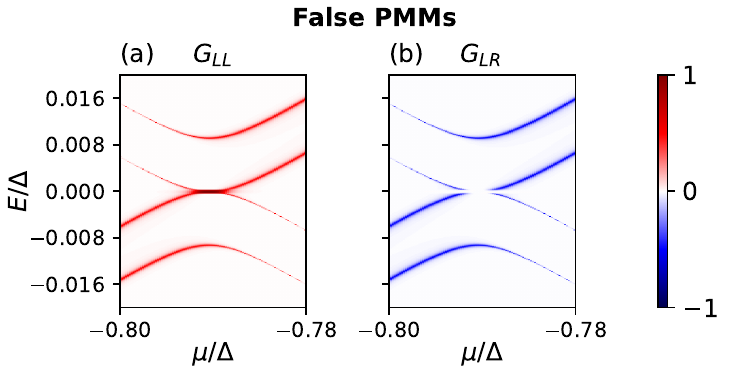}
	\caption{The parameters are as given in Fig.~\ref{fig:NSN false no U no fish 3 energy_pol_charge}, i.e., $t/\Delta=0.15$, $\Phi_\mathrm{SOI} = 0.34 \pi$, $U=0$, $\Delta_Z/\Delta = 0.8$, $\mu_M/\Delta=1.793$, and 
		$t_l/\Delta =  3 \times 10^{-4}$.
	}
	\label{fig:finite_energy_conductance_both_QD_last_fig}
\end{figure}

\FloatBarrier

}

\section{\label{app:analytics}Analytics of the trivial states in the TR}
In this Appendix, we explain the trivial fourfold degenerate zero-energy state at $\Delta_Z/\Delta=0.8$ in Fig.~\ref{fig:energy_vs_delta_z_combined}(f). Although many of the false PMMs are related to such a fourfold degenerate zero-energy state, we emphasize that this is not necessary for false PMMs.

Let us consider a double QD system without superconductivity and where only one of the QDs has a Zeeman term. Using the basis $\Psi=(d_{1,\up}, d_{1,\dn}, d_{2,\up}, d_{2,\dn})^T$, the double QD system is described by the Hamiltonian
\begin{subequations}
	\begin{align} 
		&H_\mathrm{2QD} = \Psi^\dagger \mathcal{H}_\mathrm{2QD} \Psi,\\
		&\!\!\! \mathcal{H}_\mathrm{2QD} \! = \!\!  
		\left( \!\!
		\begin{smallmatrix}
			\mu + \Delta_Z & 0 & t \cos \left( \frac{\Phi_\mathrm{SOI}}{2} \right) & t \sin \left( \frac{\Phi_\mathrm{SOI}}{2} \right) \\
			0 & \mu - \Delta_Z & -t \sin \left( \frac{\Phi_\mathrm{SOI}}{2} \right)& t \cos \left( \frac{\Phi_\mathrm{SOI}}{2} \right) \\
			t \cos \left( \frac{\Phi_\mathrm{SOI}}{2} \right) & -t \sin \left( \frac{\Phi_\mathrm{SOI}}{2} \right) & \mu_M & 0 \\
			t \sin \left( \frac{\Phi_\mathrm{SOI}}{2} \right) & t \cos \left( \frac{\Phi_\mathrm{SOI}}{2} \right) & 0 & \mu_M
		\end{smallmatrix}
		\!\! \right) \!
		.
	\end{align}
\end{subequations}
Using $\mu/\Delta= 0.593$, $\mu_M/\Delta=-3.836$, $t/\Delta=0.99$, and $\Phi_\mathrm{SOI}=0.44\pi$, the parameters for Fig.~\ref{fig:energy_vs_delta_z_combined}(f), we find that $H_\mathrm{2QD}$ has a zero-energy state at $\Delta_Z/\Delta = 0.849$. The corresponding eigenstate is
\begin{align}
	\Psi = (
	0, 0.968, -0.159, 0.193
	)^T,
\end{align}
which is mostly localized on the first QD. As a result of this, adding a superconducting pairing potential to the second dot does not significantly change the solution, except that it becomes doubly degenerate due to particle-hole symmetry. Additionally, adding more QDs to the right of the second QD has little effect on this highly localized zero-energy state. Therefore, this explains the appearance of a highly localized zero-energy state on the left end of the long chain. One can argue equivalently for the right end of the long chain, and therefore the degeneracy doubles, leading to a fourfold degenerate, highly localized zero-energy state in a long chain.
In Fig.~\ref{fig:energy_vs_delta_z_combined}(f), we find the fourfold degenerate zero-energy state at $\Delta_Z/\Delta \approx 0.8$, not at $\Delta_Z/\Delta = 0.849$. We explain this difference by the fact that the eigenstate is not perfectly localized to only the first QD, which means that adding superconductivity and more QDs affects the state slightly.

\section{\label{app:SNSNS} False PMM examples of chains starting and ending with a superconductor}

{\raggedbottom
\begin{figure}[H]
	\centering
	\includegraphics[width=\linewidth]{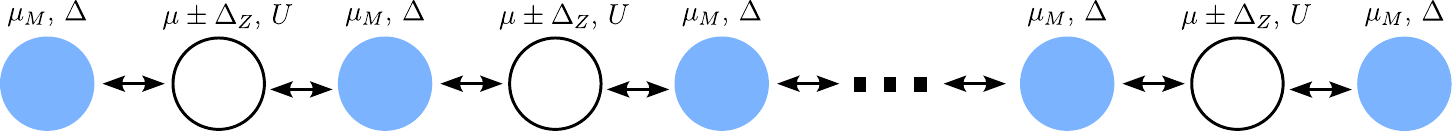}
	\caption{Slightly adapted long chain limit, where, compared to the setup in Fig.~\ref{fig:setup}(b), the chain starts and ends with a superconducting QD. Therefore, each normal QD is now directly coupled to two superconducting QDs.
	}
	\label{fig:setup_long_chain_SNSNS}
\end{figure}

We have shown that twofold degenerate energy levels split away from the bulk states because the outer QDs are fundamentally different compared to the QDs in the bulk. This fundamental difference stems from the fact that the two outer normal QDs are directly coupled to only one superconducting QD, whereas the normal QDs couple to two superconducting QDs.
If one adds additional superconducting QDs to the left and right end of the chain (see Fig.~\ref{fig:setup_long_chain_SNSNS}), then it appears that all normal QDs are equivalent, since now each normal QD couples directly two two superconducting QDs. This is, however, not the case because higher-order tunneling terms still distinguish between the QDs at the ends of the chain and in the bulk. Therefore, we still find localized states on these sites with energies that split away from bulk states in the long chain limit.

As was the case for the chains starting and ending with a normal QD, qualitatively, the energy difference, charge difference, and Majorana polarization for a true PMM (Fig.~\ref{fig:SNSNS real 1 energy_pol_charge}) and a false PMM (Fig.~\ref{fig:SNSNS false 4 energy_pol_charge}) do not differ for a chain starting and ending with a superconducting QD. Only the energy spectrum and topological invariant in a long (infinite) chain (Figs.~\ref{fig:SNSNS real 1 zeeman} and~\ref{fig:SNSNS false 4 zeeman}) reveal whether the states are topological or trivial.

\begin{figure}[H] 
	\centering
	\includegraphics[width=\linewidth]{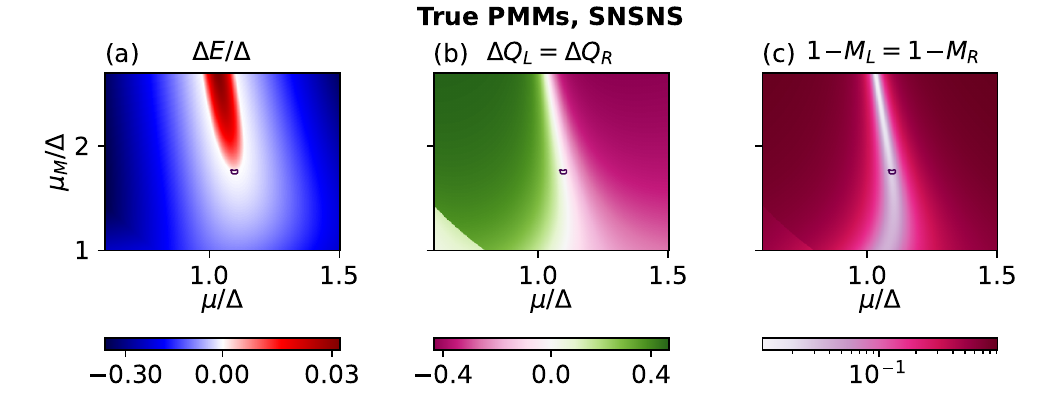}
	\caption{
		The parameters are $t/\Delta=0.6$, $\Phi_\mathrm{SOI} = 0.26 \pi$, $U=0$, and $\Delta_Z/\Delta=0.8$.
		The threshold values for the TR are $\Delta E_\mathrm{th}/\Delta = 10^{-3}$, $\Delta Q_\mathrm{th}=0.05$, $M_\mathrm{th}=0.05$, and the maximum excitation gap in the TR is $E_\mathrm{ex}/\Delta=0.154$.
	}
	\label{fig:SNSNS real 1 energy_pol_charge}
\end{figure}

\begin{figure}[H] 
	\centering
	\includegraphics[width=\linewidth]{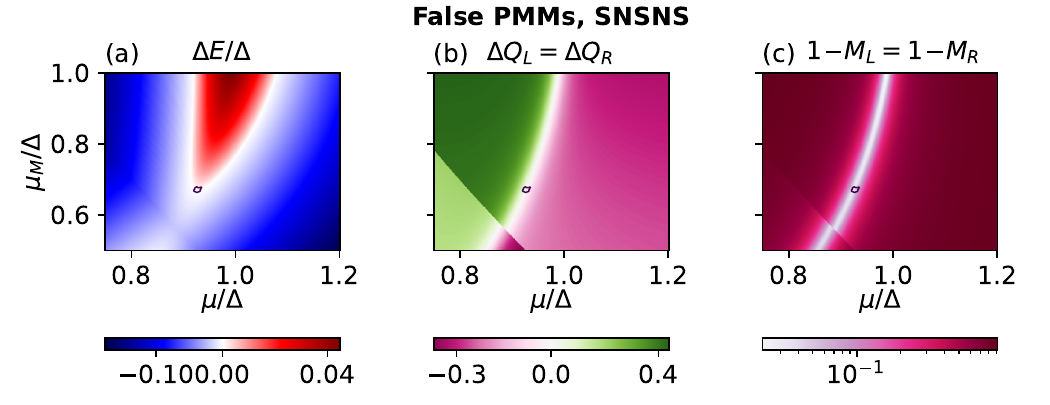}
	\caption{
		The parameters are $t/\Delta=0.5$, $\Phi_\mathrm{SOI} = 0.04 \pi$, $U=0$, and $\Delta_Z/\Delta=0.8$.
		The threshold values for the TR are $\Delta E_\mathrm{th}/\Delta = 10^{-3}$, $\Delta Q_\mathrm{th}=0.07$, $M_\mathrm{th}=0.07$, and the maximum excitation gap in the TR is $E_\mathrm{ex}/\Delta=0.042$.
	}
	\label{fig:SNSNS false 4 energy_pol_charge}
\end{figure}

\begin{figure}[H]
	\centering
	\includegraphics[width=\linewidth]{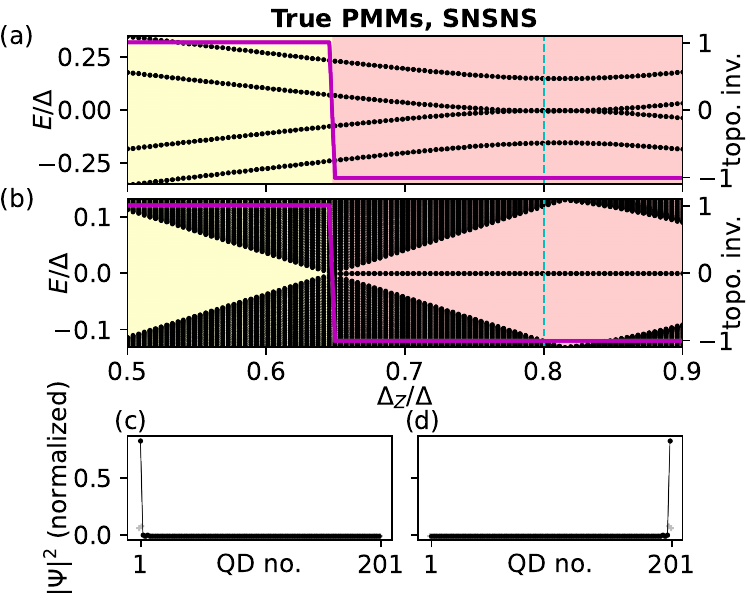}
	\caption{
		The parameters are as given in Fig.~\ref{fig:SNSNS real 1 energy_pol_charge}, i.e.,
		$t/\Delta=0.6$, $\Phi_\mathrm{SOI} = 0.26 \pi$, $U=0$,
		$\mu/\Delta = 1.096$, and $\mu_M/\Delta= 1.771$.
		We can see that these parameters in the long chain limit result in a topological MBS.
	}
	\label{fig:SNSNS real 1 zeeman}
\end{figure}

\begin{figure}[H]
	\centering
	\includegraphics[width=\linewidth]{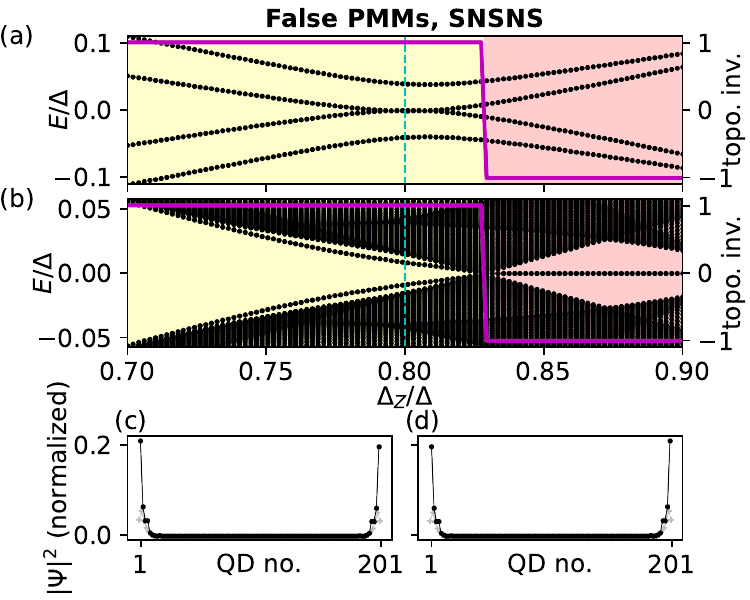}
	\caption{
		The parameters are as given in Fig.~\ref{fig:SNSNS false 4 energy_pol_charge}, i.e.,
		$t/\Delta=0.5$, $\Phi_\mathrm{SOI} = 0.04 \pi$, $U=0$, $\mu/\Delta = 0.926$, and $\mu_M/\Delta= 0.674$. We can see that these parameters in the long chain limit do not result in topological MBSs and therefore these states are false PMMs.
	}
	\label{fig:SNSNS false 4 zeeman}
\end{figure}

\subsection{Conductance}
When calculating the conductance in a long chain that starts and ends with a superconducting QD, one must first decide if the leads are attached to the superconducting QDs, or to the first normal QDs. Figures~\ref{fig:SNSNS real 1 zeeman} and~\ref{fig:SNSNS false 4 zeeman} show that the zero-energy states are mainly localized on the first normal QDs. Thus, attaching the leads to the outer superconducting QDs adds a further complication to the system, because these superconducting QDs act as insulators between the leads and the probed state. Consequently, we attached the leads to the first normal QDs.

The zero-energy conductance simulations (Figs.~\ref{fig:SNSNS real 1 zero_conductance} and~\ref{fig:SNSNS false 4 zero_conductance}) as well as the finite-energy conductance simulations where the chemical potential on one QD (Figs.~\ref{fig:SNSNS real 1 detune_one} and~\ref{fig:SNSNS false 4 detune_one}) or on two QDs (Figs.~\ref{fig:SNSNS real 1 detune_two} and~\ref{fig:SNSNS false 4 detune_two}) is tuned away from the ideal TR value, do not distinguish true PMMs from false PMMs.

\begin{figure}[H]
	\centering
	\includegraphics[width=\linewidth]{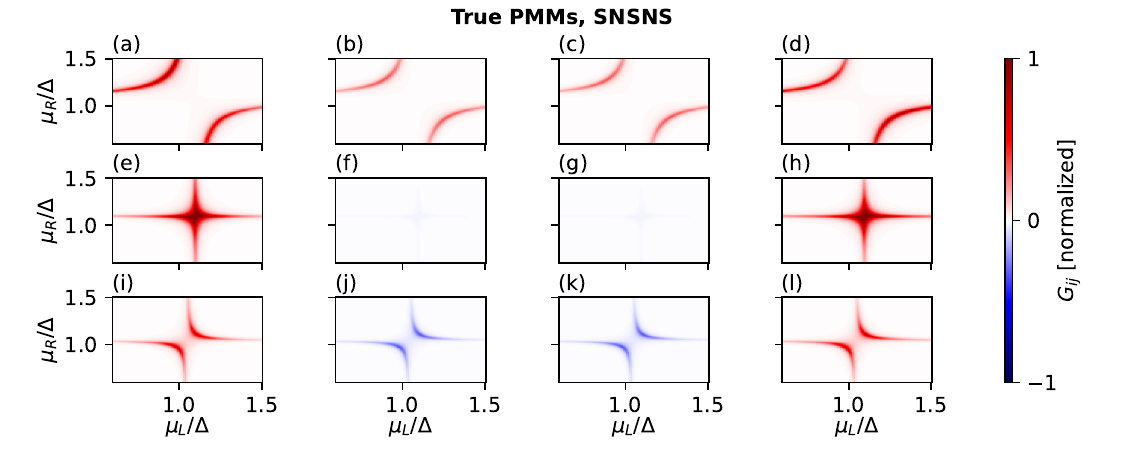}
	\caption{The parameters are as given in Fig.~\ref{fig:SNSNS real 1 energy_pol_charge}, i.e.,
		$t/\Delta=0.6$, $\Phi_\mathrm{SOI} = 0.26 \pi$, $U=0$, $\Delta_Z/\Delta=0.8$,
		$t_l/\Delta = 10^{-2}$, $\mu_{M,1}/\Delta = 1.000$, $\mu_{M,2}/\Delta = 1.771$, $\mu_{M,3}/\Delta = 2.543$.
	}
	\label{fig:SNSNS real 1 zero_conductance}
\end{figure}

\begin{figure}[H]
	\centering
	\includegraphics[width=\linewidth]{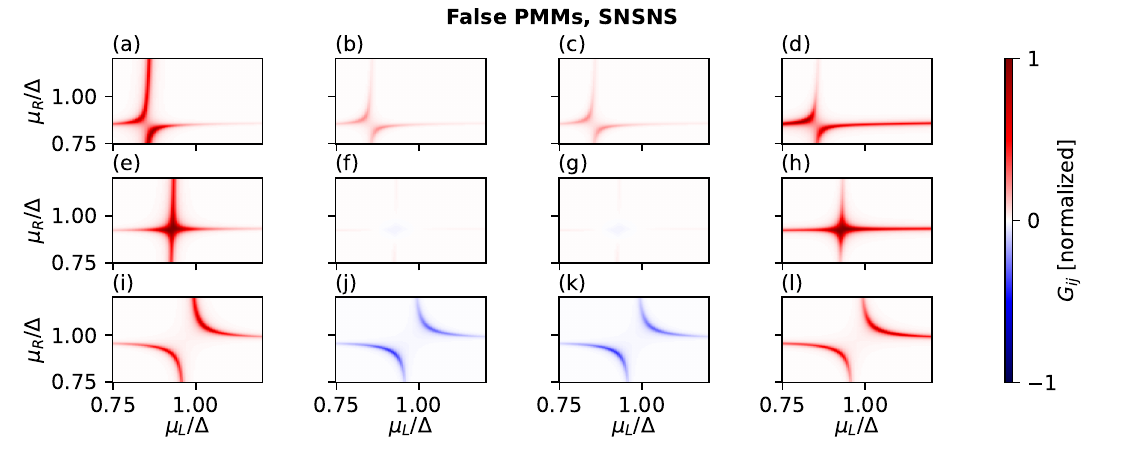}
	\caption{The parameters are as given in Fig.~\ref{fig:SNSNS false 4 energy_pol_charge}, i.e.,
		$t/\Delta=0.5$, $\Phi_\mathrm{SOI} = 0.04 \pi$, $U=0$, $\Delta_Z/\Delta=0.8$,
		$t_l/\Delta = 5 \times 10^{-3}$, $\mu_{M,1}/\Delta = 0.500$, $\mu_{M,2}/\Delta = 0.674$, $\mu_{M,3}/\Delta = 0.848$.
	}
	\label{fig:SNSNS false 4 zero_conductance}
\end{figure}

\begin{figure}[H]
	\centering
	\includegraphics[width=\linewidth]{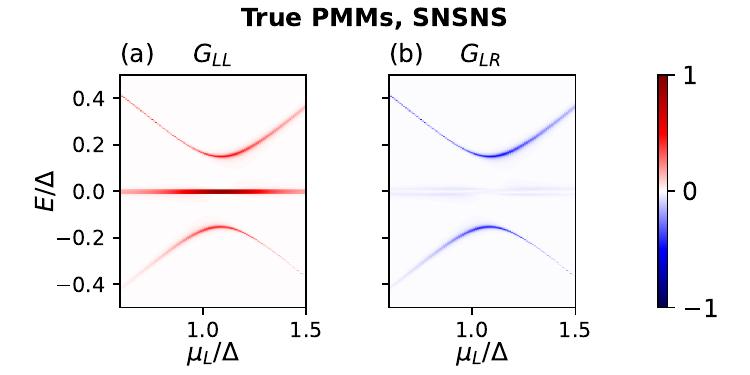}
	\caption{The parameters are as given in Fig.~\ref{fig:SNSNS real 1 energy_pol_charge}, i.e.,
		$t/\Delta=0.6$, $\Phi_\mathrm{SOI} = 0.26 \pi$, $U=0$, $\Delta_Z/\Delta=0.8$, $\mu_R/\Delta=1.096$, $\mu_M/\Delta=1.771$, and
		$t_l/\Delta = 0.01$.
	}
	\label{fig:SNSNS real 1 detune_one}
\end{figure}

\begin{figure}[H]
	\centering
	\includegraphics[width=\linewidth]{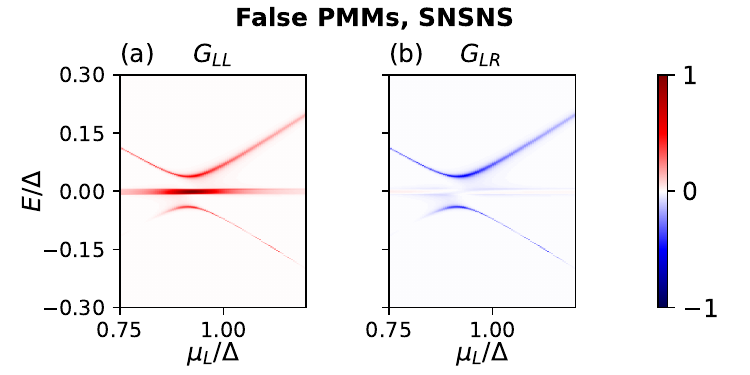}
	\caption{The parameters are as given in Fig.~\ref{fig:SNSNS false 4 energy_pol_charge}, i.e.,
		$t/\Delta=0.5$, $\Phi_\mathrm{SOI} = 0.04 \pi$, $U=0$, $\Delta_Z/\Delta=0.8$, $\mu_R/\Delta=0.926$, $\mu_M/\Delta=0.674$, and 
		$t_l/\Delta = 7 \times 10^{-3}$.
	}
	\label{fig:SNSNS false 4 detune_one}
\end{figure}

\begin{figure}[H]
	\centering
	\includegraphics[width=\linewidth]{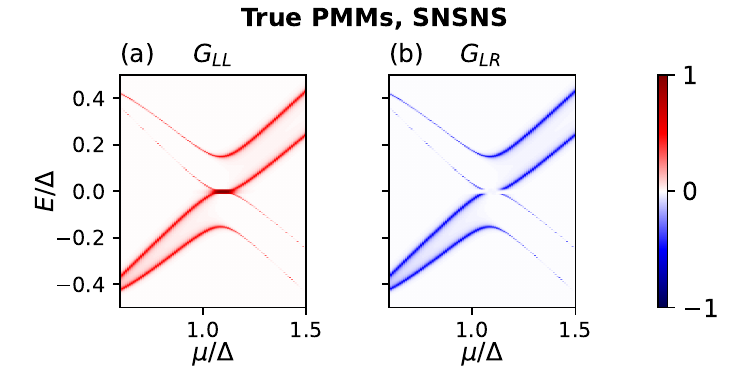}
	\caption{The parameters are as given in Fig.~\ref{fig:SNSNS real 1 energy_pol_charge}, i.e.,
		$t/\Delta=0.6$, $\Phi_\mathrm{SOI} = 0.26 \pi$, $U=0$, $\Delta_Z/\Delta=0.8$, $\mu_M/\Delta=1.771$, and 
		$t_l/\Delta = 0.01$.
	}
	\label{fig:SNSNS real 1 detune_two}
\end{figure}

\begin{figure}[H]
	\centering
	\includegraphics[width=\linewidth]{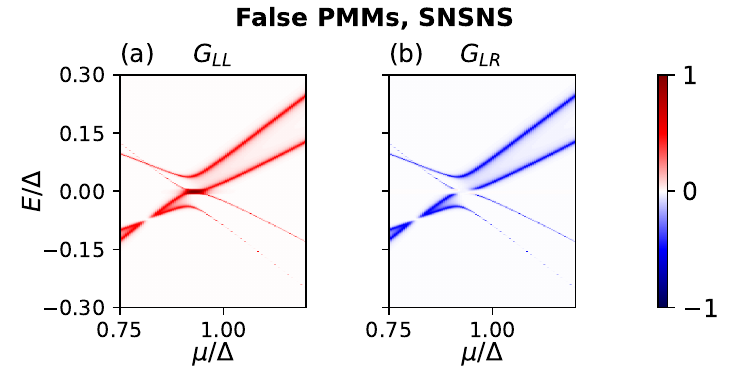}
	\caption{The parameters are as given in Fig.~\ref{fig:SNSNS false 4 energy_pol_charge}, i.e.,
		$t/\Delta=0.5$, $\Phi_\mathrm{SOI} = 0.04 \pi$, $U=0$, $\Delta_Z/\Delta=0.8$, $\mu_M/\Delta=0.674$, and 
		$t_l/\Delta = 7 \times 10^{-3}$.
	}
	\label{fig:SNSNS false 4 detune_two}
\end{figure}

}

\section{\label{app:scaled_pmms}Scaled PMMs in longer chains}
\begin{figure}
	\centering
	\includegraphics[width=\linewidth]{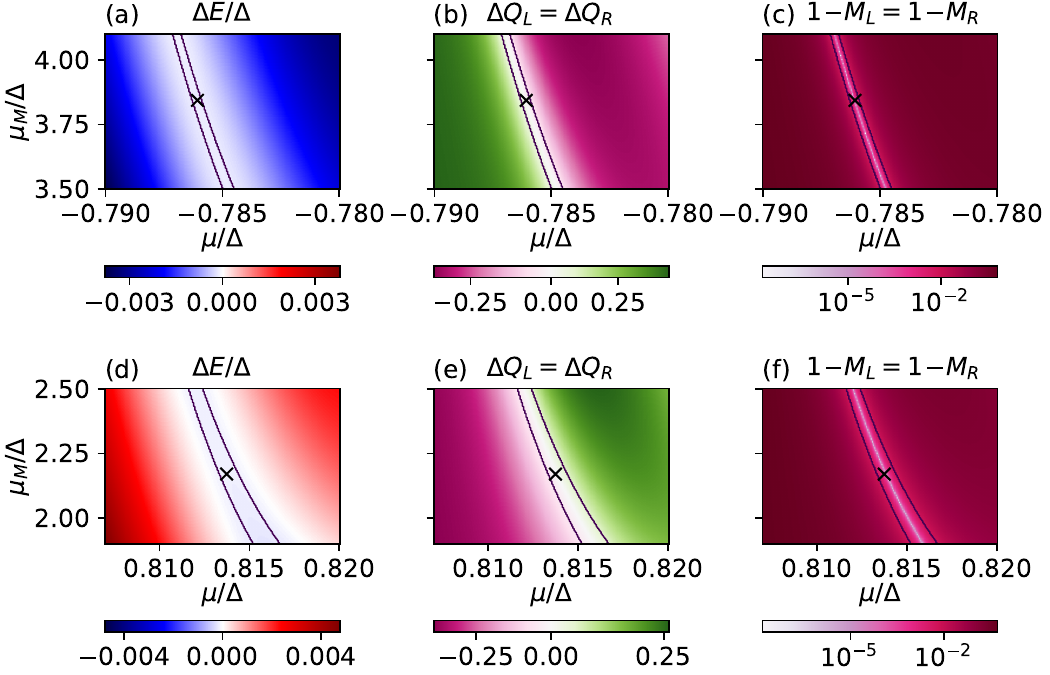}
	\caption{
		Quality measures for scaled PMMs in systems consisting of three normal QDs (and two superconducting QDs) that evolve into trivial states in the long chain limit. The parameters for (a)--(c) are $t/\Delta=0.24$, $\Phi_\mathrm{SOI}=0.42\pi$, $U=0$, $\Delta_Z/\Delta=0.8$. The cross is at $\mu/\Delta = -0.786$, $\mu_M=3.843$ and it indicates where in parameter space the long chain limit shown in Fig.~\ref{fig:zeeman_first_figure}(b) is taken. The largest excitation gap in the TR for (a)--(c) is $E_\mathrm{ex}/\Delta=0.003$. 
		The parameters for (d)--(f) are $t/\Delta=0.19$, $\Phi_\mathrm{SOI}=0.36\pi$, $U=0$, $\Delta_Z/\Delta=0.8$. The cross is at $\mu/\Delta = 0.814$, $\mu_M/\Delta=2.170$ 
		and it indicates where in parameter space the long chain limit shown in Fig.~\ref{fig:NSN fake no U fish after 3 energy vs delta_z}(b) is taken. The largest excitation gap in the TR for (d)--(f) is $E_\mathrm{ex}/\Delta=0.008$. 
		In Table~\ref{tab:scaled_pmms_four_QDs} we demonstrate that these parameters also lead to scaled PMMs in chains with four normal QDs (and three superconducting QDs).
		The threshold values for both examples are $\Delta E_\mathrm{th}/\Delta = 10^{-3}$, $M_\mathrm{th} = 0.05$, and $\Delta Q_\mathrm{th} = 0.05$. 
	}
	\label{fig:scaled_PMMs_three_QDs}
\end{figure}

In this Appendix, we consider QD chains similar to the ones shown in Fig.~\ref{fig:setup}, but consisting of more than two normal QDs. We assume uniform parameters throughout the system. The energy difference $\Delta E$, excitation gap $E_\mathrm{ex}$, charge difference $\Delta Q_j$, and Majorana polarization $M_j$ are calculated as explained in the main text, but $\Delta Q_j$ and $M_j$ are calculated for the two outermost normal QDs. We call states in these longer QD chains that fulfill the TR condition defined in Eq.~\eqref{eq:definition_ROT} scaled PMMs. 
We show two examples of scaled PMMs in a system with three normal QDs (and two superconducting QDs) that evolve into trivial states in the long chain limit in Fig.~\ref{fig:scaled_PMMs_three_QDs}. For these specific examples, both the two and three normal site chains satisfy the TR conditions, and we have shown already in Figs.~\ref{fig:zeeman_first_figure} and~\ref{fig:NSN fake no U fish after 3 energy vs delta_z} that these states evolve into trivial states in the long chain limit. We note that the excitation gaps in these examples are quite small but, since we do not optimize $E_\mathrm{ex}$ and simply require a nonzero gap, these states fulfill the TR condition. We list examples of scaled PMMs in systems with four normal QDs (and three superconducting QDs) in Table~\ref{tab:scaled_pmms_four_QDs}. All of these examples evolve into trivial states in the long chain limit.

\begin{table}
	\centering
	\resizebox{\columnwidth}{!}{
	\begin{tabular}{lcccc}
		\hline \hline
		$\Delta_Z/\Delta$ & 0.800 & 0.802 & 0.827 & 0.800 \\
		$t/\Delta$ & 0.24 & 0.7 & 1.27 & 0.19 \\
		$\Phi_\mathrm{SOI}/\pi$ & 0.42 & 0.38 & 0.38 & 0.36 \\
		$\mu/\Delta$ & -0.786 & 0.953 & 1.239 & 0.814 \\
		$\mu_M/\Delta$ & 3.843 & 3.086 & 4.110 & 2.170 \\
		$|\Delta E|/\Delta$ & $3.9 \times 10^{-7}$ & $6.7 \times 10^{-5}$ & $4.4 \times 10^{-4}$ & $3.3 \times 10^{-10}$ \\
		$E_\mathrm{ex}/\Delta$ & $1.3 \times 10^{-3}$ & $3.8 \times 10^{-2}$ & 0.11 & $4.5 \times 10^{-3}$ \\
		$|\Delta Q_L| = |\Delta Q_R|$ & $4.6 \times 10^{-3}$ & $2.9 \times 10^{-2}$ & $7.6 \times 10^{-3}$ & $2.9 \times 10^{-4}$ \\
		$1-M_L = 1-M_R$ & $4.2 \times 10^{-5}$ & $2.1 \times 10^{-3}$ & $5.7 \times 10^{-3}$ & $1.9 \times 10^{-7}$ \\
		Figure of $E(\Delta_Z)$  in & \multirow{ 2}{*}{\ref{fig:zeeman_first_figure} } & \multirow{ 2}{*}{\ref{fig:NSN fake no U fish after 1 energy vs delta_z}} & \multirow{ 2}{*}{\ref{fig:NSN fake no U fish after 2 energy vs delta_z}} & \multirow{ 2}{*}{\ref{fig:NSN fake no U fish after 3 energy vs delta_z}} \\
		long chain limit &  & & & \\
		\hline \hline
	\end{tabular}
	}
	\caption{
		Parameter values, energy difference $\Delta E$, excitation gap $E_\mathrm{ex}$, charge difference $\Delta Q_j$, and Majorana polarization $M_j$ for scaled PMMs in systems with four normal QDs (and three superconducting QDs). All of these states are highly localized near-zero-energy states that evolve into trivial states in the long chain limit, as shown in the figures referred to in the last row of the table. For simplicity, we set $U=0$ in all cases.
	}
	\label{tab:scaled_pmms_four_QDs}
\end{table}

\section{\label{app:spectrum_vs_N}Energy spectrum as a function of chain length}
\begin{figure}
	\centering
	\includegraphics[width=\linewidth]{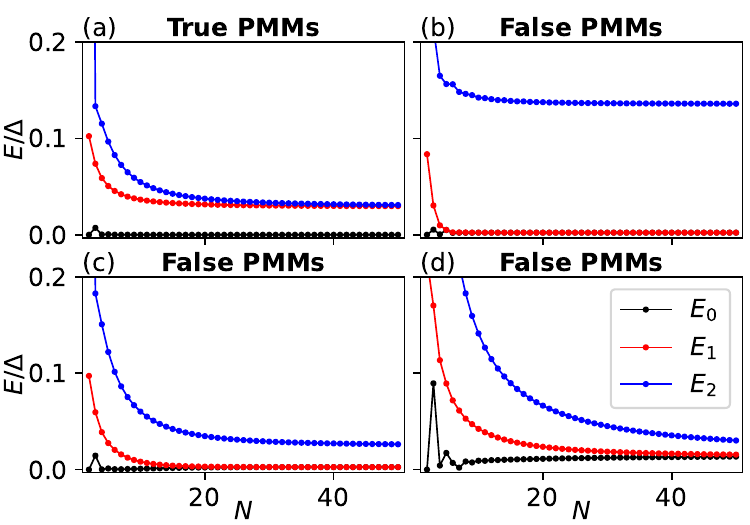}
	\caption{
		Energy levels of the BdG formulation of the Hamiltonian given in Eq.~\eqref{eq:full_hamiltonian}. The spectrum is symmetric around zero energy and the energy levels are labeled as $\pm E_0, \pm E_1, \pm E_2, \dots$, with $0 \leq E_0 \leq E_1 \leq E_2, \dots$.
		For both true and false PMMs, one finds examples where $E_0$ stays at, or close to, zero energy as the number of normal sites $N$ increases. The first excited energy level $E_1$, however, tends towards a finite value for true PMMs, indicating the topological gap. For false PMMs, $E_1$ tends towards $E_0$ as $N$ increases, thus indicating the fourfold degenerate zero-energy states found in association with false PMMs. The second excited energy level $E_2$ tends towards $E_1$ for true PMMs and towards a finite value away from $E_1$ for false PMMs.
		In some cases, a few sites are sufficient to distinguish the two cases [(b)], in other cases [(c) and (d)], many sites are required.
		For (a), the same parameters as for Fig.~\ref{fig:energy_vs_delta_z_combined}(a) are used, i.e.,
		$t/\Delta=0.42$, $\Phi_\mathrm{SOI} =  0.26 \pi$, $U=0$, $\mu/\Delta = 0.884$, and $\mu_M/\Delta= 1.275$.
		For (b), the same parameters as for Fig.~\ref{fig:energy_vs_delta_z_combined}(e) are used,
		i.e.,
		$t/\Delta= 0.99$, $\Phi_\mathrm{SOI} =  0.44\pi$, $U=0$, $\mu/\Delta = 0.592$, and $\mu_M/\Delta= -3.836$.
		For (c), the same parameters as for Fig.~\ref{fig:NSN fake no U fish after 1 energy vs delta_z} are used, i.e.,
		$t/\Delta=0.7$, $\Phi_\mathrm{SOI} = 0.38 \pi$, $U=0$,
		$\mu/\Delta = 0.953$, and $\mu_M/\Delta= 3.086$.
		For (d), the same parameters as for Fig.~\ref{fig:NSN fake no U fish after 2 energy vs delta_z} are used, i.e.,
		$t/\Delta=1.27$, $\Phi_\mathrm{SOI} = 0.38 \pi$, $U=0$, 
		$\mu/\Delta = 1.239$, and $\mu_M/\Delta= 4.110$.
		In all four panels, $\Delta_Z/\Delta=0.8$.
	}
	\label{fig:spectrum_vs_N}
\end{figure}

In the case $U=0$, the system can be described by the BdG formulation of the Hamiltonian given in Eq.~\eqref{eq:full_hamiltonian}. In this case, the energy spectrum is symmetric around zero energy and we label the energy levels as $\pm E_0, \pm E_1, \pm E_2, \dots$, with $0 \leq E_0 \leq E_1 \leq E_2, \dots$. Analyzing these energy levels as a function of the number of normal sites $N$ (with one superconducting site between two normal sites, i.e., there are $2N-1$ QDs in total) can help to distinguish between true and false PMMs, see Fig.~\ref{fig:spectrum_vs_N}. In all examples shown, $E_0$ stays at, or close to, zero energy. However, in the case of true PMMs, the first excitation energy, $E_1$, tends towards a finite value, whereas for false PMMs, $E_1$ tends towards $E_0$, thus indicating the fourfold degenerate zero-energy states observed for some false PMMs. The second excited energy level, $E_2$ tends towards $E_1$ in the case of a true PMMs, whereas it tends towards a finite value away from $E_1 \approx E_0$ for false PMMs.
In some cases, a few normal sites are sufficient to distinguish between the two cases, see Fig.~\ref{fig:spectrum_vs_N}(b). In other cases, chains with tens of normal sites are requires, see Figs.~\ref{fig:spectrum_vs_N}(c) and~\ref{fig:spectrum_vs_N}(d).

\FloatBarrier

\bibliography{bibliography}

\begin{thebibliography}{96}%
\makeatletter
\providecommand \@ifxundefined [1]{%
 \@ifx{#1\undefined}
}%
\providecommand \@ifnum [1]{%
 \ifnum #1\expandafter \@firstoftwo
 \else \expandafter \@secondoftwo
 \fi
}%
\providecommand \@ifx [1]{%
 \ifx #1\expandafter \@firstoftwo
 \else \expandafter \@secondoftwo
 \fi
}%
\providecommand \natexlab [1]{#1}%
\providecommand \enquote  [1]{``#1''}%
\providecommand \bibnamefont  [1]{#1}%
\providecommand \bibfnamefont [1]{#1}%
\providecommand \citenamefont [1]{#1}%
\providecommand \href@noop [0]{\@secondoftwo}%
\providecommand \href [0]{\begingroup \@sanitize@url \@href}%
\providecommand \@href[1]{\@@startlink{#1}\@@href}%
\providecommand \@@href[1]{\endgroup#1\@@endlink}%
\providecommand \@sanitize@url [0]{\catcode `\\12\catcode `\$12\catcode
  `\&12\catcode `\#12\catcode `\^12\catcode `\_12\catcode `\%12\relax}%
\providecommand \@@startlink[1]{}%
\providecommand \@@endlink[0]{}%
\providecommand \url  [0]{\begingroup\@sanitize@url \@url }%
\providecommand \@url [1]{\endgroup\@href {#1}{\urlprefix }}%
\providecommand \urlprefix  [0]{URL }%
\providecommand \Eprint [0]{\href }%
\providecommand \doibase [0]{https://doi.org/}%
\providecommand \selectlanguage [0]{\@gobble}%
\providecommand \bibinfo  [0]{\@secondoftwo}%
\providecommand \bibfield  [0]{\@secondoftwo}%
\providecommand \translation [1]{[#1]}%
\providecommand \BibitemOpen [0]{}%
\providecommand \bibitemStop [0]{}%
\providecommand \bibitemNoStop [0]{.\EOS\space}%
\providecommand \EOS [0]{\spacefactor3000\relax}%
\providecommand \BibitemShut  [1]{\csname bibitem#1\endcsname}%
\let\auto@bib@innerbib\@empty
\bibitem [{\citenamefont {Kitaev}(2001)}]{kitaev2001unpaired}%
  \BibitemOpen
  \bibfield  {author} {\bibinfo {author} {\bibfnamefont {A.~Y.}\ \bibnamefont
  {Kitaev}},\ }\bibfield  {title} {\bibinfo {title} {{Unpaired Majorana
  fermions in quantum wires}},\ }\href
  {https://doi.org/10.1070/1063-7869/44/10S/S29} {\bibfield  {journal}
  {\bibinfo  {journal} {Physics-uspekhi}\ }\textbf {\bibinfo {volume} {44}},\
  \bibinfo {pages} {131} (\bibinfo {year} {2001})}\BibitemShut {NoStop}%
\bibitem [{\citenamefont {Ivanov}(2001)}]{ivanov2001non}%
  \BibitemOpen
  \bibfield  {author} {\bibinfo {author} {\bibfnamefont {D.~A.}\ \bibnamefont
  {Ivanov}},\ }\bibfield  {title} {\bibinfo {title} {{Non-Abelian Statistics of
  Half-Quantum Vortices in $\mathit{p}$-Wave Superconductors}},\ }\href
  {https://doi.org/10.1103/PhysRevLett.86.268} {\bibfield  {journal} {\bibinfo
  {journal} {Phys. Rev. Lett.}\ }\textbf {\bibinfo {volume} {86}},\ \bibinfo
  {pages} {268} (\bibinfo {year} {2001})}\BibitemShut {NoStop}%
\bibitem [{\citenamefont {Kitaev}(2003)}]{kitaev2003fault}%
  \BibitemOpen
  \bibfield  {author} {\bibinfo {author} {\bibfnamefont {A.}~\bibnamefont
  {Kitaev}},\ }\bibfield  {title} {\bibinfo {title} {{Fault-tolerant quantum
  computation by anyons}},\ }\href
  {https://doi.org/https://doi.org/10.1016/S0003-4916(02)00018-0} {\bibfield
  {journal} {\bibinfo  {journal} {Annals of Physics}\ }\textbf {\bibinfo
  {volume} {303}},\ \bibinfo {pages} {2} (\bibinfo {year} {2003})}\BibitemShut
  {NoStop}%
\bibitem [{\citenamefont {Nayak}\ \emph {et~al.}(2008)\citenamefont {Nayak},
  \citenamefont {Simon}, \citenamefont {Stern}, \citenamefont {Freedman},\ and\
  \citenamefont {Das~Sarma}}]{nayak2008non}%
  \BibitemOpen
  \bibfield  {author} {\bibinfo {author} {\bibfnamefont {C.}~\bibnamefont
  {Nayak}}, \bibinfo {author} {\bibfnamefont {S.~H.}\ \bibnamefont {Simon}},
  \bibinfo {author} {\bibfnamefont {A.}~\bibnamefont {Stern}}, \bibinfo
  {author} {\bibfnamefont {M.}~\bibnamefont {Freedman}},\ and\ \bibinfo
  {author} {\bibfnamefont {S.}~\bibnamefont {Das~Sarma}},\ }\bibfield  {title}
  {\bibinfo {title} {{Non-Abelian anyons and topological quantum
  computation}},\ }\href {https://doi.org/10.1103/RevModPhys.80.1083}
  {\bibfield  {journal} {\bibinfo  {journal} {Rev. Mod. Phys.}\ }\textbf
  {\bibinfo {volume} {80}},\ \bibinfo {pages} {1083} (\bibinfo {year}
  {2008})}\BibitemShut {NoStop}%
\bibitem [{\citenamefont {Elliott}\ and\ \citenamefont
  {Franz}(2015)}]{elliot2015colloquium}%
  \BibitemOpen
  \bibfield  {author} {\bibinfo {author} {\bibfnamefont {S.~R.}\ \bibnamefont
  {Elliott}}\ and\ \bibinfo {author} {\bibfnamefont {M.}~\bibnamefont
  {Franz}},\ }\bibfield  {title} {\bibinfo {title} {{Colloquium: Majorana
  fermions in nuclear, particle, and solid-state physics}},\ }\href
  {https://doi.org/10.1103/RevModPhys.87.137} {\bibfield  {journal} {\bibinfo
  {journal} {Rev. Mod. Phys.}\ }\textbf {\bibinfo {volume} {87}},\ \bibinfo
  {pages} {137} (\bibinfo {year} {2015})}\BibitemShut {NoStop}%
\bibitem [{\citenamefont {Lutchyn}\ \emph {et~al.}(2010)\citenamefont
  {Lutchyn}, \citenamefont {Sau},\ and\ \citenamefont
  {Das~Sarma}}]{lutchyn2010majorana}%
  \BibitemOpen
  \bibfield  {author} {\bibinfo {author} {\bibfnamefont {R.~M.}\ \bibnamefont
  {Lutchyn}}, \bibinfo {author} {\bibfnamefont {J.~D.}\ \bibnamefont {Sau}},\
  and\ \bibinfo {author} {\bibfnamefont {S.}~\bibnamefont {Das~Sarma}},\
  }\bibfield  {title} {\bibinfo {title} {{Majorana Fermions and a Topological
  Phase Transition in Semiconductor-Superconductor Heterostructures}},\ }\href
  {https://doi.org/10.1103/PhysRevLett.105.077001} {\bibfield  {journal}
  {\bibinfo  {journal} {Phys. Rev. Lett.}\ }\textbf {\bibinfo {volume} {105}},\
  \bibinfo {pages} {077001} (\bibinfo {year} {2010})}\BibitemShut {NoStop}%
\bibitem [{\citenamefont {Oreg}\ \emph {et~al.}(2010)\citenamefont {Oreg},
  \citenamefont {Refael},\ and\ \citenamefont {von Oppen}}]{oreg2010helical}%
  \BibitemOpen
  \bibfield  {author} {\bibinfo {author} {\bibfnamefont {Y.}~\bibnamefont
  {Oreg}}, \bibinfo {author} {\bibfnamefont {G.}~\bibnamefont {Refael}},\ and\
  \bibinfo {author} {\bibfnamefont {F.}~\bibnamefont {von Oppen}},\ }\bibfield
  {title} {\bibinfo {title} {{Helical Liquids and Majorana Bound States in
  Quantum Wires}},\ }\href {https://doi.org/10.1103/PhysRevLett.105.177002}
  {\bibfield  {journal} {\bibinfo  {journal} {Phys. Rev. Lett.}\ }\textbf
  {\bibinfo {volume} {105}},\ \bibinfo {pages} {177002} (\bibinfo {year}
  {2010})}\BibitemShut {NoStop}%
\bibitem [{\citenamefont {Stanescu}\ \emph {et~al.}(2011)\citenamefont
  {Stanescu}, \citenamefont {Lutchyn},\ and\ \citenamefont
  {Das~Sarma}}]{stanescu2011majorana}%
  \BibitemOpen
  \bibfield  {author} {\bibinfo {author} {\bibfnamefont {T.~D.}\ \bibnamefont
  {Stanescu}}, \bibinfo {author} {\bibfnamefont {R.~M.}\ \bibnamefont
  {Lutchyn}},\ and\ \bibinfo {author} {\bibfnamefont {S.}~\bibnamefont
  {Das~Sarma}},\ }\bibfield  {title} {\bibinfo {title} {{Majorana fermions in
  semiconductor nanowires}},\ }\href
  {https://doi.org/10.1103/PhysRevB.84.144522} {\bibfield  {journal} {\bibinfo
  {journal} {Phys. Rev. B}\ }\textbf {\bibinfo {volume} {84}},\ \bibinfo
  {pages} {144522} (\bibinfo {year} {2011})}\BibitemShut {NoStop}%
\bibitem [{\citenamefont {Mourik}\ \emph {et~al.}(2012)\citenamefont {Mourik},
  \citenamefont {Zuo}, \citenamefont {Frolov}, \citenamefont {Plissard},
  \citenamefont {Bakkers},\ and\ \citenamefont
  {Kouwenhoven}}]{mourik2012signatures}%
  \BibitemOpen
  \bibfield  {author} {\bibinfo {author} {\bibfnamefont {V.}~\bibnamefont
  {Mourik}}, \bibinfo {author} {\bibfnamefont {K.}~\bibnamefont {Zuo}},
  \bibinfo {author} {\bibfnamefont {S.~M.}\ \bibnamefont {Frolov}}, \bibinfo
  {author} {\bibfnamefont {S.~R.}\ \bibnamefont {Plissard}}, \bibinfo {author}
  {\bibfnamefont {E.~P. A.~M.}\ \bibnamefont {Bakkers}},\ and\ \bibinfo
  {author} {\bibfnamefont {L.~P.}\ \bibnamefont {Kouwenhoven}},\ }\bibfield
  {title} {\bibinfo {title} {{Signatures of Majorana Fermions in Hybrid
  Superconductor-Semiconductor Nanowire Devices}},\ }\href
  {https://doi.org/10.1126/science.1222360} {\bibfield  {journal} {\bibinfo
  {journal} {Science}\ }\textbf {\bibinfo {volume} {336}},\ \bibinfo {pages}
  {1003} (\bibinfo {year} {2012})}\BibitemShut {NoStop}%
\bibitem [{\citenamefont {Das}\ \emph {et~al.}(2012)\citenamefont {Das},
  \citenamefont {Ronen}, \citenamefont {Most}, \citenamefont {Oreg},
  \citenamefont {Heiblum},\ and\ \citenamefont {Shtrikman}}]{das2012zero}%
  \BibitemOpen
  \bibfield  {author} {\bibinfo {author} {\bibfnamefont {A.}~\bibnamefont
  {Das}}, \bibinfo {author} {\bibfnamefont {Y.}~\bibnamefont {Ronen}}, \bibinfo
  {author} {\bibfnamefont {Y.}~\bibnamefont {Most}}, \bibinfo {author}
  {\bibfnamefont {Y.}~\bibnamefont {Oreg}}, \bibinfo {author} {\bibfnamefont
  {M.}~\bibnamefont {Heiblum}},\ and\ \bibinfo {author} {\bibfnamefont
  {H.}~\bibnamefont {Shtrikman}},\ }\bibfield  {title} {\bibinfo {title}
  {{Zero-bias peaks and splitting in an Al--InAs nanowire topological
  superconductor as a signature of Majorana fermions}},\ }\href
  {https://doi.org/10.1038/nphys2479} {\bibfield  {journal} {\bibinfo
  {journal} {Nature Physics}\ }\textbf {\bibinfo {volume} {8}},\ \bibinfo
  {pages} {887} (\bibinfo {year} {2012})}\BibitemShut {NoStop}%
\bibitem [{\citenamefont {Deng}\ \emph {et~al.}(2012)\citenamefont {Deng},
  \citenamefont {Yu}, \citenamefont {Huang}, \citenamefont {Larsson},
  \citenamefont {Caroff},\ and\ \citenamefont {Xu}}]{deng2012anomalous}%
  \BibitemOpen
  \bibfield  {author} {\bibinfo {author} {\bibfnamefont {M.~T.}\ \bibnamefont
  {Deng}}, \bibinfo {author} {\bibfnamefont {C.~L.}\ \bibnamefont {Yu}},
  \bibinfo {author} {\bibfnamefont {G.~Y.}\ \bibnamefont {Huang}}, \bibinfo
  {author} {\bibfnamefont {M.}~\bibnamefont {Larsson}}, \bibinfo {author}
  {\bibfnamefont {P.}~\bibnamefont {Caroff}},\ and\ \bibinfo {author}
  {\bibfnamefont {H.~Q.}\ \bibnamefont {Xu}},\ }\bibfield  {title} {\bibinfo
  {title} {{Anomalous Zero-Bias Conductance Peak in a Nb–InSb Nanowire–Nb
  Hybrid Device}},\ }\href {https://doi.org/10.1021/nl303758w} {\bibfield
  {journal} {\bibinfo  {journal} {Nano Letters}\ }\textbf {\bibinfo {volume}
  {12}},\ \bibinfo {pages} {6414} (\bibinfo {year} {2012})}\BibitemShut
  {NoStop}%
\bibitem [{\citenamefont {Laubscher}\ and\ \citenamefont
  {Klinovaja}(2021)}]{laubscher2021majorana}%
  \BibitemOpen
  \bibfield  {author} {\bibinfo {author} {\bibfnamefont {K.}~\bibnamefont
  {Laubscher}}\ and\ \bibinfo {author} {\bibfnamefont {J.}~\bibnamefont
  {Klinovaja}},\ }\bibfield  {title} {\bibinfo {title} {{Majorana bound states
  in semiconducting nanostructures}},\ }\href
  {https://doi.org/10.1063/5.0055997} {\bibfield  {journal} {\bibinfo
  {journal} {Journal of Applied Physics}\ }\textbf {\bibinfo {volume} {130}},\
  \bibinfo {pages} {081101} (\bibinfo {year} {2021})}\BibitemShut {NoStop}%
\bibitem [{\citenamefont {Kells}\ \emph {et~al.}(2012)\citenamefont {Kells},
  \citenamefont {Meidan},\ and\ \citenamefont {Brouwer}}]{kells2012near}%
  \BibitemOpen
  \bibfield  {author} {\bibinfo {author} {\bibfnamefont {G.}~\bibnamefont
  {Kells}}, \bibinfo {author} {\bibfnamefont {D.}~\bibnamefont {Meidan}},\ and\
  \bibinfo {author} {\bibfnamefont {P.~W.}\ \bibnamefont {Brouwer}},\
  }\bibfield  {title} {\bibinfo {title} {{Near-zero-energy end states in
  topologically trivial spin-orbit coupled superconducting nanowires with a
  smooth confinement}},\ }\href {https://doi.org/10.1103/PhysRevB.86.100503}
  {\bibfield  {journal} {\bibinfo  {journal} {Phys. Rev. B}\ }\textbf {\bibinfo
  {volume} {86}},\ \bibinfo {pages} {100503} (\bibinfo {year}
  {2012})}\BibitemShut {NoStop}%
\bibitem [{\citenamefont {Lee}\ \emph {et~al.}(2012)\citenamefont {Lee},
  \citenamefont {Jiang}, \citenamefont {Aguado}, \citenamefont {Katsaros},
  \citenamefont {Lieber},\ and\ \citenamefont {De~Franceschi}}]{lee2012zero}%
  \BibitemOpen
  \bibfield  {author} {\bibinfo {author} {\bibfnamefont {E.~J.~H.}\
  \bibnamefont {Lee}}, \bibinfo {author} {\bibfnamefont {X.}~\bibnamefont
  {Jiang}}, \bibinfo {author} {\bibfnamefont {R.}~\bibnamefont {Aguado}},
  \bibinfo {author} {\bibfnamefont {G.}~\bibnamefont {Katsaros}}, \bibinfo
  {author} {\bibfnamefont {C.~M.}\ \bibnamefont {Lieber}},\ and\ \bibinfo
  {author} {\bibfnamefont {S.}~\bibnamefont {De~Franceschi}},\ }\bibfield
  {title} {\bibinfo {title} {{Zero-Bias Anomaly in a Nanowire Quantum Dot
  Coupled to Superconductors}},\ }\href
  {https://doi.org/10.1103/PhysRevLett.109.186802} {\bibfield  {journal}
  {\bibinfo  {journal} {Phys. Rev. Lett.}\ }\textbf {\bibinfo {volume} {109}},\
  \bibinfo {pages} {186802} (\bibinfo {year} {2012})}\BibitemShut {NoStop}%
\bibitem [{\citenamefont {Rainis}\ \emph {et~al.}(2013)\citenamefont {Rainis},
  \citenamefont {Trifunovic}, \citenamefont {Klinovaja},\ and\ \citenamefont
  {Loss}}]{rainis2013towards}%
  \BibitemOpen
  \bibfield  {author} {\bibinfo {author} {\bibfnamefont {D.}~\bibnamefont
  {Rainis}}, \bibinfo {author} {\bibfnamefont {L.}~\bibnamefont {Trifunovic}},
  \bibinfo {author} {\bibfnamefont {J.}~\bibnamefont {Klinovaja}},\ and\
  \bibinfo {author} {\bibfnamefont {D.}~\bibnamefont {Loss}},\ }\bibfield
  {title} {\bibinfo {title} {{Towards a realistic transport modeling in a
  superconducting nanowire with Majorana fermions}},\ }\href
  {https://doi.org/10.1103/PhysRevB.87.024515} {\bibfield  {journal} {\bibinfo
  {journal} {Phys. Rev. B}\ }\textbf {\bibinfo {volume} {87}},\ \bibinfo
  {pages} {024515} (\bibinfo {year} {2013})}\BibitemShut {NoStop}%
\bibitem [{\citenamefont {Roy}\ \emph {et~al.}(2013)\citenamefont {Roy},
  \citenamefont {Bondyopadhaya},\ and\ \citenamefont
  {Tewari}}]{roy2013topologically}%
  \BibitemOpen
  \bibfield  {author} {\bibinfo {author} {\bibfnamefont {D.}~\bibnamefont
  {Roy}}, \bibinfo {author} {\bibfnamefont {N.}~\bibnamefont {Bondyopadhaya}},\
  and\ \bibinfo {author} {\bibfnamefont {S.}~\bibnamefont {Tewari}},\
  }\bibfield  {title} {\bibinfo {title} {{Topologically trivial zero-bias
  conductance peak in semiconductor Majorana wires from boundary effects}},\
  }\href {https://doi.org/10.1103/PhysRevB.88.020502} {\bibfield  {journal}
  {\bibinfo  {journal} {Phys. Rev. B}\ }\textbf {\bibinfo {volume} {88}},\
  \bibinfo {pages} {020502} (\bibinfo {year} {2013})}\BibitemShut {NoStop}%
\bibitem [{\citenamefont {Ptok}\ \emph {et~al.}(2017)\citenamefont {Ptok},
  \citenamefont {Kobia\l{}ka},\ and\ \citenamefont {Doma\ifmmode~\acute{n}\else
  \'{n}\fi{}ski}}]{ptok2017controlling}%
  \BibitemOpen
  \bibfield  {author} {\bibinfo {author} {\bibfnamefont {A.}~\bibnamefont
  {Ptok}}, \bibinfo {author} {\bibfnamefont {A.}~\bibnamefont {Kobia\l{}ka}},\
  and\ \bibinfo {author} {\bibfnamefont {T.}~\bibnamefont
  {Doma\ifmmode~\acute{n}\else \'{n}\fi{}ski}},\ }\bibfield  {title} {\bibinfo
  {title} {{Controlling the bound states in a quantum-dot hybrid nanowire}},\
  }\href {https://doi.org/10.1103/PhysRevB.96.195430} {\bibfield  {journal}
  {\bibinfo  {journal} {Phys. Rev. B}\ }\textbf {\bibinfo {volume} {96}},\
  \bibinfo {pages} {195430} (\bibinfo {year} {2017})}\BibitemShut {NoStop}%
\bibitem [{\citenamefont {Liu}\ \emph {et~al.}(2017)\citenamefont {Liu},
  \citenamefont {Sau}, \citenamefont {Stanescu},\ and\ \citenamefont
  {Das~Sarma}}]{liu2017andreev}%
  \BibitemOpen
  \bibfield  {author} {\bibinfo {author} {\bibfnamefont {C.-X.}\ \bibnamefont
  {Liu}}, \bibinfo {author} {\bibfnamefont {J.~D.}\ \bibnamefont {Sau}},
  \bibinfo {author} {\bibfnamefont {T.~D.}\ \bibnamefont {Stanescu}},\ and\
  \bibinfo {author} {\bibfnamefont {S.}~\bibnamefont {Das~Sarma}},\ }\bibfield
  {title} {\bibinfo {title} {{Andreev bound states versus Majorana bound states
  in quantum dot-nanowire-superconductor hybrid structures: Trivial versus
  topological zero-bias conductance peaks}},\ }\href
  {https://doi.org/10.1103/PhysRevB.96.075161} {\bibfield  {journal} {\bibinfo
  {journal} {Phys. Rev. B}\ }\textbf {\bibinfo {volume} {96}},\ \bibinfo
  {pages} {075161} (\bibinfo {year} {2017})}\BibitemShut {NoStop}%
\bibitem [{\citenamefont {Moore}\ \emph
  {et~al.}(2018{\natexlab{a}})\citenamefont {Moore}, \citenamefont {Stanescu},\
  and\ \citenamefont {Tewari}}]{moore2018two}%
  \BibitemOpen
  \bibfield  {author} {\bibinfo {author} {\bibfnamefont {C.}~\bibnamefont
  {Moore}}, \bibinfo {author} {\bibfnamefont {T.~D.}\ \bibnamefont
  {Stanescu}},\ and\ \bibinfo {author} {\bibfnamefont {S.}~\bibnamefont
  {Tewari}},\ }\bibfield  {title} {\bibinfo {title} {{Two-terminal charge
  tunneling: Disentangling Majorana zero modes from partially separated Andreev
  bound states in semiconductor-superconductor heterostructures}},\ }\href
  {https://doi.org/10.1103/PhysRevB.97.165302} {\bibfield  {journal} {\bibinfo
  {journal} {Phys. Rev. B}\ }\textbf {\bibinfo {volume} {97}},\ \bibinfo
  {pages} {165302} (\bibinfo {year} {2018}{\natexlab{a}})}\BibitemShut
  {NoStop}%
\bibitem [{\citenamefont {Moore}\ \emph
  {et~al.}(2018{\natexlab{b}})\citenamefont {Moore}, \citenamefont {Zeng},
  \citenamefont {Stanescu},\ and\ \citenamefont {Tewari}}]{moore2018quantized}%
  \BibitemOpen
  \bibfield  {author} {\bibinfo {author} {\bibfnamefont {C.}~\bibnamefont
  {Moore}}, \bibinfo {author} {\bibfnamefont {C.}~\bibnamefont {Zeng}},
  \bibinfo {author} {\bibfnamefont {T.~D.}\ \bibnamefont {Stanescu}},\ and\
  \bibinfo {author} {\bibfnamefont {S.}~\bibnamefont {Tewari}},\ }\bibfield
  {title} {\bibinfo {title} {{Quantized zero-bias conductance plateau in
  semiconductor-superconductor heterostructures without topological Majorana
  zero modes}},\ }\href {https://doi.org/10.1103/PhysRevB.98.155314} {\bibfield
   {journal} {\bibinfo  {journal} {Phys. Rev. B}\ }\textbf {\bibinfo {volume}
  {98}},\ \bibinfo {pages} {155314} (\bibinfo {year}
  {2018}{\natexlab{b}})}\BibitemShut {NoStop}%
\bibitem [{\citenamefont {Reeg}\ \emph {et~al.}(2018)\citenamefont {Reeg},
  \citenamefont {Dmytruk}, \citenamefont {Chevallier}, \citenamefont {Loss},\
  and\ \citenamefont {Klinovaja}}]{reeg2018zero}%
  \BibitemOpen
  \bibfield  {author} {\bibinfo {author} {\bibfnamefont {C.}~\bibnamefont
  {Reeg}}, \bibinfo {author} {\bibfnamefont {O.}~\bibnamefont {Dmytruk}},
  \bibinfo {author} {\bibfnamefont {D.}~\bibnamefont {Chevallier}}, \bibinfo
  {author} {\bibfnamefont {D.}~\bibnamefont {Loss}},\ and\ \bibinfo {author}
  {\bibfnamefont {J.}~\bibnamefont {Klinovaja}},\ }\bibfield  {title} {\bibinfo
  {title} {{Zero-energy Andreev bound states from quantum dots in proximitized
  Rashba nanowires}},\ }\href {https://doi.org/10.1103/PhysRevB.98.245407}
  {\bibfield  {journal} {\bibinfo  {journal} {Phys. Rev. B}\ }\textbf {\bibinfo
  {volume} {98}},\ \bibinfo {pages} {245407} (\bibinfo {year}
  {2018})}\BibitemShut {NoStop}%
\bibitem [{\citenamefont {Vuik}\ \emph {et~al.}(2019)\citenamefont {Vuik},
  \citenamefont {Nijholt}, \citenamefont {Akhmerov},\ and\ \citenamefont
  {Wimmer}}]{vuik2019reproducing}%
  \BibitemOpen
  \bibfield  {author} {\bibinfo {author} {\bibfnamefont {A.}~\bibnamefont
  {Vuik}}, \bibinfo {author} {\bibfnamefont {B.}~\bibnamefont {Nijholt}},
  \bibinfo {author} {\bibfnamefont {A.~R.}\ \bibnamefont {Akhmerov}},\ and\
  \bibinfo {author} {\bibfnamefont {M.}~\bibnamefont {Wimmer}},\ }\bibfield
  {title} {\bibinfo {title} {{Reproducing topological properties with
  quasi-Majorana states}},\ }\href
  {https://doi.org/10.21468/SciPostPhys.7.5.061} {\bibfield  {journal}
  {\bibinfo  {journal} {SciPost Phys.}\ }\textbf {\bibinfo {volume} {7}},\
  \bibinfo {pages} {061} (\bibinfo {year} {2019})}\BibitemShut {NoStop}%
\bibitem [{\citenamefont {Stanescu}\ and\ \citenamefont
  {Tewari}(2019)}]{stanescu2019robust}%
  \BibitemOpen
  \bibfield  {author} {\bibinfo {author} {\bibfnamefont {T.~D.}\ \bibnamefont
  {Stanescu}}\ and\ \bibinfo {author} {\bibfnamefont {S.}~\bibnamefont
  {Tewari}},\ }\bibfield  {title} {\bibinfo {title} {{Robust low-energy Andreev
  bound states in semiconductor-superconductor structures: Importance of
  partial separation of component Majorana bound states}},\ }\href
  {https://doi.org/10.1103/PhysRevB.100.155429} {\bibfield  {journal} {\bibinfo
   {journal} {Phys. Rev. B}\ }\textbf {\bibinfo {volume} {100}},\ \bibinfo
  {pages} {155429} (\bibinfo {year} {2019})}\BibitemShut {NoStop}%
\bibitem [{\citenamefont {Woods}\ \emph {et~al.}(2019)\citenamefont {Woods},
  \citenamefont {Chen}, \citenamefont {Frolov},\ and\ \citenamefont
  {Stanescu}}]{woods2019zero}%
  \BibitemOpen
  \bibfield  {author} {\bibinfo {author} {\bibfnamefont {B.~D.}\ \bibnamefont
  {Woods}}, \bibinfo {author} {\bibfnamefont {J.}~\bibnamefont {Chen}},
  \bibinfo {author} {\bibfnamefont {S.~M.}\ \bibnamefont {Frolov}},\ and\
  \bibinfo {author} {\bibfnamefont {T.~D.}\ \bibnamefont {Stanescu}},\
  }\bibfield  {title} {\bibinfo {title} {{Zero-energy pinning of topologically
  trivial bound states in multiband semiconductor-superconductor nanowires}},\
  }\href {https://doi.org/10.1103/PhysRevB.100.125407} {\bibfield  {journal}
  {\bibinfo  {journal} {Phys. Rev. B}\ }\textbf {\bibinfo {volume} {100}},\
  \bibinfo {pages} {125407} (\bibinfo {year} {2019})}\BibitemShut {NoStop}%
\bibitem [{\citenamefont {Chen}\ \emph {et~al.}(2019)\citenamefont {Chen},
  \citenamefont {Woods}, \citenamefont {Yu}, \citenamefont {Hocevar},
  \citenamefont {Car}, \citenamefont {Plissard}, \citenamefont {Bakkers},
  \citenamefont {Stanescu},\ and\ \citenamefont {Frolov}}]{chen2019ubiquitous}%
  \BibitemOpen
  \bibfield  {author} {\bibinfo {author} {\bibfnamefont {J.}~\bibnamefont
  {Chen}}, \bibinfo {author} {\bibfnamefont {B.~D.}\ \bibnamefont {Woods}},
  \bibinfo {author} {\bibfnamefont {P.}~\bibnamefont {Yu}}, \bibinfo {author}
  {\bibfnamefont {M.}~\bibnamefont {Hocevar}}, \bibinfo {author} {\bibfnamefont
  {D.}~\bibnamefont {Car}}, \bibinfo {author} {\bibfnamefont {S.~R.}\
  \bibnamefont {Plissard}}, \bibinfo {author} {\bibfnamefont {E.~P. A.~M.}\
  \bibnamefont {Bakkers}}, \bibinfo {author} {\bibfnamefont {T.~D.}\
  \bibnamefont {Stanescu}},\ and\ \bibinfo {author} {\bibfnamefont {S.~M.}\
  \bibnamefont {Frolov}},\ }\bibfield  {title} {\bibinfo {title} {{Ubiquitous
  Non-Majorana Zero-Bias Conductance Peaks in Nanowire Devices}},\ }\href
  {https://doi.org/10.1103/PhysRevLett.123.107703} {\bibfield  {journal}
  {\bibinfo  {journal} {Phys. Rev. Lett.}\ }\textbf {\bibinfo {volume} {123}},\
  \bibinfo {pages} {107703} (\bibinfo {year} {2019})}\BibitemShut {NoStop}%
\bibitem [{\citenamefont {Awoga}\ \emph {et~al.}(2019)\citenamefont {Awoga},
  \citenamefont {Cayao},\ and\ \citenamefont
  {Black-Schaffer}}]{awoga2019supercurrent}%
  \BibitemOpen
  \bibfield  {author} {\bibinfo {author} {\bibfnamefont {O.~A.}\ \bibnamefont
  {Awoga}}, \bibinfo {author} {\bibfnamefont {J.}~\bibnamefont {Cayao}},\ and\
  \bibinfo {author} {\bibfnamefont {A.~M.}\ \bibnamefont {Black-Schaffer}},\
  }\bibfield  {title} {\bibinfo {title} {{Supercurrent Detection of
  Topologically Trivial Zero-Energy States in Nanowire Junctions}},\ }\href
  {https://doi.org/10.1103/PhysRevLett.123.117001} {\bibfield  {journal}
  {\bibinfo  {journal} {Phys. Rev. Lett.}\ }\textbf {\bibinfo {volume} {123}},\
  \bibinfo {pages} {117001} (\bibinfo {year} {2019})}\BibitemShut {NoStop}%
\bibitem [{\citenamefont {Prada}\ \emph {et~al.}(2020)\citenamefont {Prada},
  \citenamefont {San-Jose}, \citenamefont {de~Moor}, \citenamefont {Geresdi},
  \citenamefont {Lee}, \citenamefont {Klinovaja}, \citenamefont {Loss},
  \citenamefont {Nyg{\aa}rd}, \citenamefont {Aguado},\ and\ \citenamefont
  {Kouwenhoven}}]{prada2020andreev}%
  \BibitemOpen
  \bibfield  {author} {\bibinfo {author} {\bibfnamefont {E.}~\bibnamefont
  {Prada}}, \bibinfo {author} {\bibfnamefont {P.}~\bibnamefont {San-Jose}},
  \bibinfo {author} {\bibfnamefont {M.~W.}\ \bibnamefont {de~Moor}}, \bibinfo
  {author} {\bibfnamefont {A.}~\bibnamefont {Geresdi}}, \bibinfo {author}
  {\bibfnamefont {E.~J.}\ \bibnamefont {Lee}}, \bibinfo {author} {\bibfnamefont
  {J.}~\bibnamefont {Klinovaja}}, \bibinfo {author} {\bibfnamefont
  {D.}~\bibnamefont {Loss}}, \bibinfo {author} {\bibfnamefont {J.}~\bibnamefont
  {Nyg{\aa}rd}}, \bibinfo {author} {\bibfnamefont {R.}~\bibnamefont {Aguado}},\
  and\ \bibinfo {author} {\bibfnamefont {L.~P.}\ \bibnamefont {Kouwenhoven}},\
  }\bibfield  {title} {\bibinfo {title} {{From Andreev to Majorana bound states
  in hybrid superconductor--semiconductor nanowires}},\ }\href
  {https://doi.org/10.1038/s42254-020-0228-y} {\bibfield  {journal} {\bibinfo
  {journal} {Nature Reviews Physics}\ }\textbf {\bibinfo {volume} {2}},\
  \bibinfo {pages} {575} (\bibinfo {year} {2020})}\BibitemShut {NoStop}%
\bibitem [{\citenamefont {Yu}\ \emph {et~al.}(2021)\citenamefont {Yu},
  \citenamefont {Chen}, \citenamefont {Gomanko}, \citenamefont {Badawy},
  \citenamefont {Bakkers}, \citenamefont {Zuo}, \citenamefont {Mourik},\ and\
  \citenamefont {Frolov}}]{yu2021non}%
  \BibitemOpen
  \bibfield  {author} {\bibinfo {author} {\bibfnamefont {P.}~\bibnamefont
  {Yu}}, \bibinfo {author} {\bibfnamefont {J.}~\bibnamefont {Chen}}, \bibinfo
  {author} {\bibfnamefont {M.}~\bibnamefont {Gomanko}}, \bibinfo {author}
  {\bibfnamefont {G.}~\bibnamefont {Badawy}}, \bibinfo {author} {\bibfnamefont
  {E.}~\bibnamefont {Bakkers}}, \bibinfo {author} {\bibfnamefont
  {K.}~\bibnamefont {Zuo}}, \bibinfo {author} {\bibfnamefont {V.}~\bibnamefont
  {Mourik}},\ and\ \bibinfo {author} {\bibfnamefont {S.}~\bibnamefont
  {Frolov}},\ }\bibfield  {title} {\bibinfo {title} {{Non-Majorana states yield
  nearly quantized conductance in proximatized nanowires}},\ }\href
  {https://doi.org/10.1038/s41567-020-01107-w} {\bibfield  {journal} {\bibinfo
  {journal} {Nature Physics}\ }\textbf {\bibinfo {volume} {17}},\ \bibinfo
  {pages} {482} (\bibinfo {year} {2021})}\BibitemShut {NoStop}%
\bibitem [{\citenamefont {Das~Sarma}\ and\ \citenamefont
  {Pan}(2021)}]{sarma2021disorder}%
  \BibitemOpen
  \bibfield  {author} {\bibinfo {author} {\bibfnamefont {S.}~\bibnamefont
  {Das~Sarma}}\ and\ \bibinfo {author} {\bibfnamefont {H.}~\bibnamefont
  {Pan}},\ }\bibfield  {title} {\bibinfo {title} {{Disorder-induced zero-bias
  peaks in Majorana nanowires}},\ }\href
  {https://doi.org/10.1103/PhysRevB.103.195158} {\bibfield  {journal} {\bibinfo
   {journal} {Phys. Rev. B}\ }\textbf {\bibinfo {volume} {103}},\ \bibinfo
  {pages} {195158} (\bibinfo {year} {2021})}\BibitemShut {NoStop}%
\bibitem [{\citenamefont {Valentini}\ \emph {et~al.}(2021)\citenamefont
  {Valentini}, \citenamefont {Pe{\~n}aranda}, \citenamefont {Hofmann},
  \citenamefont {Brauns}, \citenamefont {Hauschild}, \citenamefont {Krogstrup},
  \citenamefont {San-Jose}, \citenamefont {Prada}, \citenamefont {Aguado},\
  and\ \citenamefont {Katsaros}}]{valentini2021nontopological}%
  \BibitemOpen
  \bibfield  {author} {\bibinfo {author} {\bibfnamefont {M.}~\bibnamefont
  {Valentini}}, \bibinfo {author} {\bibfnamefont {F.}~\bibnamefont
  {Pe{\~n}aranda}}, \bibinfo {author} {\bibfnamefont {A.}~\bibnamefont
  {Hofmann}}, \bibinfo {author} {\bibfnamefont {M.}~\bibnamefont {Brauns}},
  \bibinfo {author} {\bibfnamefont {R.}~\bibnamefont {Hauschild}}, \bibinfo
  {author} {\bibfnamefont {P.}~\bibnamefont {Krogstrup}}, \bibinfo {author}
  {\bibfnamefont {P.}~\bibnamefont {San-Jose}}, \bibinfo {author}
  {\bibfnamefont {E.}~\bibnamefont {Prada}}, \bibinfo {author} {\bibfnamefont
  {R.}~\bibnamefont {Aguado}},\ and\ \bibinfo {author} {\bibfnamefont
  {G.}~\bibnamefont {Katsaros}},\ }\bibfield  {title} {\bibinfo {title}
  {{Nontopological zero-bias peaks in full-shell nanowires induced by
  flux-tunable Andreev states}},\ }\href
  {https://doi.org/10.1126/science.abf1513} {\bibfield  {journal} {\bibinfo
  {journal} {Science}\ }\textbf {\bibinfo {volume} {373}},\ \bibinfo {pages}
  {82} (\bibinfo {year} {2021})}\BibitemShut {NoStop}%
\bibitem [{\citenamefont {Hess}\ \emph {et~al.}(2021)\citenamefont {Hess},
  \citenamefont {Legg}, \citenamefont {Loss},\ and\ \citenamefont
  {Klinovaja}}]{hess2021local}%
  \BibitemOpen
  \bibfield  {author} {\bibinfo {author} {\bibfnamefont {R.}~\bibnamefont
  {Hess}}, \bibinfo {author} {\bibfnamefont {H.~F.}\ \bibnamefont {Legg}},
  \bibinfo {author} {\bibfnamefont {D.}~\bibnamefont {Loss}},\ and\ \bibinfo
  {author} {\bibfnamefont {J.}~\bibnamefont {Klinovaja}},\ }\bibfield  {title}
  {\bibinfo {title} {{Local and nonlocal quantum transport due to Andreev bound
  states in finite Rashba nanowires with superconducting and normal
  sections}},\ }\href {https://doi.org/10.1103/PhysRevB.104.075405} {\bibfield
  {journal} {\bibinfo  {journal} {Phys. Rev. B}\ }\textbf {\bibinfo {volume}
  {104}},\ \bibinfo {pages} {075405} (\bibinfo {year} {2021})}\BibitemShut
  {NoStop}%
\bibitem [{\citenamefont {Hess}\ \emph {et~al.}(2023)\citenamefont {Hess},
  \citenamefont {Legg}, \citenamefont {Loss},\ and\ \citenamefont
  {Klinovaja}}]{hess2023trivial}%
  \BibitemOpen
  \bibfield  {author} {\bibinfo {author} {\bibfnamefont {R.}~\bibnamefont
  {Hess}}, \bibinfo {author} {\bibfnamefont {H.~F.}\ \bibnamefont {Legg}},
  \bibinfo {author} {\bibfnamefont {D.}~\bibnamefont {Loss}},\ and\ \bibinfo
  {author} {\bibfnamefont {J.}~\bibnamefont {Klinovaja}},\ }\bibfield  {title}
  {\bibinfo {title} {{Trivial Andreev Band Mimicking Topological Bulk Gap
  Reopening in the Nonlocal Conductance of Long Rashba Nanowires}},\ }\href
  {https://doi.org/10.1103/PhysRevLett.130.207001} {\bibfield  {journal}
  {\bibinfo  {journal} {Phys. Rev. Lett.}\ }\textbf {\bibinfo {volume} {130}},\
  \bibinfo {pages} {207001} (\bibinfo {year} {2023})}\BibitemShut {NoStop}%
\bibitem [{\citenamefont {Hess}\ \emph {et~al.}(2022)\citenamefont {Hess},
  \citenamefont {Legg}, \citenamefont {Loss},\ and\ \citenamefont
  {Klinovaja}}]{hess2022prevalence}%
  \BibitemOpen
  \bibfield  {author} {\bibinfo {author} {\bibfnamefont {R.}~\bibnamefont
  {Hess}}, \bibinfo {author} {\bibfnamefont {H.~F.}\ \bibnamefont {Legg}},
  \bibinfo {author} {\bibfnamefont {D.}~\bibnamefont {Loss}},\ and\ \bibinfo
  {author} {\bibfnamefont {J.}~\bibnamefont {Klinovaja}},\ }\bibfield  {title}
  {\bibinfo {title} {{Prevalence of trivial zero-energy subgap states in
  nonuniform helical spin chains on the surface of superconductors}},\ }\href
  {https://doi.org/10.1103/PhysRevB.106.104503} {\bibfield  {journal} {\bibinfo
   {journal} {Phys. Rev. B}\ }\textbf {\bibinfo {volume} {106}},\ \bibinfo
  {pages} {104503} (\bibinfo {year} {2022})}\BibitemShut {NoStop}%
\bibitem [{\citenamefont {Sau}\ and\ \citenamefont
  {Sarma}(2012)}]{sau2012realizing}%
  \BibitemOpen
  \bibfield  {author} {\bibinfo {author} {\bibfnamefont {J.~D.}\ \bibnamefont
  {Sau}}\ and\ \bibinfo {author} {\bibfnamefont {S.~D.}\ \bibnamefont
  {Sarma}},\ }\bibfield  {title} {\bibinfo {title} {{Realizing a robust
  practical Majorana chain in a quantum-dot-superconductor linear array}},\
  }\href {https://doi.org/10.1038/ncomms1966} {\bibfield  {journal} {\bibinfo
  {journal} {Nature communications}\ }\textbf {\bibinfo {volume} {3}},\
  \bibinfo {pages} {964} (\bibinfo {year} {2012})}\BibitemShut {NoStop}%
\bibitem [{\citenamefont {Leijnse}\ and\ \citenamefont
  {Flensberg}(2012)}]{leijnse2012parity}%
  \BibitemOpen
  \bibfield  {author} {\bibinfo {author} {\bibfnamefont {M.}~\bibnamefont
  {Leijnse}}\ and\ \bibinfo {author} {\bibfnamefont {K.}~\bibnamefont
  {Flensberg}},\ }\bibfield  {title} {\bibinfo {title} {{Parity qubits and poor
  man's Majorana bound states in double quantum dots}},\ }\href
  {https://doi.org/10.1103/PhysRevB.86.134528} {\bibfield  {journal} {\bibinfo
  {journal} {Phys. Rev. B}\ }\textbf {\bibinfo {volume} {86}},\ \bibinfo
  {pages} {134528} (\bibinfo {year} {2012})}\BibitemShut {NoStop}%
\bibitem [{\citenamefont {Fulga}\ \emph {et~al.}(2013)\citenamefont {Fulga},
  \citenamefont {Haim}, \citenamefont {Akhmerov},\ and\ \citenamefont
  {Oreg}}]{fulga2013adaptive}%
  \BibitemOpen
  \bibfield  {author} {\bibinfo {author} {\bibfnamefont {I.~C.}\ \bibnamefont
  {Fulga}}, \bibinfo {author} {\bibfnamefont {A.}~\bibnamefont {Haim}},
  \bibinfo {author} {\bibfnamefont {A.~R.}\ \bibnamefont {Akhmerov}},\ and\
  \bibinfo {author} {\bibfnamefont {Y.}~\bibnamefont {Oreg}},\ }\bibfield
  {title} {\bibinfo {title} {{Adaptive tuning of Majorana fermions in a quantum
  dot chain}},\ }\href {https://doi.org/10.1088/1367-2630/15/4/045020}
  {\bibfield  {journal} {\bibinfo  {journal} {New journal of physics}\ }\textbf
  {\bibinfo {volume} {15}},\ \bibinfo {pages} {045020} (\bibinfo {year}
  {2013})}\BibitemShut {NoStop}%
\bibitem [{\citenamefont {Sothmann}\ \emph {et~al.}(2013)\citenamefont
  {Sothmann}, \citenamefont {Li},\ and\ \citenamefont
  {Büttiker}}]{sothmann2013fractional}%
  \BibitemOpen
  \bibfield  {author} {\bibinfo {author} {\bibfnamefont {B.}~\bibnamefont
  {Sothmann}}, \bibinfo {author} {\bibfnamefont {J.}~\bibnamefont {Li}},\ and\
  \bibinfo {author} {\bibfnamefont {M.}~\bibnamefont {Büttiker}},\ }\bibfield
  {title} {\bibinfo {title} {{Fractional Josephson effect in a quadruple
  quantum dot}},\ }\href {https://doi.org/10.1088/1367-2630/15/8/085018}
  {\bibfield  {journal} {\bibinfo  {journal} {New Journal of Physics}\ }\textbf
  {\bibinfo {volume} {15}},\ \bibinfo {pages} {085018} (\bibinfo {year}
  {2013})}\BibitemShut {NoStop}%
\bibitem [{\citenamefont {Li}\ \emph {et~al.}(2014)\citenamefont {Li},
  \citenamefont {Kundu}, \citenamefont {Zhong},\ and\ \citenamefont
  {Seradjeh}}]{li2014tunable}%
  \BibitemOpen
  \bibfield  {author} {\bibinfo {author} {\bibfnamefont {Y.}~\bibnamefont
  {Li}}, \bibinfo {author} {\bibfnamefont {A.}~\bibnamefont {Kundu}}, \bibinfo
  {author} {\bibfnamefont {F.}~\bibnamefont {Zhong}},\ and\ \bibinfo {author}
  {\bibfnamefont {B.}~\bibnamefont {Seradjeh}},\ }\bibfield  {title} {\bibinfo
  {title} {{Tunable Floquet Majorana fermions in driven coupled quantum
  dots}},\ }\href {https://doi.org/10.1103/PhysRevB.90.121401} {\bibfield
  {journal} {\bibinfo  {journal} {Phys. Rev. B}\ }\textbf {\bibinfo {volume}
  {90}},\ \bibinfo {pages} {121401} (\bibinfo {year} {2014})}\BibitemShut
  {NoStop}%
\bibitem [{\citenamefont {Su}\ \emph {et~al.}(2017)\citenamefont {Su},
  \citenamefont {Tacla}, \citenamefont {Hocevar}, \citenamefont {Car},
  \citenamefont {Plissard}, \citenamefont {Bakkers}, \citenamefont {Daley},
  \citenamefont {Pekker},\ and\ \citenamefont {Frolov}}]{su2017andreev}%
  \BibitemOpen
  \bibfield  {author} {\bibinfo {author} {\bibfnamefont {Z.}~\bibnamefont
  {Su}}, \bibinfo {author} {\bibfnamefont {A.~B.}\ \bibnamefont {Tacla}},
  \bibinfo {author} {\bibfnamefont {M.}~\bibnamefont {Hocevar}}, \bibinfo
  {author} {\bibfnamefont {D.}~\bibnamefont {Car}}, \bibinfo {author}
  {\bibfnamefont {S.~R.}\ \bibnamefont {Plissard}}, \bibinfo {author}
  {\bibfnamefont {E.~P.}\ \bibnamefont {Bakkers}}, \bibinfo {author}
  {\bibfnamefont {A.~J.}\ \bibnamefont {Daley}}, \bibinfo {author}
  {\bibfnamefont {D.}~\bibnamefont {Pekker}},\ and\ \bibinfo {author}
  {\bibfnamefont {S.~M.}\ \bibnamefont {Frolov}},\ }\bibfield  {title}
  {\bibinfo {title} {{Andreev molecules in semiconductor nanowire double
  quantum dots}},\ }\href {https://doi.org/10.1038/s41467-017-00665-7}
  {\bibfield  {journal} {\bibinfo  {journal} {Nature communications}\ }\textbf
  {\bibinfo {volume} {8}},\ \bibinfo {pages} {585} (\bibinfo {year}
  {2017})}\BibitemShut {NoStop}%
\bibitem [{\citenamefont {Tsintzis}\ \emph {et~al.}(2022)\citenamefont
  {Tsintzis}, \citenamefont {Souto},\ and\ \citenamefont
  {Leijnse}}]{tsintzis2022creating}%
  \BibitemOpen
  \bibfield  {author} {\bibinfo {author} {\bibfnamefont {A.}~\bibnamefont
  {Tsintzis}}, \bibinfo {author} {\bibfnamefont {R.~S.}\ \bibnamefont
  {Souto}},\ and\ \bibinfo {author} {\bibfnamefont {M.}~\bibnamefont
  {Leijnse}},\ }\bibfield  {title} {\bibinfo {title} {{Creating and detecting
  poor man's Majorana bound states in interacting quantum dots}},\ }\href
  {https://doi.org/10.1103/PhysRevB.106.L201404} {\bibfield  {journal}
  {\bibinfo  {journal} {Phys. Rev. B}\ }\textbf {\bibinfo {volume} {106}},\
  \bibinfo {pages} {L201404} (\bibinfo {year} {2022})}\BibitemShut {NoStop}%
\bibitem [{\citenamefont {Liu}\ \emph {et~al.}(2022)\citenamefont {Liu},
  \citenamefont {Wang}, \citenamefont {Dvir},\ and\ \citenamefont
  {Wimmer}}]{liu2022tunable}%
  \BibitemOpen
  \bibfield  {author} {\bibinfo {author} {\bibfnamefont {C.-X.}\ \bibnamefont
  {Liu}}, \bibinfo {author} {\bibfnamefont {G.}~\bibnamefont {Wang}}, \bibinfo
  {author} {\bibfnamefont {T.}~\bibnamefont {Dvir}},\ and\ \bibinfo {author}
  {\bibfnamefont {M.}~\bibnamefont {Wimmer}},\ }\bibfield  {title} {\bibinfo
  {title} {{Tunable Superconducting Coupling of Quantum Dots via Andreev Bound
  States in Semiconductor-Superconductor Nanowires}},\ }\href
  {https://doi.org/10.1103/PhysRevLett.129.267701} {\bibfield  {journal}
  {\bibinfo  {journal} {Phys. Rev. Lett.}\ }\textbf {\bibinfo {volume} {129}},\
  \bibinfo {pages} {267701} (\bibinfo {year} {2022})}\BibitemShut {NoStop}%
\bibitem [{\citenamefont {Dvir}\ \emph {et~al.}(2023)\citenamefont {Dvir},
  \citenamefont {Wang}, \citenamefont {van Loo}, \citenamefont {Liu},
  \citenamefont {Mazur}, \citenamefont {Bordin}, \citenamefont {Ten~Haaf},
  \citenamefont {Wang}, \citenamefont {van Driel}, \citenamefont {Zatelli},
  \citenamefont {Li}, \citenamefont {Malinowski}, \citenamefont {Gazibegovic},
  \citenamefont {Badawy}, \citenamefont {Bakkers}, \citenamefont {Wimmer},\
  and\ \citenamefont {Kouwenhoven}}]{dvir2023realization}%
  \BibitemOpen
  \bibfield  {author} {\bibinfo {author} {\bibfnamefont {T.}~\bibnamefont
  {Dvir}}, \bibinfo {author} {\bibfnamefont {G.}~\bibnamefont {Wang}}, \bibinfo
  {author} {\bibfnamefont {N.}~\bibnamefont {van Loo}}, \bibinfo {author}
  {\bibfnamefont {C.-X.}\ \bibnamefont {Liu}}, \bibinfo {author} {\bibfnamefont
  {G.~P.}\ \bibnamefont {Mazur}}, \bibinfo {author} {\bibfnamefont
  {A.}~\bibnamefont {Bordin}}, \bibinfo {author} {\bibfnamefont {S.~L.}\
  \bibnamefont {Ten~Haaf}}, \bibinfo {author} {\bibfnamefont {J.-Y.}\
  \bibnamefont {Wang}}, \bibinfo {author} {\bibfnamefont {D.}~\bibnamefont {van
  Driel}}, \bibinfo {author} {\bibfnamefont {F.}~\bibnamefont {Zatelli}},
  \bibinfo {author} {\bibfnamefont {X.}~\bibnamefont {Li}}, \bibinfo {author}
  {\bibfnamefont {F.~K.}\ \bibnamefont {Malinowski}}, \bibinfo {author}
  {\bibfnamefont {S.}~\bibnamefont {Gazibegovic}}, \bibinfo {author}
  {\bibfnamefont {G.}~\bibnamefont {Badawy}}, \bibinfo {author} {\bibfnamefont
  {E.~P. A.~M.}\ \bibnamefont {Bakkers}}, \bibinfo {author} {\bibfnamefont
  {M.}~\bibnamefont {Wimmer}},\ and\ \bibinfo {author} {\bibfnamefont {L.~P.}\
  \bibnamefont {Kouwenhoven}},\ }\bibfield  {title} {\bibinfo {title}
  {{Realization of a minimal Kitaev chain in coupled quantum dots}},\ }\href
  {https://doi.org/10.1038/s41586-022-05585-1} {\bibfield  {journal} {\bibinfo
  {journal} {Nature}\ }\textbf {\bibinfo {volume} {614}},\ \bibinfo {pages}
  {445} (\bibinfo {year} {2023})}\BibitemShut {NoStop}%
\bibitem [{\citenamefont {Sanches}\ \emph {et~al.}(2023)\citenamefont
  {Sanches}, \citenamefont {Lustosa}, \citenamefont {Ricco}, \citenamefont
  {Shelykh}, \citenamefont {de~Souza}, \citenamefont {Figueira},\ and\
  \citenamefont {Seridonio}}]{sanches2023fractionalization}%
  \BibitemOpen
  \bibfield  {author} {\bibinfo {author} {\bibfnamefont {J.~E.}\ \bibnamefont
  {Sanches}}, \bibinfo {author} {\bibfnamefont {L.~T.}\ \bibnamefont
  {Lustosa}}, \bibinfo {author} {\bibfnamefont {L.~S.}\ \bibnamefont {Ricco}},
  \bibinfo {author} {\bibfnamefont {I.~A.}\ \bibnamefont {Shelykh}}, \bibinfo
  {author} {\bibfnamefont {M.}~\bibnamefont {de~Souza}}, \bibinfo {author}
  {\bibfnamefont {M.~S.}\ \bibnamefont {Figueira}},\ and\ \bibinfo {author}
  {\bibfnamefont {A.~C.}\ \bibnamefont {Seridonio}},\ }\bibfield  {title}
  {\bibinfo {title} {{Fractionalization of Majorana-Ising-type quasiparticle
  excitations}},\ }\href {https://doi.org/10.1103/PhysRevB.107.155144}
  {\bibfield  {journal} {\bibinfo  {journal} {Phys. Rev. B}\ }\textbf {\bibinfo
  {volume} {107}},\ \bibinfo {pages} {155144} (\bibinfo {year}
  {2023})}\BibitemShut {NoStop}%
\bibitem [{\citenamefont {Liu}\ \emph {et~al.}(2023)\citenamefont {Liu},
  \citenamefont {Pan}, \citenamefont {Setiawan}, \citenamefont {Wimmer},\ and\
  \citenamefont {Sau}}]{chunxiao2023fusion}%
  \BibitemOpen
  \bibfield  {author} {\bibinfo {author} {\bibfnamefont {C.-X.}\ \bibnamefont
  {Liu}}, \bibinfo {author} {\bibfnamefont {H.}~\bibnamefont {Pan}}, \bibinfo
  {author} {\bibfnamefont {F.}~\bibnamefont {Setiawan}}, \bibinfo {author}
  {\bibfnamefont {M.}~\bibnamefont {Wimmer}},\ and\ \bibinfo {author}
  {\bibfnamefont {J.~D.}\ \bibnamefont {Sau}},\ }\bibfield  {title} {\bibinfo
  {title} {{Fusion protocol for Majorana modes in coupled quantum dots}},\
  }\href {https://doi.org/10.1103/PhysRevB.108.085437} {\bibfield  {journal}
  {\bibinfo  {journal} {Phys. Rev. B}\ }\textbf {\bibinfo {volume} {108}},\
  \bibinfo {pages} {085437} (\bibinfo {year} {2023})}\BibitemShut {NoStop}%
\bibitem [{\citenamefont {Koch}\ \emph {et~al.}(2023)\citenamefont {Koch},
  \citenamefont {van Driel}, \citenamefont {Bordin}, \citenamefont {Lado},\
  and\ \citenamefont {Greplova}}]{koch2023adversarial}%
  \BibitemOpen
  \bibfield  {author} {\bibinfo {author} {\bibfnamefont {R.}~\bibnamefont
  {Koch}}, \bibinfo {author} {\bibfnamefont {D.}~\bibnamefont {van Driel}},
  \bibinfo {author} {\bibfnamefont {A.}~\bibnamefont {Bordin}}, \bibinfo
  {author} {\bibfnamefont {J.~L.}\ \bibnamefont {Lado}},\ and\ \bibinfo
  {author} {\bibfnamefont {E.}~\bibnamefont {Greplova}},\ }\bibfield  {title}
  {\bibinfo {title} {{Adversarial Hamiltonian learning of quantum dots in a
  minimal Kitaev chain}},\ }\href
  {https://doi.org/10.1103/PhysRevApplied.20.044081} {\bibfield  {journal}
  {\bibinfo  {journal} {Phys. Rev. Appl.}\ }\textbf {\bibinfo {volume} {20}},\
  \bibinfo {pages} {044081} (\bibinfo {year} {2023})}\BibitemShut {NoStop}%
\bibitem [{\citenamefont {Souto}\ \emph {et~al.}(2023)\citenamefont {Souto},
  \citenamefont {Tsintzis}, \citenamefont {Leijnse},\ and\ \citenamefont
  {Danon}}]{souto2023probing}%
  \BibitemOpen
  \bibfield  {author} {\bibinfo {author} {\bibfnamefont {R.~S.}\ \bibnamefont
  {Souto}}, \bibinfo {author} {\bibfnamefont {A.}~\bibnamefont {Tsintzis}},
  \bibinfo {author} {\bibfnamefont {M.}~\bibnamefont {Leijnse}},\ and\ \bibinfo
  {author} {\bibfnamefont {J.}~\bibnamefont {Danon}},\ }\bibfield  {title}
  {\bibinfo {title} {{Probing Majorana localization in minimal Kitaev chains
  through a quantum dot}},\ }\href
  {https://doi.org/10.1103/PhysRevResearch.5.043182} {\bibfield  {journal}
  {\bibinfo  {journal} {Phys. Rev. Res.}\ }\textbf {\bibinfo {volume} {5}},\
  \bibinfo {pages} {043182} (\bibinfo {year} {2023})}\BibitemShut {NoStop}%
\bibitem [{\citenamefont {Zatelli}\ \emph {et~al.}(2023)\citenamefont
  {Zatelli}, \citenamefont {van Driel}, \citenamefont {Xu}, \citenamefont
  {Wang}, \citenamefont {Liu}, \citenamefont {Bordin}, \citenamefont {Roovers},
  \citenamefont {Mazur}, \citenamefont {van Loo}, \citenamefont {Wolff},
  \citenamefont {Bozkurt}, \citenamefont {Badawy}, \citenamefont {Gazibegovic},
  \citenamefont {Bakkers}, \citenamefont {Wimmer}, \citenamefont
  {Kouwenhoven},\ and\ \citenamefont {Dvir}}]{zatelli2023robust}%
  \BibitemOpen
  \bibfield  {author} {\bibinfo {author} {\bibfnamefont {F.}~\bibnamefont
  {Zatelli}}, \bibinfo {author} {\bibfnamefont {D.}~\bibnamefont {van Driel}},
  \bibinfo {author} {\bibfnamefont {D.}~\bibnamefont {Xu}}, \bibinfo {author}
  {\bibfnamefont {G.}~\bibnamefont {Wang}}, \bibinfo {author} {\bibfnamefont
  {C.-X.}\ \bibnamefont {Liu}}, \bibinfo {author} {\bibfnamefont
  {A.}~\bibnamefont {Bordin}}, \bibinfo {author} {\bibfnamefont
  {B.}~\bibnamefont {Roovers}}, \bibinfo {author} {\bibfnamefont {G.~P.}\
  \bibnamefont {Mazur}}, \bibinfo {author} {\bibfnamefont {N.}~\bibnamefont
  {van Loo}}, \bibinfo {author} {\bibfnamefont {J.~C.}\ \bibnamefont {Wolff}},
  \bibinfo {author} {\bibfnamefont {A.~M.}\ \bibnamefont {Bozkurt}}, \bibinfo
  {author} {\bibfnamefont {G.}~\bibnamefont {Badawy}}, \bibinfo {author}
  {\bibfnamefont {S.}~\bibnamefont {Gazibegovic}}, \bibinfo {author}
  {\bibfnamefont {E.~P. A.~M.}\ \bibnamefont {Bakkers}}, \bibinfo {author}
  {\bibfnamefont {M.}~\bibnamefont {Wimmer}}, \bibinfo {author} {\bibfnamefont
  {L.~P.}\ \bibnamefont {Kouwenhoven}},\ and\ \bibinfo {author} {\bibfnamefont
  {T.}~\bibnamefont {Dvir}},\ }\bibfield  {title} {\bibinfo {title} {{Robust
  poor man's Majorana zero modes using Yu-Shiba-Rusinov states}},\ }\href
  {https://arxiv.org/abs/2311.03193} {\  (\bibinfo {year} {2023})},\ \Eprint
  {https://arxiv.org/abs/2311.03193} {arXiv:2311.03193 [cond-mat.mes-hall]}
  \BibitemShut {NoStop}%
\bibitem [{\citenamefont {Samuelson}\ \emph {et~al.}(2024)\citenamefont
  {Samuelson}, \citenamefont {Svensson},\ and\ \citenamefont
  {Leijnse}}]{samuelson2023minimal}%
  \BibitemOpen
  \bibfield  {author} {\bibinfo {author} {\bibfnamefont {W.}~\bibnamefont
  {Samuelson}}, \bibinfo {author} {\bibfnamefont {V.}~\bibnamefont
  {Svensson}},\ and\ \bibinfo {author} {\bibfnamefont {M.}~\bibnamefont
  {Leijnse}},\ }\bibfield  {title} {\bibinfo {title} {{Minimal quantum dot
  based Kitaev chain with only local superconducting proximity effect}},\
  }\href {https://doi.org/10.1103/PhysRevB.109.035415} {\bibfield  {journal}
  {\bibinfo  {journal} {Phys. Rev. B}\ }\textbf {\bibinfo {volume} {109}},\
  \bibinfo {pages} {035415} (\bibinfo {year} {2024})}\BibitemShut {NoStop}%
\bibitem [{\citenamefont {Tsintzis}\ \emph {et~al.}(2024)\citenamefont
  {Tsintzis}, \citenamefont {Souto}, \citenamefont {Flensberg}, \citenamefont
  {Danon},\ and\ \citenamefont {Leijnse}}]{tsintzis2023roadmap}%
  \BibitemOpen
  \bibfield  {author} {\bibinfo {author} {\bibfnamefont {A.}~\bibnamefont
  {Tsintzis}}, \bibinfo {author} {\bibfnamefont {R.~S.}\ \bibnamefont {Souto}},
  \bibinfo {author} {\bibfnamefont {K.}~\bibnamefont {Flensberg}}, \bibinfo
  {author} {\bibfnamefont {J.}~\bibnamefont {Danon}},\ and\ \bibinfo {author}
  {\bibfnamefont {M.}~\bibnamefont {Leijnse}},\ }\bibfield  {title} {\bibinfo
  {title} {{Majorana Qubits and Non-Abelian Physics in Quantum Dot--Based
  Minimal Kitaev Chains}},\ }\href
  {https://doi.org/10.1103/PRXQuantum.5.010323} {\bibfield  {journal} {\bibinfo
   {journal} {PRX Quantum}\ }\textbf {\bibinfo {volume} {5}},\ \bibinfo {pages}
  {010323} (\bibinfo {year} {2024})}\BibitemShut {NoStop}%
\bibitem [{\citenamefont {Pino}\ \emph
  {et~al.}(2024{\natexlab{a}})\citenamefont {Pino}, \citenamefont {Souto},\
  and\ \citenamefont {Aguado}}]{pino2024minimal}%
  \BibitemOpen
  \bibfield  {author} {\bibinfo {author} {\bibfnamefont {D.~M.}\ \bibnamefont
  {Pino}}, \bibinfo {author} {\bibfnamefont {R.~S.}\ \bibnamefont {Souto}},\
  and\ \bibinfo {author} {\bibfnamefont {R.}~\bibnamefont {Aguado}},\
  }\bibfield  {title} {\bibinfo {title} {{Minimal Kitaev-transmon qubit based
  on double quantum dots}},\ }\href
  {https://doi.org/10.1103/PhysRevB.109.075101} {\bibfield  {journal} {\bibinfo
   {journal} {Phys. Rev. B}\ }\textbf {\bibinfo {volume} {109}},\ \bibinfo
  {pages} {075101} (\bibinfo {year} {2024}{\natexlab{a}})}\BibitemShut
  {NoStop}%
\bibitem [{\citenamefont {Luna}\ \emph {et~al.}(2024)\citenamefont {Luna},
  \citenamefont {Bozkurt}, \citenamefont {Wimmer},\ and\ \citenamefont
  {Liu}}]{luna2024fluxtunable}%
  \BibitemOpen
  \bibfield  {author} {\bibinfo {author} {\bibfnamefont {J.~D.~T.}\
  \bibnamefont {Luna}}, \bibinfo {author} {\bibfnamefont {A.~M.}\ \bibnamefont
  {Bozkurt}}, \bibinfo {author} {\bibfnamefont {M.}~\bibnamefont {Wimmer}},\
  and\ \bibinfo {author} {\bibfnamefont {C.-X.}\ \bibnamefont {Liu}},\
  }\bibfield  {title} {\bibinfo {title} {{Flux-tunable Kitaev chain in a
  quantum dot array}},\ }\href {https://arxiv.org/abs/2402.07575} {\  (\bibinfo
  {year} {2024})},\ \Eprint {https://arxiv.org/abs/2402.07575}
  {arXiv:2402.07575 [cond-mat.mes-hall]} \BibitemShut {NoStop}%
\bibitem [{\citenamefont {Bordin}\ \emph
  {et~al.}(2024{\natexlab{a}})\citenamefont {Bordin}, \citenamefont {Liu},
  \citenamefont {Dvir}, \citenamefont {Zatelli}, \citenamefont {ten Haaf},
  \citenamefont {van Driel}, \citenamefont {Wang}, \citenamefont {van Loo},
  \citenamefont {van Caekenberghe}, \citenamefont {Wolff}, \citenamefont
  {Zhang}, \citenamefont {Badawy}, \citenamefont {Gazibegovic}, \citenamefont
  {Bakkers}, \citenamefont {Wimmer}, \citenamefont {Kouwenhoven},\ and\
  \citenamefont {Mazur}}]{bordin2024signatures}%
  \BibitemOpen
  \bibfield  {author} {\bibinfo {author} {\bibfnamefont {A.}~\bibnamefont
  {Bordin}}, \bibinfo {author} {\bibfnamefont {C.-X.}\ \bibnamefont {Liu}},
  \bibinfo {author} {\bibfnamefont {T.}~\bibnamefont {Dvir}}, \bibinfo {author}
  {\bibfnamefont {F.}~\bibnamefont {Zatelli}}, \bibinfo {author} {\bibfnamefont
  {S.~L.~D.}\ \bibnamefont {ten Haaf}}, \bibinfo {author} {\bibfnamefont
  {D.}~\bibnamefont {van Driel}}, \bibinfo {author} {\bibfnamefont
  {G.}~\bibnamefont {Wang}}, \bibinfo {author} {\bibfnamefont {N.}~\bibnamefont
  {van Loo}}, \bibinfo {author} {\bibfnamefont {T.}~\bibnamefont {van
  Caekenberghe}}, \bibinfo {author} {\bibfnamefont {J.~C.}\ \bibnamefont
  {Wolff}}, \bibinfo {author} {\bibfnamefont {Y.}~\bibnamefont {Zhang}},
  \bibinfo {author} {\bibfnamefont {G.}~\bibnamefont {Badawy}}, \bibinfo
  {author} {\bibfnamefont {S.}~\bibnamefont {Gazibegovic}}, \bibinfo {author}
  {\bibfnamefont {E.~P. A.~M.}\ \bibnamefont {Bakkers}}, \bibinfo {author}
  {\bibfnamefont {M.}~\bibnamefont {Wimmer}}, \bibinfo {author} {\bibfnamefont
  {L.~P.}\ \bibnamefont {Kouwenhoven}},\ and\ \bibinfo {author} {\bibfnamefont
  {G.~P.}\ \bibnamefont {Mazur}},\ }\bibfield  {title} {\bibinfo {title}
  {{Signatures of Majorana protection in a three-site Kitaev chain}},\ }\href
  {https://arxiv.org/abs/2402.19382} {\  (\bibinfo {year}
  {2024}{\natexlab{a}})},\ \Eprint {https://arxiv.org/abs/2402.19382}
  {arXiv:2402.19382 [cond-mat.supr-con]} \BibitemShut {NoStop}%
\bibitem [{\citenamefont {Boross}\ and\ \citenamefont
  {P\'alyi}(2024)}]{boross2023braiding}%
  \BibitemOpen
  \bibfield  {author} {\bibinfo {author} {\bibfnamefont {P.}~\bibnamefont
  {Boross}}\ and\ \bibinfo {author} {\bibfnamefont {A.}~\bibnamefont
  {P\'alyi}},\ }\bibfield  {title} {\bibinfo {title} {{Braiding-based quantum
  control of a Majorana qubit built from quantum dots}},\ }\href
  {https://doi.org/10.1103/PhysRevB.109.125410} {\bibfield  {journal} {\bibinfo
   {journal} {Phys. Rev. B}\ }\textbf {\bibinfo {volume} {109}},\ \bibinfo
  {pages} {125410} (\bibinfo {year} {2024})}\BibitemShut {NoStop}%
\bibitem [{\citenamefont {Liu}\ \emph {et~al.}(2024{\natexlab{a}})\citenamefont
  {Liu}, \citenamefont {Zeng},\ and\ \citenamefont {Xu}}]{liu2024coupling}%
  \BibitemOpen
  \bibfield  {author} {\bibinfo {author} {\bibfnamefont {Z.-H.}\ \bibnamefont
  {Liu}}, \bibinfo {author} {\bibfnamefont {C.}~\bibnamefont {Zeng}},\ and\
  \bibinfo {author} {\bibfnamefont {H.~Q.}\ \bibnamefont {Xu}},\ }\bibfield
  {title} {\bibinfo {title} {{Coupling of quantum-dot states via
  elastic-cotunneling and crossed Andreev reflection in a minimal Kitaev
  chain}},\ }\href {https://arxiv.org/abs/2403.08636} {\  (\bibinfo {year}
  {2024}{\natexlab{a}})},\ \Eprint {https://arxiv.org/abs/2403.08636}
  {arXiv:2403.08636 [cond-mat.mes-hall]} \BibitemShut {NoStop}%
\bibitem [{\citenamefont {Vilkelis}\ \emph {et~al.}(2024)\citenamefont
  {Vilkelis}, \citenamefont {Manesco}, \citenamefont {Luna}, \citenamefont
  {Miles}, \citenamefont {Wimmer},\ and\ \citenamefont
  {Akhmerov}}]{vilkelis2024fermionic}%
  \BibitemOpen
  \bibfield  {author} {\bibinfo {author} {\bibfnamefont {K.}~\bibnamefont
  {Vilkelis}}, \bibinfo {author} {\bibfnamefont {A.}~\bibnamefont {Manesco}},
  \bibinfo {author} {\bibfnamefont {J.~D.~T.}\ \bibnamefont {Luna}}, \bibinfo
  {author} {\bibfnamefont {S.}~\bibnamefont {Miles}}, \bibinfo {author}
  {\bibfnamefont {M.}~\bibnamefont {Wimmer}},\ and\ \bibinfo {author}
  {\bibfnamefont {A.}~\bibnamefont {Akhmerov}},\ }\bibfield  {title} {\bibinfo
  {title} {{Fermionic quantum computation with Cooper pair splitters}},\ }\href
  {https://doi.org/10.21468/SciPostPhys.16.5.135} {\bibfield  {journal}
  {\bibinfo  {journal} {SciPost Phys.}\ }\textbf {\bibinfo {volume} {16}},\
  \bibinfo {pages} {135} (\bibinfo {year} {2024})}\BibitemShut {NoStop}%
\bibitem [{\citenamefont {Ezawa}(2024)}]{ezawa2024even}%
  \BibitemOpen
  \bibfield  {author} {\bibinfo {author} {\bibfnamefont {M.}~\bibnamefont
  {Ezawa}},\ }\bibfield  {title} {\bibinfo {title} {{Even-odd effect on
  robustness of Majorana edge states in short Kitaev chains}},\ }\href
  {https://doi.org/10.1103/PhysRevB.109.L161404} {\bibfield  {journal}
  {\bibinfo  {journal} {Phys. Rev. B}\ }\textbf {\bibinfo {volume} {109}},\
  \bibinfo {pages} {L161404} (\bibinfo {year} {2024})}\BibitemShut {NoStop}%
\bibitem [{\citenamefont {Bozkurt}\ \emph {et~al.}(2024)\citenamefont
  {Bozkurt}, \citenamefont {Miles}, \citenamefont {ten Haaf}, \citenamefont
  {Liu}, \citenamefont {Hassler},\ and\ \citenamefont
  {Wimmer}}]{bozkurt2024interaction}%
  \BibitemOpen
  \bibfield  {author} {\bibinfo {author} {\bibfnamefont {A.~M.}\ \bibnamefont
  {Bozkurt}}, \bibinfo {author} {\bibfnamefont {S.}~\bibnamefont {Miles}},
  \bibinfo {author} {\bibfnamefont {S.~L.~D.}\ \bibnamefont {ten Haaf}},
  \bibinfo {author} {\bibfnamefont {C.-X.}\ \bibnamefont {Liu}}, \bibinfo
  {author} {\bibfnamefont {F.}~\bibnamefont {Hassler}},\ and\ \bibinfo {author}
  {\bibfnamefont {M.}~\bibnamefont {Wimmer}},\ }\bibfield  {title} {\bibinfo
  {title} {{Interaction-induced strong zero modes in short quantum dot chains
  with time-reversal symmetry}},\ }\href {https://arxiv.org/abs/2405.14940} {\
  (\bibinfo {year} {2024})},\ \Eprint {https://arxiv.org/abs/2405.14940}
  {arXiv:2405.14940 [cond-mat.mes-hall]} \BibitemShut {NoStop}%
\bibitem [{\citenamefont {van Driel}\ \emph {et~al.}(2024)\citenamefont {van
  Driel}, \citenamefont {Koch}, \citenamefont {Sietses}, \citenamefont {ten
  Haaf}, \citenamefont {Liu}, \citenamefont {Zatelli}, \citenamefont {Roovers},
  \citenamefont {Bordin}, \citenamefont {van Loo}, \citenamefont {Wang},
  \citenamefont {Wolff}, \citenamefont {Mazur}, \citenamefont {Dvir},
  \citenamefont {Kulesh}, \citenamefont {Wang}, \citenamefont {Bozkurt},
  \citenamefont {Gazibegovic}, \citenamefont {Badawy}, \citenamefont {Bakkers},
  \citenamefont {Wimmer}, \citenamefont {Goswami}, \citenamefont {Lado},
  \citenamefont {Kouwenhoven},\ and\ \citenamefont
  {Greplova}}]{vandriel2024crossplatform}%
  \BibitemOpen
  \bibfield  {author} {\bibinfo {author} {\bibfnamefont {D.}~\bibnamefont {van
  Driel}}, \bibinfo {author} {\bibfnamefont {R.}~\bibnamefont {Koch}}, \bibinfo
  {author} {\bibfnamefont {V.~P.~M.}\ \bibnamefont {Sietses}}, \bibinfo
  {author} {\bibfnamefont {S.~L.~D.}\ \bibnamefont {ten Haaf}}, \bibinfo
  {author} {\bibfnamefont {C.-X.}\ \bibnamefont {Liu}}, \bibinfo {author}
  {\bibfnamefont {F.}~\bibnamefont {Zatelli}}, \bibinfo {author} {\bibfnamefont
  {B.}~\bibnamefont {Roovers}}, \bibinfo {author} {\bibfnamefont
  {A.}~\bibnamefont {Bordin}}, \bibinfo {author} {\bibfnamefont
  {N.}~\bibnamefont {van Loo}}, \bibinfo {author} {\bibfnamefont
  {G.}~\bibnamefont {Wang}}, \bibinfo {author} {\bibfnamefont {J.~C.}\
  \bibnamefont {Wolff}}, \bibinfo {author} {\bibfnamefont {G.~P.}\ \bibnamefont
  {Mazur}}, \bibinfo {author} {\bibfnamefont {T.}~\bibnamefont {Dvir}},
  \bibinfo {author} {\bibfnamefont {I.}~\bibnamefont {Kulesh}}, \bibinfo
  {author} {\bibfnamefont {Q.}~\bibnamefont {Wang}}, \bibinfo {author}
  {\bibfnamefont {A.~M.}\ \bibnamefont {Bozkurt}}, \bibinfo {author}
  {\bibfnamefont {S.}~\bibnamefont {Gazibegovic}}, \bibinfo {author}
  {\bibfnamefont {G.}~\bibnamefont {Badawy}}, \bibinfo {author} {\bibfnamefont
  {E.~P. A.~M.}\ \bibnamefont {Bakkers}}, \bibinfo {author} {\bibfnamefont
  {M.}~\bibnamefont {Wimmer}}, \bibinfo {author} {\bibfnamefont
  {S.}~\bibnamefont {Goswami}}, \bibinfo {author} {\bibfnamefont {J.~L.}\
  \bibnamefont {Lado}}, \bibinfo {author} {\bibfnamefont {L.~P.}\ \bibnamefont
  {Kouwenhoven}},\ and\ \bibinfo {author} {\bibfnamefont {E.}~\bibnamefont
  {Greplova}},\ }\bibfield  {title} {\bibinfo {title} {{Cross-Platform
  Autonomous Control of Minimal Kitaev Chains}},\ }\href
  {https://arxiv.org/abs/2405.04596} {\  (\bibinfo {year} {2024})},\ \Eprint
  {https://arxiv.org/abs/2405.04596} {arXiv:2405.04596 [cond-mat.mes-hall]}
  \BibitemShut {NoStop}%
\bibitem [{\citenamefont {Pino}\ \emph
  {et~al.}(2024{\natexlab{b}})\citenamefont {Pino}, \citenamefont {Meir},\ and\
  \citenamefont {Aguado}}]{pino2024thermodynamics}%
  \BibitemOpen
  \bibfield  {author} {\bibinfo {author} {\bibfnamefont {D.~M.}\ \bibnamefont
  {Pino}}, \bibinfo {author} {\bibfnamefont {Y.}~\bibnamefont {Meir}},\ and\
  \bibinfo {author} {\bibfnamefont {R.}~\bibnamefont {Aguado}},\ }\href
  {https://arxiv.org/abs/2405.02387} {\bibinfo {title} {{Thermodynamics of
  Non-Hermitian Josephson junctions with exceptional points}}} (\bibinfo {year}
  {2024}{\natexlab{b}}),\ \Eprint {https://arxiv.org/abs/2405.02387}
  {arXiv:2405.02387 [cond-mat.mes-hall]} \BibitemShut {NoStop}%
\bibitem [{\citenamefont {Geier}\ \emph {et~al.}(2024)\citenamefont {Geier},
  \citenamefont {Souto}, \citenamefont {Schulenborg}, \citenamefont {Asaad},
  \citenamefont {Leijnse},\ and\ \citenamefont
  {Flensberg}}]{geier2023fermionparity}%
  \BibitemOpen
  \bibfield  {author} {\bibinfo {author} {\bibfnamefont {M.}~\bibnamefont
  {Geier}}, \bibinfo {author} {\bibfnamefont {R.~S.}\ \bibnamefont {Souto}},
  \bibinfo {author} {\bibfnamefont {J.}~\bibnamefont {Schulenborg}}, \bibinfo
  {author} {\bibfnamefont {S.}~\bibnamefont {Asaad}}, \bibinfo {author}
  {\bibfnamefont {M.}~\bibnamefont {Leijnse}},\ and\ \bibinfo {author}
  {\bibfnamefont {K.}~\bibnamefont {Flensberg}},\ }\bibfield  {title} {\bibinfo
  {title} {{Fermion-parity qubit in a proximitized double quantum dot}},\
  }\href {https://doi.org/10.1103/PhysRevResearch.6.023281} {\bibfield
  {journal} {\bibinfo  {journal} {Phys. Rev. Res.}\ }\textbf {\bibinfo {volume}
  {6}},\ \bibinfo {pages} {023281} (\bibinfo {year} {2024})}\BibitemShut
  {NoStop}%
\bibitem [{\citenamefont {Ten~Haaf}\ \emph {et~al.}(2024)\citenamefont
  {Ten~Haaf}, \citenamefont {Wang}, \citenamefont {Bozkurt}, \citenamefont
  {Liu}, \citenamefont {Kulesh}, \citenamefont {Kim}, \citenamefont {Xiao},
  \citenamefont {Thomas}, \citenamefont {Manfra}, \citenamefont {Dvir},
  \citenamefont {Wimmer},\ and\ \citenamefont {Goswami}}]{haaf2023engineering}%
  \BibitemOpen
  \bibfield  {author} {\bibinfo {author} {\bibfnamefont {S.~L.}\ \bibnamefont
  {Ten~Haaf}}, \bibinfo {author} {\bibfnamefont {Q.}~\bibnamefont {Wang}},
  \bibinfo {author} {\bibfnamefont {A.~M.}\ \bibnamefont {Bozkurt}}, \bibinfo
  {author} {\bibfnamefont {C.-X.}\ \bibnamefont {Liu}}, \bibinfo {author}
  {\bibfnamefont {I.}~\bibnamefont {Kulesh}}, \bibinfo {author} {\bibfnamefont
  {P.}~\bibnamefont {Kim}}, \bibinfo {author} {\bibfnamefont {D.}~\bibnamefont
  {Xiao}}, \bibinfo {author} {\bibfnamefont {C.}~\bibnamefont {Thomas}},
  \bibinfo {author} {\bibfnamefont {M.~J.}\ \bibnamefont {Manfra}}, \bibinfo
  {author} {\bibfnamefont {T.}~\bibnamefont {Dvir}}, \bibinfo {author}
  {\bibfnamefont {M.}~\bibnamefont {Wimmer}},\ and\ \bibinfo {author}
  {\bibfnamefont {S.}~\bibnamefont {Goswami}},\ }\bibfield  {title} {\bibinfo
  {title} {{A two-site Kitaev chain in a two-dimensional electron gas}},\
  }\href {https://doi.org/10.1038/s41586-024-07434-9} {\bibfield  {journal}
  {\bibinfo  {journal} {Nature}\ }\textbf {\bibinfo {volume} {630}},\ \bibinfo
  {pages} {329} (\bibinfo {year} {2024})}\BibitemShut {NoStop}%
\bibitem [{\citenamefont {Kocsis}\ \emph {et~al.}(2024)\citenamefont {Kocsis},
  \citenamefont {Scher\"ubl}, \citenamefont {F\"ul\"op}, \citenamefont {Makk},\
  and\ \citenamefont {Csonka}}]{kocsis2024strong}%
  \BibitemOpen
  \bibfield  {author} {\bibinfo {author} {\bibfnamefont {M.}~\bibnamefont
  {Kocsis}}, \bibinfo {author} {\bibfnamefont {Z.}~\bibnamefont {Scher\"ubl}},
  \bibinfo {author} {\bibfnamefont {G.~m.~H.}\ \bibnamefont {F\"ul\"op}},
  \bibinfo {author} {\bibfnamefont {P.}~\bibnamefont {Makk}},\ and\ \bibinfo
  {author} {\bibfnamefont {S.}~\bibnamefont {Csonka}},\ }\bibfield  {title}
  {\bibinfo {title} {{Strong nonlocal tuning of the current-phase relation of a
  quantum dot based Andreev molecule}},\ }\href
  {https://doi.org/10.1103/PhysRevB.109.245133} {\bibfield  {journal} {\bibinfo
   {journal} {Phys. Rev. B}\ }\textbf {\bibinfo {volume} {109}},\ \bibinfo
  {pages} {245133} (\bibinfo {year} {2024})}\BibitemShut {NoStop}%
\bibitem [{\citenamefont {Cayao}(2024)}]{cayao2024emergent}%
  \BibitemOpen
  \bibfield  {author} {\bibinfo {author} {\bibfnamefont {J.}~\bibnamefont
  {Cayao}},\ }\href {https://arxiv.org/abs/2406.17508} {\bibinfo {title}
  {{Emergent pair symmetries in systems with poor man's Majorana modes}}}
  (\bibinfo {year} {2024}),\ \Eprint {https://arxiv.org/abs/2406.17508}
  {arXiv:2406.17508 [cond-mat.mes-hall]} \BibitemShut {NoStop}%
\bibitem [{\citenamefont {Miles}\ \emph {et~al.}(2024)\citenamefont {Miles},
  \citenamefont {van Driel}, \citenamefont {Wimmer},\ and\ \citenamefont
  {Liu}}]{miles2024kitaev}%
  \BibitemOpen
  \bibfield  {author} {\bibinfo {author} {\bibfnamefont {S.}~\bibnamefont
  {Miles}}, \bibinfo {author} {\bibfnamefont {D.}~\bibnamefont {van Driel}},
  \bibinfo {author} {\bibfnamefont {M.}~\bibnamefont {Wimmer}},\ and\ \bibinfo
  {author} {\bibfnamefont {C.-X.}\ \bibnamefont {Liu}},\ }\bibfield  {title}
  {\bibinfo {title} {{Kitaev chain in an alternating quantum dot-Andreev bound
  state array}},\ }\href {https://doi.org/10.1103/PhysRevB.110.024520}
  {\bibfield  {journal} {\bibinfo  {journal} {Phys. Rev. B}\ }\textbf {\bibinfo
  {volume} {110}},\ \bibinfo {pages} {024520} (\bibinfo {year}
  {2024})}\BibitemShut {NoStop}%
\bibitem [{\citenamefont {Cayao}\ and\ \citenamefont
  {Aguado}(2024)}]{cayao2024nonhermitian}%
  \BibitemOpen
  \bibfield  {author} {\bibinfo {author} {\bibfnamefont {J.}~\bibnamefont
  {Cayao}}\ and\ \bibinfo {author} {\bibfnamefont {R.}~\bibnamefont {Aguado}},\
  }\href {https://arxiv.org/abs/2406.18974} {\bibinfo {title} {{Non-Hermitian
  minimal Kitaev chains}}} (\bibinfo {year} {2024}),\ \Eprint
  {https://arxiv.org/abs/2406.18974} {arXiv:2406.18974 [cond-mat.mes-hall]}
  \BibitemShut {NoStop}%
\bibitem [{\citenamefont {Liu}\ \emph {et~al.}(2024{\natexlab{b}})\citenamefont
  {Liu}, \citenamefont {Miles}, \citenamefont {Bordin}, \citenamefont {ten
  Haaf}, \citenamefont {Bozkurt},\ and\ \citenamefont
  {Wimmer}}]{liu2024protocol}%
  \BibitemOpen
  \bibfield  {author} {\bibinfo {author} {\bibfnamefont {C.-X.}\ \bibnamefont
  {Liu}}, \bibinfo {author} {\bibfnamefont {S.}~\bibnamefont {Miles}}, \bibinfo
  {author} {\bibfnamefont {A.}~\bibnamefont {Bordin}}, \bibinfo {author}
  {\bibfnamefont {S.~L.~D.}\ \bibnamefont {ten Haaf}}, \bibinfo {author}
  {\bibfnamefont {A.~M.}\ \bibnamefont {Bozkurt}},\ and\ \bibinfo {author}
  {\bibfnamefont {M.}~\bibnamefont {Wimmer}},\ }\href
  {https://arxiv.org/abs/2407.04630} {\bibinfo {title} {{Protocol for scaling
  up a sign-ordered Kitaev chain without magnetic flux control}}} (\bibinfo
  {year} {2024}{\natexlab{b}}),\ \Eprint {https://arxiv.org/abs/2407.04630}
  {arXiv:2407.04630 [cond-mat.mes-hall]} \BibitemShut {NoStop}%
\bibitem [{\citenamefont {Alvarado}\ \emph {et~al.}(2024)\citenamefont
  {Alvarado}, \citenamefont {Yeyati}, \citenamefont {Aguado},\ and\
  \citenamefont {Souto}}]{alvarado2024interplay}%
  \BibitemOpen
  \bibfield  {author} {\bibinfo {author} {\bibfnamefont {M.}~\bibnamefont
  {Alvarado}}, \bibinfo {author} {\bibfnamefont {A.~L.}\ \bibnamefont
  {Yeyati}}, \bibinfo {author} {\bibfnamefont {R.}~\bibnamefont {Aguado}},\
  and\ \bibinfo {author} {\bibfnamefont {R.~S.}\ \bibnamefont {Souto}},\ }\href
  {https://arxiv.org/abs/2407.07050} {\bibinfo {title} {{Interplay between
  Majorana and Shiba states in a minimal Kitaev chain coupled to a
  superconductor}}} (\bibinfo {year} {2024}),\ \Eprint
  {https://arxiv.org/abs/2407.07050} {arXiv:2407.07050 [cond-mat.mes-hall]}
  \BibitemShut {NoStop}%
\bibitem [{\citenamefont {Svensson}\ and\ \citenamefont
  {Leijnse}(2024)}]{svensson2024quantum}%
  \BibitemOpen
  \bibfield  {author} {\bibinfo {author} {\bibfnamefont {V.}~\bibnamefont
  {Svensson}}\ and\ \bibinfo {author} {\bibfnamefont {M.}~\bibnamefont
  {Leijnse}},\ }\href {https://arxiv.org/abs/2407.09211} {\bibinfo {title}
  {{Quantum-dot-based Kitaev chains: Majorana quality measures and scaling with
  increasing chain length}}} (\bibinfo {year} {2024}),\ \Eprint
  {https://arxiv.org/abs/2407.09211} {arXiv:2407.09211 [cond-mat.mes-hall]}
  \BibitemShut {NoStop}%
\bibitem [{\citenamefont {Álvaro Gómez-León}\ \emph
  {et~al.}(2024)\citenamefont {Álvaro Gómez-León}, \citenamefont {Schirò},\
  and\ \citenamefont {Dmytruk}}]{gomez2024high}%
  \BibitemOpen
  \bibfield  {author} {\bibinfo {author} {\bibnamefont {Álvaro Gómez-León}},
  \bibinfo {author} {\bibfnamefont {M.}~\bibnamefont {Schirò}},\ and\ \bibinfo
  {author} {\bibfnamefont {O.}~\bibnamefont {Dmytruk}},\ }\href
  {https://arxiv.org/abs/2407.12088} {\bibinfo {title} {{High-quality poor
  man's Majorana bound states from cavity embedding}}} (\bibinfo {year}
  {2024}),\ \Eprint {https://arxiv.org/abs/2407.12088} {arXiv:2407.12088
  [cond-mat.mes-hall]} \BibitemShut {NoStop}%
\bibitem [{\citenamefont {Pan}\ \emph {et~al.}(2024)\citenamefont {Pan},
  \citenamefont {Sarma},\ and\ \citenamefont {Liu}}]{pan2024rabi}%
  \BibitemOpen
  \bibfield  {author} {\bibinfo {author} {\bibfnamefont {H.}~\bibnamefont
  {Pan}}, \bibinfo {author} {\bibfnamefont {S.~D.}\ \bibnamefont {Sarma}},\
  and\ \bibinfo {author} {\bibfnamefont {C.-X.}\ \bibnamefont {Liu}},\ }\href
  {https://arxiv.org/abs/2407.16750} {\bibinfo {title} {{Rabi and Ramsey
  oscillations of a Majorana qubit in a quantum dot-superconductor array}}}
  (\bibinfo {year} {2024}),\ \Eprint {https://arxiv.org/abs/2407.16750}
  {arXiv:2407.16750 [cond-mat.mes-hall]} \BibitemShut {NoStop}%
\bibitem [{\citenamefont {Pandey}\ and\ \citenamefont
  {Dagotto}(2024)}]{pandey2024dynamics}%
  \BibitemOpen
  \bibfield  {author} {\bibinfo {author} {\bibfnamefont {B.}~\bibnamefont
  {Pandey}}\ and\ \bibinfo {author} {\bibfnamefont {E.}~\bibnamefont
  {Dagotto}},\ }\href {https://arxiv.org/abs/2407.20783} {\bibinfo {title}
  {{Dynamics and Fusion of Majorana Zero Modes in Quantum-dot based Interacting
  Kitaev Chains}}} (\bibinfo {year} {2024}),\ \Eprint
  {https://arxiv.org/abs/2407.20783} {arXiv:2407.20783 [cond-mat.supr-con]}
  \BibitemShut {NoStop}%
\bibitem [{\citenamefont {Liu}\ \emph {et~al.}(2024{\natexlab{c}})\citenamefont
  {Liu}, \citenamefont {Bozkurt}, \citenamefont {Zatelli}, \citenamefont {ten
  Haaf}, \citenamefont {Dvir},\ and\ \citenamefont
  {Wimmer}}]{liu2023enhancing}%
  \BibitemOpen
  \bibfield  {author} {\bibinfo {author} {\bibfnamefont {C.-X.}\ \bibnamefont
  {Liu}}, \bibinfo {author} {\bibfnamefont {A.~M.}\ \bibnamefont {Bozkurt}},
  \bibinfo {author} {\bibfnamefont {F.}~\bibnamefont {Zatelli}}, \bibinfo
  {author} {\bibfnamefont {S.~L.}\ \bibnamefont {ten Haaf}}, \bibinfo {author}
  {\bibfnamefont {T.}~\bibnamefont {Dvir}},\ and\ \bibinfo {author}
  {\bibfnamefont {M.}~\bibnamefont {Wimmer}},\ }\bibfield  {title} {\bibinfo
  {title} {{Enhancing the excitation gap of a quantum-dot-based Kitaev
  chain}},\ }\href {https://doi.org/10.1038/s42005-024-01715-5} {\bibfield
  {journal} {\bibinfo  {journal} {Communications Physics}\ }\textbf {\bibinfo
  {volume} {7}},\ \bibinfo {pages} {235} (\bibinfo {year}
  {2024}{\natexlab{c}})}\BibitemShut {NoStop}%
\bibitem [{\citenamefont {Benestad}\ \emph {et~al.}(2024)\citenamefont
  {Benestad}, \citenamefont {Tsintzis}, \citenamefont {Souto}, \citenamefont
  {Leijnse}, \citenamefont {van Nieuwenburg},\ and\ \citenamefont
  {Danon}}]{benestad2024machine}%
  \BibitemOpen
  \bibfield  {author} {\bibinfo {author} {\bibfnamefont {J.}~\bibnamefont
  {Benestad}}, \bibinfo {author} {\bibfnamefont {A.}~\bibnamefont {Tsintzis}},
  \bibinfo {author} {\bibfnamefont {R.~S.}\ \bibnamefont {Souto}}, \bibinfo
  {author} {\bibfnamefont {M.}~\bibnamefont {Leijnse}}, \bibinfo {author}
  {\bibfnamefont {E.}~\bibnamefont {van Nieuwenburg}},\ and\ \bibinfo {author}
  {\bibfnamefont {J.}~\bibnamefont {Danon}},\ }\bibfield  {title} {\bibinfo
  {title} {{Machine-learned tuning of artificial Kitaev chains from tunneling
  spectroscopy measurements}},\ }\href
  {https://doi.org/10.1103/PhysRevB.110.075402} {\bibfield  {journal} {\bibinfo
   {journal} {Phys. Rev. B}\ }\textbf {\bibinfo {volume} {110}},\ \bibinfo
  {pages} {075402} (\bibinfo {year} {2024})}\BibitemShut {NoStop}%
\bibitem [{\citenamefont {Luethi}\ \emph {et~al.}(2024)\citenamefont {Luethi},
  \citenamefont {Legg}, \citenamefont {Loss},\ and\ \citenamefont
  {Klinovaja}}]{luethi2024perfect}%
  \BibitemOpen
  \bibfield  {author} {\bibinfo {author} {\bibfnamefont {M.}~\bibnamefont
  {Luethi}}, \bibinfo {author} {\bibfnamefont {H.~F.}\ \bibnamefont {Legg}},
  \bibinfo {author} {\bibfnamefont {D.}~\bibnamefont {Loss}},\ and\ \bibinfo
  {author} {\bibfnamefont {J.}~\bibnamefont {Klinovaja}},\ }\bibfield  {title}
  {\bibinfo {title} {{From perfect to imperfect poor man's Majoranas in minimal
  Kitaev chains}},\ }\href {https://doi.org/10.1103/PhysRevB.110.245412}
  {\bibfield  {journal} {\bibinfo  {journal} {Phys. Rev. B}\ }\textbf {\bibinfo
  {volume} {110}},\ \bibinfo {pages} {245412} (\bibinfo {year}
  {2024})}\BibitemShut {NoStop}%
\bibitem [{\citenamefont {ten Haaf}\ \emph {et~al.}(2024)\citenamefont {ten
  Haaf}, \citenamefont {Zhang}, \citenamefont {Wang}, \citenamefont {Bordin},
  \citenamefont {Liu}, \citenamefont {Kulesh}, \citenamefont {Sietses},
  \citenamefont {Prosko}, \citenamefont {Xiao}, \citenamefont {Thomas} \emph
  {et~al.}}]{ten2024edge}%
  \BibitemOpen
  \bibfield  {author} {\bibinfo {author} {\bibfnamefont {S.~L.}\ \bibnamefont
  {ten Haaf}}, \bibinfo {author} {\bibfnamefont {Y.}~\bibnamefont {Zhang}},
  \bibinfo {author} {\bibfnamefont {Q.}~\bibnamefont {Wang}}, \bibinfo {author}
  {\bibfnamefont {A.}~\bibnamefont {Bordin}}, \bibinfo {author} {\bibfnamefont
  {C.-X.}\ \bibnamefont {Liu}}, \bibinfo {author} {\bibfnamefont
  {I.}~\bibnamefont {Kulesh}}, \bibinfo {author} {\bibfnamefont {V.~P.}\
  \bibnamefont {Sietses}}, \bibinfo {author} {\bibfnamefont {C.~G.}\
  \bibnamefont {Prosko}}, \bibinfo {author} {\bibfnamefont {D.}~\bibnamefont
  {Xiao}}, \bibinfo {author} {\bibfnamefont {C.}~\bibnamefont {Thomas}}, \emph
  {et~al.},\ }\bibfield  {title} {\bibinfo {title} {{Edge and bulk states in a
  three-site Kitaev chain}},\ }\href {https://arxiv.org/abs/2410.00658}
  {\bibfield  {journal} {\bibinfo  {journal} {arXiv:2410.00658}\ } (\bibinfo
  {year} {2024})}\BibitemShut {NoStop}%
\bibitem [{\citenamefont {Recher}\ \emph {et~al.}(2001)\citenamefont {Recher},
  \citenamefont {Sukhorukov},\ and\ \citenamefont {Loss}}]{recher2001andreev}%
  \BibitemOpen
  \bibfield  {author} {\bibinfo {author} {\bibfnamefont {P.}~\bibnamefont
  {Recher}}, \bibinfo {author} {\bibfnamefont {E.~V.}\ \bibnamefont
  {Sukhorukov}},\ and\ \bibinfo {author} {\bibfnamefont {D.}~\bibnamefont
  {Loss}},\ }\bibfield  {title} {\bibinfo {title} {{Andreev tunneling, Coulomb
  blockade, and resonant transport of nonlocal spin-entangled electrons}},\
  }\href {https://doi.org/10.1103/PhysRevB.63.165314} {\bibfield  {journal}
  {\bibinfo  {journal} {Phys. Rev. B}\ }\textbf {\bibinfo {volume} {63}},\
  \bibinfo {pages} {165314} (\bibinfo {year} {2001})}\BibitemShut {NoStop}%
\bibitem [{\citenamefont {Lesovik}\ \emph {et~al.}(2001)\citenamefont
  {Lesovik}, \citenamefont {Martin},\ and\ \citenamefont
  {Blatter}}]{lesovik2001electronic}%
  \BibitemOpen
  \bibfield  {author} {\bibinfo {author} {\bibfnamefont {G.~B.}\ \bibnamefont
  {Lesovik}}, \bibinfo {author} {\bibfnamefont {T.}~\bibnamefont {Martin}},\
  and\ \bibinfo {author} {\bibfnamefont {G.}~\bibnamefont {Blatter}},\
  }\bibfield  {title} {\bibinfo {title} {{Electronic entanglement in the
  vicinity of a superconductor}},\ }\href
  {https://doi.org/10.1007/s10051-001-8675-4} {\bibfield  {journal} {\bibinfo
  {journal} {The European Physical Journal B-Condensed Matter and Complex
  Systems}\ }\textbf {\bibinfo {volume} {24}},\ \bibinfo {pages} {287}
  (\bibinfo {year} {2001})}\BibitemShut {NoStop}%
\bibitem [{\citenamefont {Falci}\ \emph {et~al.}(2001)\citenamefont {Falci},
  \citenamefont {Feinberg},\ and\ \citenamefont
  {Hekking}}]{falci2001correlated}%
  \BibitemOpen
  \bibfield  {author} {\bibinfo {author} {\bibfnamefont {G.}~\bibnamefont
  {Falci}}, \bibinfo {author} {\bibfnamefont {D.}~\bibnamefont {Feinberg}},\
  and\ \bibinfo {author} {\bibfnamefont {F.~W.~J.}\ \bibnamefont {Hekking}},\
  }\bibfield  {title} {\bibinfo {title} {{Correlated tunneling into a
  superconductor in a multiprobe hybrid structure}},\ }\href
  {https://doi.org/10.1209/epl/i2001-00303-0} {\bibfield  {journal} {\bibinfo
  {journal} {Europhysics Letters}\ }\textbf {\bibinfo {volume} {54}},\ \bibinfo
  {pages} {255} (\bibinfo {year} {2001})}\BibitemShut {NoStop}%
\bibitem [{\citenamefont {Bouchiat}\ \emph {et~al.}(2002)\citenamefont
  {Bouchiat}, \citenamefont {Chtchelkatchev}, \citenamefont {Feinberg},
  \citenamefont {Lesovik}, \citenamefont {Martin},\ and\ \citenamefont
  {Torrès}}]{bouchiat2002single}%
  \BibitemOpen
  \bibfield  {author} {\bibinfo {author} {\bibfnamefont {V.}~\bibnamefont
  {Bouchiat}}, \bibinfo {author} {\bibfnamefont {N.}~\bibnamefont
  {Chtchelkatchev}}, \bibinfo {author} {\bibfnamefont {D.}~\bibnamefont
  {Feinberg}}, \bibinfo {author} {\bibfnamefont {G.~B.}\ \bibnamefont
  {Lesovik}}, \bibinfo {author} {\bibfnamefont {T.}~\bibnamefont {Martin}},\
  and\ \bibinfo {author} {\bibfnamefont {J.}~\bibnamefont {Torrès}},\
  }\bibfield  {title} {\bibinfo {title} {{Single-walled carbon
  nanotube–superconductor entangler: noise correlations and
  Einstein–Podolsky–Rosen states}},\ }\href
  {https://doi.org/10.1088/0957-4484/14/1/318} {\bibfield  {journal} {\bibinfo
  {journal} {Nanotechnology}\ }\textbf {\bibinfo {volume} {14}},\ \bibinfo
  {pages} {77} (\bibinfo {year} {2002})}\BibitemShut {NoStop}%
\bibitem [{\citenamefont {Feinberg}(2003)}]{feinberg2003andreev}%
  \BibitemOpen
  \bibfield  {author} {\bibinfo {author} {\bibfnamefont {D.}~\bibnamefont
  {Feinberg}},\ }\bibfield  {title} {\bibinfo {title} {{Andreev scattering and
  cotunneling between two superconductor-normal metal interfaces: the dirty
  limit}},\ }\href {https://doi.org/10.1140/epjb/e2003-00361-6} {\bibfield
  {journal} {\bibinfo  {journal} {The European Physical Journal B-Condensed
  Matter and Complex Systems}\ }\textbf {\bibinfo {volume} {36}},\ \bibinfo
  {pages} {419} (\bibinfo {year} {2003})}\BibitemShut {NoStop}%
\bibitem [{\citenamefont {Hofstetter}\ \emph {et~al.}(2009)\citenamefont
  {Hofstetter}, \citenamefont {Csonka}, \citenamefont {Nyg{\aa}rd},\ and\
  \citenamefont {Sch{\"o}nenberger}}]{hofstetter2009cooper}%
  \BibitemOpen
  \bibfield  {author} {\bibinfo {author} {\bibfnamefont {L.}~\bibnamefont
  {Hofstetter}}, \bibinfo {author} {\bibfnamefont {S.}~\bibnamefont {Csonka}},
  \bibinfo {author} {\bibfnamefont {J.}~\bibnamefont {Nyg{\aa}rd}},\ and\
  \bibinfo {author} {\bibfnamefont {C.}~\bibnamefont {Sch{\"o}nenberger}},\
  }\bibfield  {title} {\bibinfo {title} {{Cooper pair splitter realized in a
  two-quantum-dot Y-junction}},\ }\href {https://doi.org/10.1038/nature08432}
  {\bibfield  {journal} {\bibinfo  {journal} {Nature}\ }\textbf {\bibinfo
  {volume} {461}},\ \bibinfo {pages} {960} (\bibinfo {year}
  {2009})}\BibitemShut {NoStop}%
\bibitem [{\citenamefont {Herrmann}\ \emph {et~al.}(2010)\citenamefont
  {Herrmann}, \citenamefont {Portier}, \citenamefont {Roche}, \citenamefont
  {Yeyati}, \citenamefont {Kontos},\ and\ \citenamefont
  {Strunk}}]{herrmann2010carbon}%
  \BibitemOpen
  \bibfield  {author} {\bibinfo {author} {\bibfnamefont {L.~G.}\ \bibnamefont
  {Herrmann}}, \bibinfo {author} {\bibfnamefont {F.}~\bibnamefont {Portier}},
  \bibinfo {author} {\bibfnamefont {P.}~\bibnamefont {Roche}}, \bibinfo
  {author} {\bibfnamefont {A.~L.}\ \bibnamefont {Yeyati}}, \bibinfo {author}
  {\bibfnamefont {T.}~\bibnamefont {Kontos}},\ and\ \bibinfo {author}
  {\bibfnamefont {C.}~\bibnamefont {Strunk}},\ }\bibfield  {title} {\bibinfo
  {title} {{Carbon Nanotubes as Cooper-Pair Beam Splitters}},\ }\href
  {https://doi.org/10.1103/PhysRevLett.104.026801} {\bibfield  {journal}
  {\bibinfo  {journal} {Phys. Rev. Lett.}\ }\textbf {\bibinfo {volume} {104}},\
  \bibinfo {pages} {026801} (\bibinfo {year} {2010})}\BibitemShut {NoStop}%
\bibitem [{\citenamefont {Wang}\ \emph {et~al.}(2022)\citenamefont {Wang},
  \citenamefont {Dvir}, \citenamefont {Mazur}, \citenamefont {Liu},
  \citenamefont {van Loo}, \citenamefont {Ten~Haaf}, \citenamefont {Bordin},
  \citenamefont {Gazibegovic}, \citenamefont {Badawy}, \citenamefont {Bakkers},
  \citenamefont {Wimmer},\ and\ \citenamefont {P}}]{wang2022singlet}%
  \BibitemOpen
  \bibfield  {author} {\bibinfo {author} {\bibfnamefont {G.}~\bibnamefont
  {Wang}}, \bibinfo {author} {\bibfnamefont {T.}~\bibnamefont {Dvir}}, \bibinfo
  {author} {\bibfnamefont {G.~P.}\ \bibnamefont {Mazur}}, \bibinfo {author}
  {\bibfnamefont {C.-X.}\ \bibnamefont {Liu}}, \bibinfo {author} {\bibfnamefont
  {N.}~\bibnamefont {van Loo}}, \bibinfo {author} {\bibfnamefont {S.~L.}\
  \bibnamefont {Ten~Haaf}}, \bibinfo {author} {\bibfnamefont {A.}~\bibnamefont
  {Bordin}}, \bibinfo {author} {\bibfnamefont {S.}~\bibnamefont {Gazibegovic}},
  \bibinfo {author} {\bibfnamefont {G.}~\bibnamefont {Badawy}}, \bibinfo
  {author} {\bibfnamefont {E.~P.}\ \bibnamefont {Bakkers}}, \bibinfo {author}
  {\bibfnamefont {M.}~\bibnamefont {Wimmer}},\ and\ \bibinfo {author}
  {\bibfnamefont {K.~L.}\ \bibnamefont {P}},\ }\bibfield  {title} {\bibinfo
  {title} {{Singlet and triplet Cooper pair splitting in hybrid superconducting
  nanowires}},\ }\href {https://doi.org/10.1038/s41586-022-05352-2} {\bibfield
  {journal} {\bibinfo  {journal} {Nature}\ }\textbf {\bibinfo {volume} {612}},\
  \bibinfo {pages} {448} (\bibinfo {year} {2022})}\BibitemShut {NoStop}%
\bibitem [{\citenamefont {K{\"u}rt{\"o}ssy}\ \emph {et~al.}(2022)\citenamefont
  {K{\"u}rt{\"o}ssy}, \citenamefont {Scher{\"u}bl}, \citenamefont
  {F{\"u}l{\"o}p}, \citenamefont {Luk{\'a}cs}, \citenamefont {Kanne},
  \citenamefont {Nyg{\aa}rd}, \citenamefont {Makk},\ and\ \citenamefont
  {Csonka}}]{kurtossy2022parallel}%
  \BibitemOpen
  \bibfield  {author} {\bibinfo {author} {\bibfnamefont {O.}~\bibnamefont
  {K{\"u}rt{\"o}ssy}}, \bibinfo {author} {\bibfnamefont {Z.}~\bibnamefont
  {Scher{\"u}bl}}, \bibinfo {author} {\bibfnamefont {G.}~\bibnamefont
  {F{\"u}l{\"o}p}}, \bibinfo {author} {\bibfnamefont {I.~E.}\ \bibnamefont
  {Luk{\'a}cs}}, \bibinfo {author} {\bibfnamefont {T.}~\bibnamefont {Kanne}},
  \bibinfo {author} {\bibfnamefont {J.}~\bibnamefont {Nyg{\aa}rd}}, \bibinfo
  {author} {\bibfnamefont {P.}~\bibnamefont {Makk}},\ and\ \bibinfo {author}
  {\bibfnamefont {S.}~\bibnamefont {Csonka}},\ }\bibfield  {title} {\bibinfo
  {title} {{Parallel InAs nanowires for Cooper pair splitters with Coulomb
  repulsion}},\ }\href@noop {} {\bibfield  {journal} {\bibinfo  {journal} {npj
  Quantum Materials}\ }\textbf {\bibinfo {volume} {7}},\ \bibinfo {pages} {88}
  (\bibinfo {year} {2022})}\BibitemShut {NoStop}%
\bibitem [{\citenamefont {de~Jong}\ \emph {et~al.}(2023)\citenamefont
  {de~Jong}, \citenamefont {Prosko}, \citenamefont {Han}, \citenamefont
  {Malinowski}, \citenamefont {Liu}, \citenamefont {Kouwenhoven},\ and\
  \citenamefont {Pfaff}}]{dejong2023controllable}%
  \BibitemOpen
  \bibfield  {author} {\bibinfo {author} {\bibfnamefont {D.}~\bibnamefont
  {de~Jong}}, \bibinfo {author} {\bibfnamefont {C.~G.}\ \bibnamefont {Prosko}},
  \bibinfo {author} {\bibfnamefont {L.}~\bibnamefont {Han}}, \bibinfo {author}
  {\bibfnamefont {F.~K.}\ \bibnamefont {Malinowski}}, \bibinfo {author}
  {\bibfnamefont {Y.}~\bibnamefont {Liu}}, \bibinfo {author} {\bibfnamefont
  {L.~P.}\ \bibnamefont {Kouwenhoven}},\ and\ \bibinfo {author} {\bibfnamefont
  {W.}~\bibnamefont {Pfaff}},\ }\bibfield  {title} {\bibinfo {title}
  {{Controllable Single Cooper Pair Splitting in Hybrid Quantum Dot Systems}},\
  }\href {https://doi.org/10.1103/PhysRevLett.131.157001} {\bibfield  {journal}
  {\bibinfo  {journal} {Phys. Rev. Lett.}\ }\textbf {\bibinfo {volume} {131}},\
  \bibinfo {pages} {157001} (\bibinfo {year} {2023})}\BibitemShut {NoStop}%
\bibitem [{\citenamefont {Bordin}\ \emph {et~al.}(2023)\citenamefont {Bordin},
  \citenamefont {Wang}, \citenamefont {Liu}, \citenamefont {ten Haaf},
  \citenamefont {van Loo}, \citenamefont {Mazur}, \citenamefont {Xu},
  \citenamefont {van Driel}, \citenamefont {Zatelli}, \citenamefont
  {Gazibegovic}, \citenamefont {Badawy}, \citenamefont {Bakkers}, \citenamefont
  {Wimmer}, \citenamefont {Kouwenhoven},\ and\ \citenamefont
  {Dvir}}]{bordin2023tunable}%
  \BibitemOpen
  \bibfield  {author} {\bibinfo {author} {\bibfnamefont {A.}~\bibnamefont
  {Bordin}}, \bibinfo {author} {\bibfnamefont {G.}~\bibnamefont {Wang}},
  \bibinfo {author} {\bibfnamefont {C.-X.}\ \bibnamefont {Liu}}, \bibinfo
  {author} {\bibfnamefont {S.~L.~D.}\ \bibnamefont {ten Haaf}}, \bibinfo
  {author} {\bibfnamefont {N.}~\bibnamefont {van Loo}}, \bibinfo {author}
  {\bibfnamefont {G.~P.}\ \bibnamefont {Mazur}}, \bibinfo {author}
  {\bibfnamefont {D.}~\bibnamefont {Xu}}, \bibinfo {author} {\bibfnamefont
  {D.}~\bibnamefont {van Driel}}, \bibinfo {author} {\bibfnamefont
  {F.}~\bibnamefont {Zatelli}}, \bibinfo {author} {\bibfnamefont
  {S.}~\bibnamefont {Gazibegovic}}, \bibinfo {author} {\bibfnamefont
  {G.}~\bibnamefont {Badawy}}, \bibinfo {author} {\bibfnamefont {E.~P. A.~M.}\
  \bibnamefont {Bakkers}}, \bibinfo {author} {\bibfnamefont {M.}~\bibnamefont
  {Wimmer}}, \bibinfo {author} {\bibfnamefont {L.~P.}\ \bibnamefont
  {Kouwenhoven}},\ and\ \bibinfo {author} {\bibfnamefont {T.}~\bibnamefont
  {Dvir}},\ }\bibfield  {title} {\bibinfo {title} {{Tunable Crossed Andreev
  Reflection and Elastic Cotunneling in Hybrid Nanowires}},\ }\href
  {https://doi.org/10.1103/PhysRevX.13.031031} {\bibfield  {journal} {\bibinfo
  {journal} {Phys. Rev. X}\ }\textbf {\bibinfo {volume} {13}},\ \bibinfo
  {pages} {031031} (\bibinfo {year} {2023})}\BibitemShut {NoStop}%
\bibitem [{\citenamefont {Wang}\ \emph {et~al.}(2023)\citenamefont {Wang},
  \citenamefont {Ten~Haaf}, \citenamefont {Kulesh}, \citenamefont {Xiao},
  \citenamefont {Thomas}, \citenamefont {Manfra},\ and\ \citenamefont
  {Goswami}}]{wang2023triplet}%
  \BibitemOpen
  \bibfield  {author} {\bibinfo {author} {\bibfnamefont {Q.}~\bibnamefont
  {Wang}}, \bibinfo {author} {\bibfnamefont {S.~L.}\ \bibnamefont {Ten~Haaf}},
  \bibinfo {author} {\bibfnamefont {I.}~\bibnamefont {Kulesh}}, \bibinfo
  {author} {\bibfnamefont {D.}~\bibnamefont {Xiao}}, \bibinfo {author}
  {\bibfnamefont {C.}~\bibnamefont {Thomas}}, \bibinfo {author} {\bibfnamefont
  {M.~J.}\ \bibnamefont {Manfra}},\ and\ \bibinfo {author} {\bibfnamefont
  {S.}~\bibnamefont {Goswami}},\ }\bibfield  {title} {\bibinfo {title}
  {{Triplet correlations in Cooper pair splitters realized in a two-dimensional
  electron gas}},\ }\href {https://doi.org/10.1038/s41467-023-40551-z}
  {\bibfield  {journal} {\bibinfo  {journal} {Nature Communications}\ }\textbf
  {\bibinfo {volume} {14}},\ \bibinfo {pages} {4876} (\bibinfo {year}
  {2023})}\BibitemShut {NoStop}%
\bibitem [{\citenamefont {Bordin}\ \emph
  {et~al.}(2024{\natexlab{b}})\citenamefont {Bordin}, \citenamefont {Li},
  \citenamefont {van Driel}, \citenamefont {Wolff}, \citenamefont {Wang},
  \citenamefont {ten Haaf}, \citenamefont {Wang}, \citenamefont {van Loo},
  \citenamefont {Kouwenhoven},\ and\ \citenamefont {Dvir}}]{bordin2023crossed}%
  \BibitemOpen
  \bibfield  {author} {\bibinfo {author} {\bibfnamefont {A.}~\bibnamefont
  {Bordin}}, \bibinfo {author} {\bibfnamefont {X.}~\bibnamefont {Li}}, \bibinfo
  {author} {\bibfnamefont {D.}~\bibnamefont {van Driel}}, \bibinfo {author}
  {\bibfnamefont {J.~C.}\ \bibnamefont {Wolff}}, \bibinfo {author}
  {\bibfnamefont {Q.}~\bibnamefont {Wang}}, \bibinfo {author} {\bibfnamefont
  {S.~L.~D.}\ \bibnamefont {ten Haaf}}, \bibinfo {author} {\bibfnamefont
  {G.}~\bibnamefont {Wang}}, \bibinfo {author} {\bibfnamefont {N.}~\bibnamefont
  {van Loo}}, \bibinfo {author} {\bibfnamefont {L.~P.}\ \bibnamefont
  {Kouwenhoven}},\ and\ \bibinfo {author} {\bibfnamefont {T.}~\bibnamefont
  {Dvir}},\ }\bibfield  {title} {\bibinfo {title} {{Crossed Andreev Reflection
  and Elastic Cotunneling in Three Quantum Dots Coupled by Superconductors}},\
  }\href {https://doi.org/10.1103/PhysRevLett.132.056602} {\bibfield  {journal}
  {\bibinfo  {journal} {Phys. Rev. Lett.}\ }\textbf {\bibinfo {volume} {132}},\
  \bibinfo {pages} {056602} (\bibinfo {year} {2024}{\natexlab{b}})}\BibitemShut
  {NoStop}%
\bibitem [{\citenamefont {Valentini}\ \emph {et~al.}(2024)\citenamefont
  {Valentini}, \citenamefont {Souto}, \citenamefont {Borovkov}, \citenamefont
  {Krogstrup}, \citenamefont {Meir}, \citenamefont {Leijnse}, \citenamefont
  {Danon},\ and\ \citenamefont {Katsaros}}]{valentini2024subgap}%
  \BibitemOpen
  \bibfield  {author} {\bibinfo {author} {\bibfnamefont {M.}~\bibnamefont
  {Valentini}}, \bibinfo {author} {\bibfnamefont {R.~S.}\ \bibnamefont
  {Souto}}, \bibinfo {author} {\bibfnamefont {M.}~\bibnamefont {Borovkov}},
  \bibinfo {author} {\bibfnamefont {P.}~\bibnamefont {Krogstrup}}, \bibinfo
  {author} {\bibfnamefont {Y.}~\bibnamefont {Meir}}, \bibinfo {author}
  {\bibfnamefont {M.}~\bibnamefont {Leijnse}}, \bibinfo {author} {\bibfnamefont
  {J.}~\bibnamefont {Danon}},\ and\ \bibinfo {author} {\bibfnamefont
  {G.}~\bibnamefont {Katsaros}},\ }\href {https://arxiv.org/abs/2407.05195}
  {\bibinfo {title} {{Subgap-state-mediated transport in
  superconductor--semiconductor hybrid islands: Weak and strong coupling
  regimes}}} (\bibinfo {year} {2024}),\ \Eprint
  {https://arxiv.org/abs/2407.05195} {arXiv:2407.05195 [cond-mat.mes-hall]}
  \BibitemShut {NoStop}%
\bibitem [{\citenamefont {Groth}\ \emph {et~al.}(2014)\citenamefont {Groth},
  \citenamefont {Wimmer}, \citenamefont {Akhmerov},\ and\ \citenamefont
  {Waintal}}]{groth2014kwant}%
  \BibitemOpen
  \bibfield  {author} {\bibinfo {author} {\bibfnamefont {C.~W.}\ \bibnamefont
  {Groth}}, \bibinfo {author} {\bibfnamefont {M.}~\bibnamefont {Wimmer}},
  \bibinfo {author} {\bibfnamefont {A.~R.}\ \bibnamefont {Akhmerov}},\ and\
  \bibinfo {author} {\bibfnamefont {X.}~\bibnamefont {Waintal}},\ }\bibfield
  {title} {\bibinfo {title} {{Kwant: a software package for quantum
  transport}},\ }\href {https://doi.org/10.1088/1367-2630/16/6/063065}
  {\bibfield  {journal} {\bibinfo  {journal} {New Journal of Physics}\ }\textbf
  {\bibinfo {volume} {16}},\ \bibinfo {pages} {063065} (\bibinfo {year}
  {2014})}\BibitemShut {NoStop}%
\bibitem [{\citenamefont {Legg}\ \emph {et~al.}(2024)\citenamefont {Legg},
  \citenamefont {Hess}, \citenamefont {Loss},\ and\ \citenamefont
  {Klinovaja}}]{legg2024reply}%
  \BibitemOpen
  \bibfield  {author} {\bibinfo {author} {\bibfnamefont {H.~F.}\ \bibnamefont
  {Legg}}, \bibinfo {author} {\bibfnamefont {R.}~\bibnamefont {Hess}}, \bibinfo
  {author} {\bibfnamefont {D.}~\bibnamefont {Loss}},\ and\ \bibinfo {author}
  {\bibfnamefont {J.}~\bibnamefont {Klinovaja}},\ }\bibfield  {title} {\bibinfo
  {title} {{Reply to Antipov et al., Microsoft Quantum: "Comment on Hess et al.
  Phys. Rev. Lett. 130, 207001 (2023)"}},\ }\href
  {https://arxiv.org/abs/2308.10669} {\  (\bibinfo {year} {2024})},\ \Eprint
  {https://arxiv.org/abs/2308.10669} {arXiv:2308.10669 [cond-mat.mes-hall]}
  \BibitemShut {NoStop}%
\bibitem [{\citenamefont {White}(1992)}]{steven1992density}%
  \BibitemOpen
  \bibfield  {author} {\bibinfo {author} {\bibfnamefont {S.~R.}\ \bibnamefont
  {White}},\ }\bibfield  {title} {\bibinfo {title} {{Density matrix formulation
  for quantum renormalization groups}},\ }\href
  {https://doi.org/10.1103/PhysRevLett.69.2863} {\bibfield  {journal} {\bibinfo
   {journal} {Phys. Rev. Lett.}\ }\textbf {\bibinfo {volume} {69}},\ \bibinfo
  {pages} {2863} (\bibinfo {year} {1992})}\BibitemShut {NoStop}%
\bibitem [{\citenamefont {White}(1993)}]{steven1993density}%
  \BibitemOpen
  \bibfield  {author} {\bibinfo {author} {\bibfnamefont {S.~R.}\ \bibnamefont
  {White}},\ }\bibfield  {title} {\bibinfo {title} {{Density-matrix algorithms
  for quantum renormalization groups}},\ }\href
  {https://doi.org/10.1103/PhysRevB.48.10345} {\bibfield  {journal} {\bibinfo
  {journal} {Phys. Rev. B}\ }\textbf {\bibinfo {volume} {48}},\ \bibinfo
  {pages} {10345} (\bibinfo {year} {1993})}\BibitemShut {NoStop}%
\bibitem [{\citenamefont {Schollw\"ock}(2005)}]{schollwock2005density}%
  \BibitemOpen
  \bibfield  {author} {\bibinfo {author} {\bibfnamefont {U.}~\bibnamefont
  {Schollw\"ock}},\ }\bibfield  {title} {\bibinfo {title} {{The density-matrix
  renormalization group}},\ }\href {https://doi.org/10.1103/RevModPhys.77.259}
  {\bibfield  {journal} {\bibinfo  {journal} {Rev. Mod. Phys.}\ }\textbf
  {\bibinfo {volume} {77}},\ \bibinfo {pages} {259} (\bibinfo {year}
  {2005})}\BibitemShut {NoStop}%
\bibitem [{\citenamefont {Chiu}\ \emph {et~al.}(2016)\citenamefont {Chiu},
  \citenamefont {Teo}, \citenamefont {Schnyder},\ and\ \citenamefont
  {Ryu}}]{chiu2016classification}%
  \BibitemOpen
  \bibfield  {author} {\bibinfo {author} {\bibfnamefont {C.-K.}\ \bibnamefont
  {Chiu}}, \bibinfo {author} {\bibfnamefont {J.~C.~Y.}\ \bibnamefont {Teo}},
  \bibinfo {author} {\bibfnamefont {A.~P.}\ \bibnamefont {Schnyder}},\ and\
  \bibinfo {author} {\bibfnamefont {S.}~\bibnamefont {Ryu}},\ }\bibfield
  {title} {\bibinfo {title} {{Classification of topological quantum matter with
  symmetries}},\ }\href {https://doi.org/10.1103/RevModPhys.88.035005}
  {\bibfield  {journal} {\bibinfo  {journal} {Rev. Mod. Phys.}\ }\textbf
  {\bibinfo {volume} {88}},\ \bibinfo {pages} {035005} (\bibinfo {year}
  {2016})}\BibitemShut {NoStop}%
\bibitem [{\citenamefont {Fishman}\ \emph {et~al.}(2022)\citenamefont
  {Fishman}, \citenamefont {White},\ and\ \citenamefont
  {Stoudenmire}}]{ITensor}%
  \BibitemOpen
  \bibfield  {author} {\bibinfo {author} {\bibfnamefont {M.}~\bibnamefont
  {Fishman}}, \bibinfo {author} {\bibfnamefont {S.~R.}\ \bibnamefont {White}},\
  and\ \bibinfo {author} {\bibfnamefont {E.~M.}\ \bibnamefont {Stoudenmire}},\
  }\bibfield  {title} {\bibinfo {title} {{The ITensor Software Library for
  Tensor Network Calculations}},\ }\href
  {https://doi.org/10.21468/SciPostPhysCodeb.4} {\bibfield  {journal} {\bibinfo
   {journal} {SciPost Phys. Codebases}\ ,\ \bibinfo {pages} {4}} (\bibinfo
  {year} {2022})}\BibitemShut {NoStop}%
\end{thebibliography}%

\end{document}